\documentclass{aa}  
\usepackage{xcolor}
\usepackage{graphicx}
\begin{document} 

\title{Jellyfish galaxy candidates in MACS~J0717.5+3745 and 39 other clusters of the DAFT/FADA and CLASH surveys}
\titlerunning{Jellyfish galaxy candidates in medium redshift clusters}

%\subtitle{}             

\author{F. Durret\inst{1}
          \and
          S. Chiche\inst{1} 
          \and
          C. Lobo\inst{2,3}
          \and 
          M. Jauzac \inst{4,5,6,7}
\thanks{Based on archive data of the Hubble Space Telescope. This paper has made use of the NASA Extragalactic Database
and of the GAZPAR tool to apply the LePhare software. }          }
          
\institute{Sorbonne Universit\'e, CNRS, UMR 7095, Institut d'Astrophysique de Paris, 98bis Bd Arago, 75014, Paris, France
         \and
     Instituto de Astrof\'{\i}sica e Ci\^encias do Espa\c co, Universidade do Porto, CAUP, Rua das Estrelas, PT4150-762 Porto, Portugal
            \and
        Departamento de F\'{\i}sica e Astronomia, Faculdade de Ci\^encias, Universidade do Porto, Rua do Campo Alegre 687, PT4169-007 Porto, Portugal
    \and
Centre for Extragalactic Astronomy, Durham University, South Road, Durham DH1 3LE, UK
    \and
Institute for Computational Cosmology, Durham University, South Road, Durham DH1 3LE, UK
\and 
Astrophysics Research Centre, University of KwaZulu-Natal, Westville Campus, Durban 4041, South Africa
\and
School of Mathematics, Statistics \& Computer Science, University of KwaZulu-Natal, Westville Campus, Durban 4041, South Africa
   }     
             
\date{}

% \abstract{}{}{}{}{} 
% 5 {} token are mandatory
  \abstract
   {Galaxies in clusters undergo several phenomena, such as ram pressure stripping and tidal interactions, that can trigger or quench their star formation and, in some cases, lead to galaxies acquiring unusual shapes and long tails - some become jellyfish.}
   {We searched for jellyfish galaxy candidates in a sample of 40 clusters from the DAFT/FADA and CLASH surveys covering the redshift range $0.2<z<0.9$. In MACS~J0717.5+3745 (MACS0717), our large spatial coverage and abundant sampling of spectroscopic redshifts allowed us to pursue a detailed analysis of jellyfish galaxy candidates in this cluster and its extended filament.}
   {We retrieved galaxy spectroscopic redshifts in the NASA Extragalactic Database for galaxies in all the clusters of our sample (except for MACS0717 for which we had an extensive catalogue),
   looked at the Hubble Space Telescope ACS images of these objects (mainly the F606W and F814W bands), and classified them as a function of their likeliness to be jellyfish galaxies. We give catalogues of jellyfish galaxy candidates with positions, redshifts, magnitudes, and projected distance to their respective cluster centre. For MACS0717, an eight-magnitude optical and infrared catalogue covering the entire region allowed us to compute the best stellar population fits with LePhare through the GAZPAR interface. For the 31 jellyfish candidates in the other clusters belonging to the CLASH survey, we extracted up to 17 magnitudes available in the CLASH catalogues to fit their spectral energy distribution in the same way.}
   {We found 81 jellyfish galaxy candidates in the extended region around MACS0717 as well as 97 in 22 other clusters. Jellyfish galaxy candidates in MACS0717 tend to avoid the densest regions of the cluster, while this does not appear to be the case in the other clusters. 
  The best fit templates found by LePhare show that star formation is occurring. 
   Stellar masses are in the range $10^9-10^{11}$~M$_\odot$, and the star formation rates (SFRs) are in the $10^{-1}-60$~M$_\odot$~yr$^{-1}$ range for MACS0717 and in the 
   $10^{-1}-10$~M$_\odot$~yr$^{-1}$ range for the other sample. Specific star formation rates (sSFR) are notably higher in MACS0717, with more than half of the sample having values larger than $10^{-9}$~yr$^{-1}$, while in the other clusters, most galaxies have sSFR$<10^{-10}$~yr$^{-1}$. Stellar populations appear younger in MACS0717 (more than half have an age smaller than $1.5\times 10^9$~yrs), and following mid-infrared criteria two galaxies may contain an AGN. In a SFR versus stellar mass diagram, jellyfish galaxy candidates appear to have  somewhat larger SFRs than ``non-jellyfish star forming'' galaxies. For MACS0717, the mean sSFR of the 79 jellyfish galaxy candidates is 3.2 times larger than that of star-forming non jellyfish galaxies (selected with log(sSFR)$\geq -11$). } 
   {Our jellyfish galaxy candidates are star forming objects, with young ages and blue colours. Based on several arguments, the jellyfish candidates identified in MACS0717 seem to have fallen rather recently into the cluster. A very rough estimate of the proportions of jellyfish galaxies in the studied clusters is about 10\%, and this number does not seem to vary strongly with the cluster relaxation state, though this result must be confirmed with more data.
   Our sample of 97 galaxies in 22 clusters represents the basis of future works.
   }

               
\maketitle

\section{Introduction}
\label{sec:intro}

Due to the high density of galaxies found in clusters and to the presence of hot X-ray emitting gas, cluster galaxies are subject to environmental mechanisms that do not affect significantly their field counterparts. The most important ones are ram pressure stripping \citep{GunnGott72}, hereafter RPS, which affects the gas contained in galaxies, and tidal effects which affect both gas and stars and often lead to harrassment \citep{Moore+96}. A nice summary of all the physical processes taking place in clusters can be found in e.g. the introduction of \citet{Poggianti+17} and will not be repeated here. We will here concentrate on the description of observations performed these last 15 years by various teams.
Galaxies with unusual shapes and star forming properties have been found in many galaxy clusters, mostly at optical wavelengths, but with some features (mainly long tails) that are also sometimes detected in X-rays and/or at radio wavelengths. The number and depth of analyses of such objects have been increasing tremendously these last few years, in particular with the advent of VLT/MUSE (see below). To our knowledge, the first one to name some of these objects « jellyfish galaxies » was \citet{Bekki09}.

\citet{Owen+06} studied the very rich merging cluster Abell~2125 (z=0.247) at several wavelengths, and found galaxy C153 showing an X-ray plume with [OII] emission in knots, interpreted as due to ram pressure stripping. The spectral energy distribution (SED) of C153 shows that it has undergone continuous star formation for 3.5 Gyr, further supporting that it can be considered a prototype jellyfish galaxy, even if not named as such. 
\citet{Sun+07} found ESO137-001 in the massive cluster Abell~3627 (z=0.01625) with a 40 kpc H$\alpha$ tail coinciding with a 70 kpc X-ray tail. 
The H$\alpha$ emission of the galaxy itself is sharply truncated, and 49 emission line knots are distributed  along  the tail over 39 kpc. These authors attribute the origin of the tail to ram pressure stripping, and mention that heat conduction may contribute to the energy of the optical lines. 
\citet{Cortese+07} found two peculiar galaxies falling into the massive galaxy clusters Abell 1689 (z = 0.18) and Abell 2667 (z = 0.23). Their Hubble Space Telescope ({\it HST}) images show extraordinary trails composed of bright blue knots (with absolute magnitudes in the range 
$-16.5<$ M $<-11.5$) and stellar streams associated with each of these systems, one of them experiencing a strong burst of star formation (SF) while the other has recently ceased its SF activity. They interpret these results as due to the combined action of tidal interaction with the cluster potential and ram pressure stripping.
\citet{Yoshida+08} detected a string of "fireballs" (star-forming clouds with a linear stream of young stars extending towards the galaxy, detected in H$\alpha$) in the Coma cluster, "hanging" from galaxy RB~199. They showed that tidal effects alone could not account for the formation of such fireballs while the ram pressure stripping mechanism could provide a good explanation. 
In the optical and UV, \citet{Smith+10} observed 13 asymmetric star forming galaxies in the Coma cluster, due to star formation from the gas stripped from the galaxies by interaction with their environment, and long tails reaching 100 kpc. 
Also in Coma, \citet{Yagi+10} found extended H$\alpha$ clouds in 14 galaxies at the edges of the cluster, suggesting that the parent galaxies have a large velocity with respect to the Coma cluster. 

\citet{Rawle+14} analysed star formation in 53 galaxies of Abell~2744 (z=0.308), including some jellyfish galaxies. They found that the orientations of the trails, and of the material stripped from constituent galaxies, indicated that the passing shock front of the cluster merger was the trigger. 
\citet{Ebeling+14} searched for extreme cases of jellyfish galaxies at $z>0.3$ with {\it HST} Snapshots, and found six very bright objects with ${\rm M_{F606W}<-21}$. They proposed a classification of jellyfish galaxies, from J=1 (mildly affected) to J=5 (very strongly affected) that is now commonly used. This paper was followed by several others. \citet{McPartland+16} studied 63 MACS clusters, and found many more jellyfish galaxies (but with no measured redshifts), showing the presence of optical tails. Their comparison to a simple model showed that extreme RPS events are associated to cluster mergers rather than to infall along filaments, even though these do occur too. \citet{KalitaEbeling19} then detected a showcase jellyfish galaxy in Abell~1758N and analysed it in detail, with [OII] emission up to 40 kpc. \citet{EbelingKalita19} analysed the field of Abell~1758N (z=0.28) and detected 8 RPS jellyfish candidates undergoing intense SF. 

The first paper describing observations of jellyfish galaxies with VLT/MUSE was that of \citet{Fumagalli+14} on ESO~137-001 (z=0.01625). They detected a double tail reaching 80 kpc, seen previously in X-rays, and inferred that the galaxy is falling radially into the massive Norma cluster. A complementary study with APEX \citep{Jachym+14} had already uncovered an exceptionally long molecular tail in this galaxy; follow-up observations with ALMA 
\citep{Jachym+19} allowed detecting, for the first time, the molecular gas at the heads of several "fireballs" located in the complex tail structure of this spectacular galaxy.

\citet{Poggianti+16} published the first analysis of several hundred jellyfish galaxies at low redshift, based on the WINGS + OMEGAWINGS sample, selecting galaxies with various asymmetric/disturbed morphologies and knots, suggestive of triggered SF. This team then obtained a large observing program on VLT/MUSE : the GAs Stripping Phenomena survey  \citep[GASP,][]{Poggianti+17}. In this first paper of a long series, they show MUSE results on the massive galaxy JO206 (z=0.0513),
which is undergoing RPS in a poor cluster and shows a 90~kpc tail.
This paper has been followed by many others, out of which we only quote a few. \citet{Bellhouse+19} analysed the 94 kpc long tails of JO201 in Abell 85.   \citet{George+19} analysed the galaxy JO201 and showed that this galaxy, which is falling into Abell~85, is located close to the cluster centre, and undergoes both RPS and AGN feedback.  \citet{Radovich+19} analysed seven jellyfish galaxies and highlighted the importance of outflows. \citet{Poggianti+19} achieved a very complete study of JW100 with MUSE and also ALMA, VLA, UVIT, and {\it Chandra}, and studied the influence of gas stripping, gas heating and AGN. They propose that ISM heating due to interaction with the intracluster medium is responsible for the X-ray tail. 
\citet{Moretti+20} analysed ALMA observations of the jellyfish galaxy JW100, and detected a large amount of molecular gas, 30\% of which is located in the stripped gas tail out to 5 kpc from the galaxy center. They interpreted this molecular gas, which within the disk is totally displaced relatively to the stellar component, as newly born from stripped HI gas or recently condensed from stripped diﬀuse molecular gas. 

Another interesting result of the GASP survey, obtained to our knowledge for the first time, is that found by \citet{Vulcani+19}, who observed four field galaxies with increased SF and tattered H$\alpha$, making them appear similar to some of our jellyfish galaxies, although they are not members of any cluster. They attributed this increased SF to the effect of cosmic web filaments (none of the galaxies is in a cluster, three are in small groups, and all are embedded in a filament). 

At higher redshift, VLT/MUSE results were also obtained by \citet{Boselli+19}, who undertook a spectroscopic survey at redshift 0.25 < z < 0.85 in groups and clusters (the MAGIC survey: MUSE gAlaxy Groups In Cosmos, Epinat et al., in prep.). They detected two star-forming galaxies in the COSMOS cluster CGr32 at z = 0.73 with two extended (up to $\sim 100$ kpc in projected distance) tails of ionised gas without any stellar counterpart in the deep optical images. 

All these studies are devoted to the observations of rather small numbers of jellyfish galaxies per cluster. On the other hand, \citet{Roman+19} observed a large number of jellyfish galaxies (73, out of which they discarded three that were probably interacting galaxies) in a single zone: the multi-cluster system Abell 901/902 (z=0.165). In particular, they showed that the starburst phenomenon increases with jellyfish class. 

As far as numerical simulations on these specific objects are concerned, \citet{Bekki09} made hydrodynamical simulations to study the effect of RPS on galaxies in clusters and, to our knowledge, was the first to use the term « jellyfish ». His simulated galaxies look very much like the jellyfish galaxies that are observed.
More recently, \citet{Ruggiero+19} published hydrodynamical simulations to model the four structures observed in Abell~901/902 by \citet{Roman+19}. They showed that many (but not all) jellyfish galaxies are located in the vicinity of ram pressure boundaries, defined as regions where gas moving along each subcluster and gas from the remainder of the system meet. Galaxies become jellyfish when they cross a boundary within their parent subcluster, where a significant pressure increase takes place, due to the merging of the cluster gas and subcluster gas. A significant amount of jellyfish galaxies must be created by this mechanism. We can also mention the model by \citet{SafarzadehLoeb19} accounting for SF due to RPS. According to their results, jellyfish galaxies must be late infallers for their model to work, and they predict no jellyfish galaxies to be present at short clustercentric distances (smaller than (0.3-0.4) R$_{200}$, see their Fig . 3).

We present here a search for jellyfish galaxy candidates in {\it HST} images available for clusters of the 
DAFT/FADA\footnote{http://cesam.lam.fr/DAFT/index.php} and CLASH\footnote{https://archive.stsci.edu/prepds/clash/} surveys.
Though a number of such objects have been detected and thoroughly analysed these last years (as discussed before), the number of jellyfish galaxies at medium redshift is still limited, and our aim here is to  increase this number in the redshift range $0.2<z<0.9$. For the cluster MACS~J0717.5+3745 (hereafter MACS0717) we have a large spectroscopic redshift catalogue that allows us to search for jellyfish galaxies not only in the cluster core but also in its extended filament \citep{Jauzac+12,Jauzac+18a,Durret+16,Martinet+16}. This will allow a detailed study of the distribution of jellyfish galaxy candidates in this extended environment.
For the other clusters, our method will not allow us to make a statistical study since the redshift coverage of the clusters is by no means complete, but it is a
first step towards the study of these interesting objects, in particular those at relatively high redshift, and therefore closer to
the redshift of cluster formation. The list of new jellyfish galaxy candidates proposed here will hopefully be exploited later at various wavelengths by us or others.

The paper is organized as follows. We describe our initial sample of 40 clusters and the method we apply in Section~\ref{sec:sample}. We give our catalogue of 81 jellyfish candidates in the extended region of  MACS0717 in Section~\ref{sec:macs}, and 
discuss their spatial distribution and colour. The 97 jellyfish candidates in 22 other clusters (there are 17 clusters in which we found no jellyfish galaxy) 
are presented in Section~\ref{sec:jelly}. The spectral energy distribution and derived quantities (such as stellar mass, star formation rate, etc.) of all the jellyfish candidates are analysed in Section~\ref{sec:properties}. 
Finally, we summarise and discuss our results and propose  some conclusions in Sect.~\ref{sec:conclusion}.

All distances are computed with Ned Wright's calculator\footnote{http://www.astro.ucla.edu/\~wright/CosmoCalc.html},
assuming H$_0=70$~km~s$^{-1}$~Mpc$^{-1}$, $\Omega_\Lambda=0.7$ and
$\Omega _{\rm m}=0.3$. Magnitudes are quoted in AB system.

\section{Galaxy sample and identification of jellyfish candidates}
\label{sec:sample}

\subsection{Selection of jellyfish candidates}

We have considered 40 clusters from the DAFT/FADA and CLASH surveys,
which were all selected to be massive clusters (M$>2\times 10^{14}$~M$_\odot$ for DAFT/FADA, and kT$>5$~keV, corresponding to a mass in the
$(5-30)\times 10^{14}$~M$_\odot$ range, for CLASH). For all clusters except MACS0717, we retrieved
from NED\footnote{https://ned.ipac.caltech.edu/} all galaxy spectroscopic redshifts 
available in the regions covered by the {\it HST} images. The cluster list is given in
Table~\ref{tab:clusters}. The cluster redshift range covered is
$0.206\leq z \leq 0.890$. For MACS0717, the large spectroscopic redshift catalogue available covers not only the cluster core but its extended filament as well, so we dedicate an important part of this paper to its study. The redshift coverage in most clusters is quite homogeneous. 

For each cluster, we identify galaxies with spectroscopic redshifts in the range previously chosen to draw galaxy density maps \citep{Durret+16,Durret+19}, and roughly corresponding to an interval of $\pm 0.02$ around the corresponding cluster redshift. This translates into a velocity range indicated in Table~\ref{tab:clusters} for each cluster, expressed in units of the corresponding cluster velocity dispersion (computed in Section~\ref{sec:radiivel}). We can thus see that the redshift interval first chosen by \cite{Durret+16} is somewhat limited in some cases, and could make us miss fast moving galaxies. In particular for MACS0717, to which a large part of this paper is dedicated, this strategy would limit our analysis to a range of $\pm 2.2\sigma_v$ (see Fig.~5 of \citet{Durret+16}, initial value). We thus decided to extend the redshift interval of this cluster to $\pm 4\sigma_v$ (final value reported in the same Table), the sole system where we have a large enough spectroscopic and spatial coverage allowing for a more detailed analysis than what is possible for the remaining clusters.

Two of us (FD and SC) separately looked at each of the selected galaxies,
searching for objects that could be classified as jellyfish galaxy candidates,
based on several criteria: asymmetry, tidal arms, star trails. We independently
classified them between J=1 and J=5 according to their probability of
being a jellyfish, as suggested by \citet{Ebeling+14}: J=1 being
the smallest confidence index and J=5 the largest. In most cases, our classifications agreed within $\pm 1$, 
but we prefer to give both classifications, in Tables~\ref{tab:macs0717} and \ref{tab:galaxies}, 
to illustrate the relative difficulty of eye classification. We nonetheless favored eye classification since jellyfish galaxies cover a large variety of shapes, making it difficult to automatize their identification. 
Out of the 40 clusters considered, there were 17 in which we detected no jellyfish candidate. This is probably due to the fact that in some clusters we only have a small number of redshifts within the imaged area, and in those clusters none of the galaxies with a measured redshift entered this category. We are therefore left with a sample of jellyfish candidates in 22 cluster fields (besides MACS0717).

To visually identify jellyfish galaxies, we used {\it HST} images.
All clusters apart from Cl0152.7-1357 have data in the F814W filter.
Whenever possible, we also considered images in the F606W filter as well,
to compare the aspect of the galaxies in both filters.  Sometimes, F606W images were not available, so we
considered another filter, as indicated in Table~\ref{tab:galaxies}. When
possible, we show for each galaxy its images in two filters (with the
bluer image to the left and the redder to the right, at the same
scale, see Appendix). In some cases, fields covered by the two filters are not exactly 
the same, so even if a cluster is observed in two filters, an
individual galaxy may be found only in one image. In such cases, as well as for
clusters observed in a single band, only one image is shown.
For a very small number of cases, the astrometries of the two {\it HST} images
are slightly different, so images look a little displaced.

However, we must keep in mind that clusters in our sample cover a rather large redshift range, so 
the rest-frame wavelengths corresponding to the filters analysed are not all the same.
For MACS0717 (z=0.5458), the central wavelengths of F606W and F814W filters correspond to rest-frame wavelengths of 392 and 527~nm respectively. 
At the extreme redshifts of our sample, at z=0.2, the central wavelengths of F606W and F814W filters correspond to rest-frame wavelengths of 505 and 678~nm respectively, while at z=0.9 they correspond to rest frame wavelengths of 310 and 428~nm respectively.

The selection of jellyfish candidates in MAC0717 followed the same general procedure as outlined here but the dedicated catalogue of magnitudes and redshifts for this specific system introduced some differences that will be described in Section \ref{sec:macs}.
We must note that, except for MACS0717, our study relies on spectroscopic redshifts gathered from NED, which are not
complete in any way. We will therefore obtain some indications, but will not be able to
obtain statistically meaningful results, so this search is mainly a basis for future studies.

Jellyfish galaxies are mainly accounted for by hydrodynamical interactions (ram pressure stripping) with the hot intracluster gas, which causes various observable effects on the galaxy gas, such as compression of the leading edge of the galaxy, trailing tails, or even unwinding of spiral arms \citep{Bellhouse+21}. On the other hand, gravitational effects (tidal effects, harrassment) affect both gas and stars and can lead to galaxy shapes that can be reminiscent of those of jellyfish galaxies. It is therefore important to spot jellyfish candidates that may be undergoing tidal effects from a neighbour galaxy. For this, we examined all the images of jellyfish candidates and searched for galaxies located within a distance of 50~kpc. 
Out of the 90 (in MACS0717) and 103 (in the other clusters) jellyfish galaxy candidates that we initially identified, 9 (in MACS0717) and 6 (in other clusters) showed actual evidence for gravitational interaction
(tidal arms), so they were eliminated from our sample. Among the remaining ones, 5 jellyfish candidates (in MACS0717) and 18 (in the other clusters) had possible companions but without any signature of interaction with the jellyfish candidate. We kept these galaxies but mark them in Tables 2 and 3 with an asterisk. Our final sample therefore includes 81 jellyfish candidates in MACS0717, and 97 in the 22 other clusters.

\begin{table*}[ht]
\centering
\caption{Clusters studied, ordered by right ascension. Columns are:
  cluster name (where the subscript C indicates that the cluster comes
  from the CLASH survey, the other clusters belonging to the DAFT/FADA
  sample), coordinates, redshift, number of galaxies examined (i.e. galaxies in the cluster redshift range and found in the {\it HST} images that we analyzed) and
  number of jellyfish candidates. A zero in the sixth column means that none
  of the galaxies for which a spectroscopic redshift in the cluster range was available appeared to be a
  jellyfish candidate. For the clusters in which jellyfish candidates were found, we give in the last three columns the values of $r_{200}$, $\sigma_v$, and the velocity interval $\Delta v$ in which jellyfish candidates were searched in units of $\sigma_v$ (see text). The last three columns are empty for $[$MJM98$]$\_034, for which we did not find the information   }
\begin{tabular}{lrrrrrrrr}\hline
\hline
  Cluster                   &RA (J2000.0)&  Dec (J2000.0)& redshift& Obs. & Jelly & $r_{200}$ & $\sigma_v$ & $\Delta v/\sigma_v$ \\
                            & ~~(deg)    &  ~~(deg) &         &    gal.&    cand. & (kpc)  & (km/s) &\\
\hline
  Cl~0016+1609              &   4.64098 &  16.43796 & 0.5410  &  103 & 8   & 1820 &  1034 & $\pm 2.7$ \\
  Abell~209$^C$             &  22.97083 & -13.60944 & 0.2060  &  39  &    2   & 2410 &  1205& $\pm 2.6$ \\ 
  Cl~J0152.7-1357           &  28.17083 & -13.96250 & 0.8310  &  66  &    6   & 1670 &  1083& $\pm 2.4$     \\
  Abell~383$^C$             &  42.00833 &  -3.53750 & 0.1871  &  32 &    2   & 1960 &  973& $\pm 5.0$    \\
  MACS~J0416.1-2403$^C$     &  64.04125 & -24.06611 & 0.3960  &  205 &    8   & 2420 &   1313& $\pm 1.2$  \\
  MACS~J0429.6-0253$^C$     &  67.40000 &  -2.88556 & 0.3990  &  2   &    2   & 1730 &   928 & $\pm 4.1$  \\
  MACS~J0454.1-0300         &  73.54635 &  -3.01494 & 0.5500  &  92  &    5   & 2110 &   1205& $\pm 2.1$  \\
  MACS~J0717.5+3745$^C$     & 109.37886 &  37.75826 & 0.5458  &  632 &    81  & 2236 &   1288 & $\pm 4.0$  \\
  MACS~J0744.9+3927$^C$     & 116.21863 &  39.45759 & 0.6976  &  2   &    0   & &  &    \\
  Abell~611$^C$             & 120.24542 &  36.04722 & 0.2880  &  2   &    0   & &  &    \\
  Abell~0851                & 145.74167 &  46.98667 & 0.4069  &  102 &    11  & 1542 &   843 & $\pm 4.5$  \\
  LCDCS~0172                & 163.60083 & -11.77167 & 0.6972  &  45  &    9   &  870 &   526 & $\pm 2.4$  \\
  MACS~J1115.8+0129$^C$     & 168.96667 &   1.49861 & 0.3520  &  3   &    0   & &  &    \\
  MACS~J1149.5+2223$^C$     & 176.39622 &  22.40304 & 0.5444  &  106 &    4   & 2920 &   1665 & $\pm 1.4$  \\
  MACS~J1206.2-0847$^C$     & 181.55083 &  -8.80028 & 0.4400  &  64  &    3   & 2030 &   1109 & $\pm 2.9$  \\
  BMW-HRI\_J122657.3+333253$^C$& 186.74167 &   3.54836 & 0.8900  & 23 &    0    &&&   \\
  LCDCS~0541                & 188.12625 & -12.84344 & 0.5414  &  80 &    7   & 860 & 491 & $\pm 9.8$ \\
  MACS~J1311.0-0310$^C$     & 197.75792 &  -3.17667 & 0.4940  &  3   &    0   & &  &    \\
  ZwCl~1332.8+5043          & 203.58500 &  50.51806 & 0.6200  &  6         &    0   && &      \\
 $[$MJM98$]$\_034           & 203.80742 &  37.81564 & 0.3830  &  8      &    1   & &      \\
  LCDCS~0829$^C$            & 206.87750 & -11.75278 & 0.4510  &  35 &    1    & 1638 &  1475 & $\pm 1.1$  \\
  LCDCS~0853                & 208.54083 & -12.51708 & 0.7627  &  20 &    7    & 1590 &   987 & $\pm 1.0$ \\
  3C 295 CLUSTER             & 212.85167 &  52.21056 & 0.4600  &  30 &    8  & 1790 &    984 & $\pm 5.4$  \\
  MACS~J1423.8+2404$^C$     & 215.94860 &  24.07782 & 0.5431  &  7         &    0   && &      \\
  RX~J1524.6+0957           & 231.16792 &   9.96083 & 0.5160  &  2         &    0   & &&      \\
  RX~J1532.9+3021$^C$       & 233.22417 &  30.34944 & 0.3450  &  2         &    1   & 1630 &  860 & $\pm 4.8$   \\
  RCS~J1620.2+2929          & 245.05000 &  29.48333 & 0.8700  &  1         &    0   & &&      \\
  MACS~J1621.4+3810         & 245.35292 &  38.16889 & 0.4650  &  1         &    0   & & &     \\
  MS~1621.5+2640            & 245.89792 &  26.57028 & 0.4260  &  31  &    3   & 1718 &  941 & $\pm 2.3$  \\
  OC02~J1701+6412           & 255.34583 &  64.23583 & 0.4530  &  1         &    0   & &  &    \\
  RX~J1716.4+6708           & 259.20667 &  67.14167 & 0.8130  &  22        &    1   & 1685 &  1085 & $\pm 1.8$  \\
  MACS~J1720.2+3536$^C$     & 260.07000 &  35.60722 & 0.3913  &  2         &    0   & &   &   \\
  Abell~2261                & 260.61292 &  32.13389 & 0.2240  &  14        &    0   & &    &  \\
  NEP~0200                  & 269.33083 &  66.52528 & 0.6909  &  1         &    0   & &     & \\
  MACS~J1931.8-2634$^C$     & 292.95667 & -26.57611 & 0.3520  &  3         &    2   & 1930 &  1018 & $\pm 4.0$   \\
  MS~2053.7-0449            & 314.09083 &  -4.63083 & 0.5830  & 32         &    1   & 1620 &   952 & $\pm 2.2$  \\ 
  MACS~J2129.4-0741$^C$     & 322.35922 &  -7.69062 & 0.5889  &  2         &    0   & & &     \\
  MS~2137.3-2353$^C$        & 325.06333 & -23.66111 & 0.3130  &  2         &    0   & & &     \\
  RXC~J2248.7-4431$^C$      & 342.18125 & -44.52889 & 0.3475  &  42        &    5   & 2300 &   1215 & $\pm 3.0$  \\
  RX~J2328.8+1453           & 352.20792 &  14.88667 & 0.4970  &  3         &    0   & &  &    \\
  \hline
\end{tabular}
\label{tab:clusters}
\end{table*}

\subsection{Galaxy magnitudes}

In order to obtain magnitude measurements for our candidate jellyfish galaxies,
we proceeded as follows.

For all CLASH clusters except for MACS0717, we retrieved the corresponding catalogues from the CLASH website that contain up to 17 wavebands, between 225~nm and 1.6~$\mu$m. Some of these magnitudes are given in Table~\ref{tab:galaxies}, to help characterise the galaxies. The CLASH magnitudes are corrected for Galactic extinction, and can therefore be used for spectral energy distribution analyses straightforwardly (section \ref{subsubsec:SED_32}).

For galaxies from the DAFT/FADA survey that are not part of CLASH, we computed the zero points ${\rm ZP_{AB}}$ applying the {\it HST} formula:\footnote{http://www.stsci.edu/hst/instrumentation/acs/data-analysis/zeropoints}
$${\rm -2.5 * log10(PHOTFLAM) - 5 * log10 (PHOTPLAM)-2.408},$$
\noindent
where the PHOTFLAM and PHOTPLAM values were found in the image headers.  We then ran SExtractor \citep{Bertin96} on individual images
and retrieved the MAG\_AUTO magnitudes. The values given in Table~\ref{tab:clusters} are corrected for 
Galactic extinction computed from the E(B-V) maps by \citet{Schlegel98} multiplied by the R values given in Table~6 of \citet{Schlafly11}.
 
For MACS0717: the F606W and F814W magnitudes for the entire mosaic of images of MACS0717 were taken from the data by \citet{Martinet+17}
(we did not use the CLASH catalogue available in the F606W and F814W filters for this cluster
because it only covers the central region). We also used an eight-band ground-based optical and infrared catalogue for the whole zone covered by MACS0717 and its filament  with Subaru/SuprimeCam data in the B, V, R$_c$, I$_c$ and z bands, CFHT/MegaCam data in the u$^*$ band, and CFHT/WIRCAM data in the near-infrared J and K$_s$ bands from  \citet{Jauzac+12}. More details can be found in \citet{Ma+08,Ma+09}.

This eight magnitude catalog for MACS0717 as well as the 17 CLASH magnitudes for all the other CLASH clusters were used to fit the spectral energy distributions of the jellyfish galaxies, and analyse their main stellar populations, with LePhare \citep{Ilbert+06}, through the 
GAZPAR facility\footnote{https://gazpar.lam.fr/home} as reported in Sect.~\ref{sec:properties}. We also analyzed the stellar populations of non-jellyfish galaxies in MACS0717, in order to compare the properties of our jellyfish candidates with those of ``normal'' galaxies.

\subsection{Cluster radii and velocity dispersions}\label{sec:radiivel} 

In order to calculate the projected distances of jellyfish galaxy candidates in units linked to the cluster properties, we compute for each cluster its $r_{200}$ value, corresponding to the radius at which the cluster density is 200 times the mean density of the Universe. We did  this in several ways. For the seven clusters studied by \cite{Martinet+16}, we directly take $r_{200}$ from this paper.
For the other clusters, we compute $r_{200}$ from the $M_{200}$ mass.
Nine clusters have $M_{200}$ values in \cite{Umetsu+18}, and 
for the remaining clusters (except one, MJM98, for which we can not find a mass in the literature), we take the $M_{200}$  masses derived from X-ray masses by Chu et al. (A\&A submitted). 
We then calculate $r_{200}$ by applying the following formula (Biviano, private communication):
$$G \times M_{200} = 100 \times H(z)^2 \times r_{200}^3 $$
\noindent
where 
$$ H(z)=H_0 \times [\Omega_m (1+z)^3 + \Omega_\Lambda]^{(1/2)} $$
\noindent 
is the Hubble parameter at the cluster redshift, z, computed with the cosmological parameters given at the end of Section~1, and G is the gravitational constant.

We also compute the cluster velocity dispersions using equation (1) from \cite{Munari+13}:
$$\sigma_v  = 1090 \times [h(z) \times M_{200}]^{(1/3)}$$
\noindent
where $h(z) = H(z) / 100$, $M_{200}$ is expressed in units of $10^{15}$~M$_\odot$, and $\sigma_v$ is the unidimensional velocity dispersion in units of km/s.

The values of $r_{200}$ and $\sigma_v$ are given in Table~\ref{tab:clusters} for all clusters that have jellyfish candidates.

For each galaxy, we compute its velocity relative to the mean cluster velocity in units of $\sigma_v$, and give the corresponding values in Tables~\ref{tab:macs0717} and \ref{tab:galaxies}.

\section{Jellyfish candidates in MACS~J0717.5+3745 (z=0.5458)}
\label{sec:macs}

\subsection{The catalogue of jellyfish candidates in MACS~J0717.5+3745 }

\begin{table*}[h]
\centering
\setcounter{table}{1}
\caption{Jellyfish candidates in the large structure enclosing the cluster MACS~J0717.5+3745. Columns are: galaxy number, RA, DEC, redshift,
F606W and F814W magnitudes, jellyfish classifications S and F by two of the authors (SC and FD), projected distance to cluster centre in kpc and in units of r$_{200}$, and velocity relative to cluster centre divided by cluster velocity dispersion.}
\label{tab:macs0717}
\begin{tabular}{lrrrrrrrrrr}
\hline
\hline
  Galaxy number & RA & DEC & z & F606W & F814W & S & F & Dist. & Dist. & v/$\sigma _v$\\
                &    &     &   &       &       &   &     & (kpc)   & (r$_{200}$)& \\    
\hline
  1      &  109.28460 & 37.76579 & 0.5424 & 21.402 & 20.539 & 3 & 2 & 2453 & 1.097 & $-0.43$ \\
  2      &  109.29681 & 37.74413 & 0.5445 & 21.605 & 20.526 & 3 & 2 & 2178 & 0.974 & $-0.16$\\    
  3$^*$     &  109.30550 & 37.79056 & 0.5416 & 22.103 & 20.638 & 2 & 3 & 2121 & 0.948 & $-0.53$ \\ 
  4      &  109.33180 & 37.76147 & 0.5444 & 21.810 & 20.873 & 2 & 1 & 1364 & 0.610 & $-0.18$\\   
  5      &  109.33253 & 37.76822 & 0.5346 & 22.483 & 21.874 & 1 & 2 & 1372 & 0.613 & $-1.41$\\
6$^*$   &  109.33324 & 37.68276 & 0.5763 & 22.134 & 21.036 & 1 & 2 & 2134 & 0.955 & 3.76\\
  7      &  109.33848 & 37.80727 & 0.5350  & 26.239 & 21.237 & 2 & 3 &1692 & 0.757 & $-1.36$ \\ 
  8      &  109.35336 & 37.71206 & 0.5459 & 22.138 & 20.870 & 3 & 3 & 1320 & 0.590 & 0.01 \\
  9      &  109.35354 & 37.73134 & 0.5377 & -99.000 & 22.869 & 4 & 4 & 1022 & 0.457 & $-1.02$\\  
  10     &  109.37781 & 37.71374 & 0.5374 & 21.923 & 20.766 & 3 & 3  & 1007 & 0.450 & $-1.06$\\ 
11      &  109.37863 & 37.78915 & 0.5754 & 22.365 & 21.570 & 5 & 5  & 822 & 0.367 & 3.65\\ 
12     &  109.38213 & 37.72594 & 0.5315 & 22.126 & 21.245 & 4 & 4 & 710 & 0.317 & $-1.81$\\
13$^*$     &  109.38460 & 37.73306 & 0.5325 & 22.111 & 21.418 & 1 & 2 & 537 & 0.240 & $-1.68$ \\   
14    &  109.39410 & 37.79197 & 0.5757 & 22.878 & 22.084 & 2 & 2 & 840 & 0.376 & 3.68\\  
15     &  109.39528 & 37.84684 & 0.5348 & 21.759 & 20.512 & 1 & 1 & 2101 & 0.940 & $-1.39$\\ 
  16     &  109.39620 & 37.76458 & 0.5490  & 22.027 & 21.231 & 3 & 2 & 241 & 0.108 & 0.40\\ 
17     & 109.40541 & 37.70779 & 0.5290 & 22.269 & 21.554 & 1 & 1 & 1148 & 0.514 & $-2.13$ \\
  18     &  109.40749 & 37.61744 & 0.5456 & 22.097 & 21.340 & 4 & 4 & 3199 & 1.431 & $-0.03$\\   
  19     &  109.40773 & 37.62261 & 0.5459 & 21.055 & 20.251 & 2 & 2 & 3082 & 1.378 & 0.01\\    
  20$^*$     &  109.41479 & 37.71172 & 0.5395 & 21.618 & 20.502 & 2 & 4 & 1149 & 0.514 & $-0.79$\\  
  21     &  109.41522 & 37.72866 & 0.5395 & 22.709 & 21.719 & 3 & 2 & 835 & 0.373 & $-0.79$\\ 
  22     &  109.41720 & 37.70310 & 0.5420 & 22.270 & 21.353 & 4 & 4 & 1350 & 0.604 & $-0.48$ \\    
  23     &  109.41817 & 37.79837 & 0.5634 & 22.225 & 20.565 & 4 & 3 & 1168 & 0.522 & 2.18\\   
  24     &  109.42203 & 37.74784 & 0.5660  & 22.493 & 21.425 & 3 & 2 & 768 & 0.343 & 2.50\\ 
  25     &  109.42928 &  37.67791 & 0.5442 & 20.739 & 19.954 & 3 & 2 & 1992 & 0.891 & $-0.20$\\    
  26     &  109.42990 & 37.78072 & 0.5622 & 22.227 & 21.308 & 3 & 3 & 1068 & 0.478 & 2.04 \\ 
  27     &  109.43111 & 37.62285 & 0.5472 & 21.150 & 20.410 & 2 & 2 & 3189 & 1.426 & 0.18\\     
  28     &  109.43424 & 37.75050 & 0.5500   & 22.224 & 20.866 & 1 & 2 & 1005 & 0.449 & 0.52 \\ 
  29     &  109.43758 & 37.72152 & 0.5409 & 22.502 & 21.217 & 2 & 0 & 1330 & 0.595 & $-0.62$\\ 
  30     &  109.43866 & 37.79811 & 0.5384 & 21.801 & 20.199 & 1 & 0 & 1472 & 0.658 & $-0.93$\\ 
  31     &  109.44090 & 37.77155 & 0.5521 & 22.692 & 21.920 & 3 & 1 & 1209 & 0.540 & 0.79\\ 
  32     &  109.44299 & 37.72516 & 0.5429 & 20.538 & 19.928 & 1 & 0 & 1388 & 0.621 & $-0.36$\\ 
  33     &  109.44773 & 37.63008 & 0.5469 & 22.637 & 21.694 & 3 & 2 & 3168 & 1.417 & 0.14\\ 
  34     &  109.44783 & 37.68979 & 0.5577 & 21.773 & 21.141 & 2 & 3 & 2001 & 0.895 & 1.48 \\ 
  35     &  109.45184 & 37.61336 & 0.5475 & 22.227 & 21.117 & 2 & 1 & 3558 & 1.591 & 0.21\\   
  36     &  109.45511 & 37.64426 & 0.5404 & 21.156 & 20.171 & 2 & 2 & 2955 & 1.322 & $-0.68$\\
  37     &  109.45564 & 37.62384 & 0.5480  & 21.919 & 21.650 & 2 & 2 & 3375 & 1.510 & 0.28\\   
  38     &  109.45809 & 37.75248 & 0.5546 & 22.010 & 21.090 & 4 & 2 & 1548 & 0.692 & 1.10\\ 
  39     &  109.45857 & 37.61006 & 0.5511 & 21.694 & 20.937 & 5 & 4 & 3690 & 1.650 & 0.66\\ 
  40     &  109.46005 & 37.73063 & 0.5401 & 21.537 & 20.933 & 5 & 2 & 1692 & 0.757 & $-0.72$\\ 
  41     &  109.46016 & 37.58284 & 0.5439 & 21.075 & 20.242 & 4 & 1 & 4279 & 1.914 & $-0.24$ \\ 
  2     &  109.46469 & 37.61392 & 0.5495 & 22.426 & 21.308 & 3 & 3 & 3673 & 1.643 & 0.46\\  
  43     &  109.46571 & 37.79222 & 0.5515 & 22.613 & 21.842 & 2 & 2 & 1917 & 0.857 & 0.71\\ 
  44     &  109.46701 & 37.70718 & 0.5319 & 22.287 & 21.646 & 5 & 2 & 2075 & 0.928 & $-1.76$\\ 
  45     &  109.47195 & 37.61126 & 0.54703 & 21.526 & 20.278 & 1 & 0 & 3806 & 1.702 & 0.15 \\ 
  46     &  109.47451 & 37.70863 & 0.5311 & 21.922 & 20.994 & 3 & 2 & 2206 & 0.987 & $-1.86$\\ 
  47     &  109.47494 & 37.77757 & 0.5451 & 21.181 & 20.115 & 3 & 0 & 1999 & 0.894 & $-0.09$ \\ 
  48     &  109.47574 & 37.71760 & 0.5440  & 22.315 & 19.658 & 4 & 2 & 2139 & 0.956  & $-0.23$\\   
  49     &  109.48136 & 37.74231 & 0.5637 & 21.287 & 20.237 & 2 & 0 & 2104 & 0.941 & 2.22 \\ 
  50     &  109.48801 & 37.72609 & 0.5362 & 22.385 & 21.872 & 4 & 5 & 2335 & 1.044 & $-1.21$\\ 
  51     &  109.48879 & 37.55154 & 0.5293 & 22.061 & 21.185 & 4 & 3 & 5204 & 2.327 & $-2.09$\\ 
  52     &  109.49406 & 37.56277 & 0.55338 & 22.452 & 21.247 & 2 & 0 & 5028 & 2.249 & 0.95 \\ 
  53     &  109.49385 & 37.64037 & 0.5461 & 22.405 & 21.182 & 2 & 3 & 3553 & 1.589 & 0.04\\ 
  54     &  109.49413 & 37.59956 & 0.5464 & 22.503 & 20.107 & 5 & 2 & 4302 & 1.924 & 0.07 \\ 
55*   & 109.50267 & 37.65858 & 0.5772 & 22.103 & 21.021 & 4 & 3 & 3404 & 1.522 & 3.86\\ 
56     &  109.50533 & 37.63641 & 0.5412 & 20.762 & 19.552 & 3 & 1 & 3800 & 1.699 & $-0.58$ \\
\hline
\hline
\end{tabular}
\end{table*}

\begin{table*}[ht]
\centering
\setcounter{table}{1}
  \caption{Continued.}
\begin{tabular}{lrrrrrrrrrr}
  \hline
  \hline
  57     &  109.50554 & 37.61765 & 0.5426 & 22.114 & 21.392 & 3 & 3 & 4125 & 1.845 & $-0.40$ \\   
  58     &  109.50652 & 37.67290 & 0.5430  & 21.256 & 20.058 & 1 & 0 & 3269 & 1.462 & $-0.35$\\
  59     &  109.50808 & 37.64077 & 0.5420  & 21.681 & 20.739 & 3 & 2 & 3773 & 1.687 & $-0.48$ \\    
  60     &  109.50873 & 37.62831 & 0.5425 & 23.489 & 21.638 & 2 & 2 & 3989 & 1.784 & $-0.41$ \\ 
  61     &  109.50883 & 37.63732 & 0.5481 & 21.519 & 20.357 & 2 & 2 & 3841 & 1.718 & 0.29\\ 
  62     &  109.52255 & 37.55546 & 0.5465 & 22.372 & 21.748 & 4 & 2 & 5508 & 2.463 & 0.09\\
  63     &  109.53121 & 37.63280 & 0.5421 & 22.468 & 21.475 & 4 & 4 & 4288 & 1.918 & $-0.47$\\
  64     &  109.53217 & 37.58715 & 0.5447 & 21.959 & 21.064 & 5 & 4 & 5055 & 2.261 & $-0.14$\\
  65     &  109.53279 & 37.55027 & 0.5495 & 21.663 & 20.238 & 1 & 0 & 5739 & 2.567 & 0.46 \\ 
  66$^*$     &  109.53319 & 37.64305 & 0.5478 & 21.084 & 20.302 & 3 & 2 & 4172 & 1.866 & 0.25\\  
  67     &  109.53738 & 37.57487 & 0.5513 & 21.634 & 20.531 & 5 & 2 & 5349 & 2.392 & 0.69\\   
  68     &  109.53812 & 37.68567 & 0.5422 & 23.393 & 23.081 & 2 & 2 & 3749 & 1.676 & $-0.45$\\ %
  69     &  109.54137 & 37.70054 & 0.5419 & 21.532 & 20.190 & 1 & 0 & 3685 & 1.648 & $-0.49$\\ 
70   & 109.55122 & 37.64091 & 0.5701 & 22.283 & 21.443 & 3 & 2 & 4533 & 2.027 & 3.00\\
71     &  109.55586 & 37.57695 & 0.5500   & 21.673 & 20.747 & 3 & 1 & 5592 & 2.501 & 0.52 \\
  72$^*$     &  109.55758 & 37.68818 & 0.5419 & 22.206 & 21.046 & 3 & 2 & 4135 & 1.849 & $-0.49$\\  
  73     &  109.56507 & 37.57967 & 0.5488 & 21.814 & 20.612 & 4 & 2 & 5693 & 2.546 & 0.38\\
  74     &  109.57690 & 37.67111 & 0.5236 & 22.552 & 21.855 & 2 & 3 & 4698 & 2.101 & $-2.82$\\    
  75     &  109.58230 & 37.62237 & 0.5625 & 21.298 & 20.379 & 4 & 0 & 5363 & 2.398 & 2.07 \\
  76     &  109.58243 & 37.60096 & 0.5449 & 22.312 & 21.025 & 2 & 2 & 5661 & 2.532 & $-0.11$\\   
  77     &  109.58437 & 37.68565 & 0.5482 & 21.371 & 20.508 & 2 & 0 & 4732 & 2.116 & 0.30\\ 
  78     &  109.58499 & 37.61001 & 0.5489 & 22.352 & 21.587 & 5 & 4 & 5579 & 2.495 & 0.39\\   
  79     &  109.59241 & 37.65249 & 0.5378 & 24.508 & 21.100 & 4 & 3 & 5206 & 2.328 & $-1.01$\\
  80     &  109.60329 & 37.59384 & 0.5542 & 21.376 & 20.345 & 4 & 4 & 6139 & 2.746 & 1.05\\
  81     &  109.60480 & 37.64520 & 0.5225 & 22.826 & 22.159 & 3 & 2 & 5536 & 2.476 & $-2.96$\\
\hline
\hline
\end{tabular}
\end{table*}

MACS0717 is a well-known massive cluster with a large
extension/filament reaching a total of about 9 Mpc towards the south-east and studied by \citet{Jauzac+12,Jauzac+18b,Jauzac+18a} and \cite{Martinet+16}. Based on a weak lensing study, its mass was estimated to be $2.4\times 10^{15}$~M$_\odot$ within the R$_{200}$ radius by \citet{Martinet+16}. An 
extensive spectroscopic redshift catalogue, with 
646 galaxies in the redshift interval $0.5145\leq z \leq 0.5785$, allows us to search
for jellyfish galaxies in a very efficient way throughout the structure, 
which has been almost entirely covered by {\it HST/ACS} observations.

\begin{figure}[ht]
\begin{center}
  \includegraphics[width=8cm]{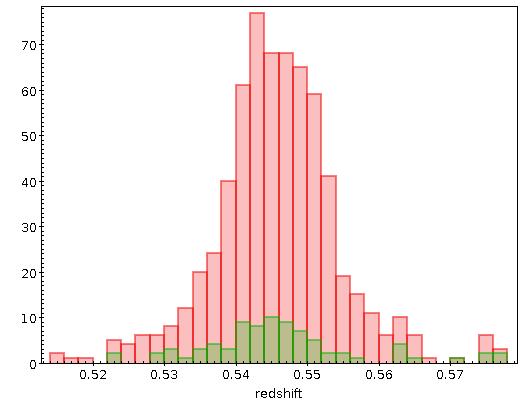} 
\end{center}
\caption{In red, redshift histogram of the 646 galaxies in all the MACS0717 region, and in green, redshift histogram of the 81 candidate jellyfish galaxies.}
\label{fig:macs0717_histoz}
\end{figure}

For this cluster, we examined 18 individual {\it HST/ACS} images in both
F606W and F814W bands.  Out of the 646 galaxies with redshifts in the cluster range,
81 were identified as jellyfish candidates.  The list of these galaxies
is given in Table~\ref{tab:macs0717}, and corresponding images are displayed in Appendix~\ref{appendix1}. The histograms of all redshifts available in the MACS0717 region and of the identified jellyfish galaxy candidates is shown in Fig.~\ref{fig:macs0717_histoz}. This plot seems to show that jellyfish candidates generally follow the velocity distribution of the cluster.

The jellyfish classification was separately
made by two of us, and the corresponding classes are given in the last
two columns (classS and classF) of Table~\ref{tab:macs0717}. It
appears that classF is often stricter than classS, so for some results shown below we made two samples of  ``strong probability'' jellyfish objects
(i.e. of types 3, 4 and 5), according to classF and to classS separately. 

\begin{figure}[ht]
\begin{center}
\includegraphics[width=8cm]{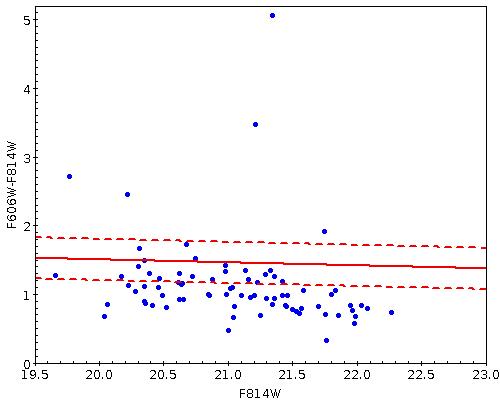} 
\end{center}
\caption{Colour-magnitude diagram for MACS0717. The blue points show the 81 candidate jellyfish galaxies belonging to the cluster. The full red line shows the position of the red sequence and the dashed lines correspond to $\pm 0.3$ on either side of the red sequence.}
\label{fig:macs0717_coulmag}
\end{figure}

The colour-magnitude diagram for MACS0717 is shown in Fig.~\ref{fig:macs0717_coulmag}.
The red sequence drawn from Subaru data in the V and I bands was: V-I=$-0.0436\times$I+2.75 \citep{Durret+16}, computed by considering the positions on this sequence of several tens of galaxies at the cluster redshift. Its width of $\pm 0.3$ was chosen to include all the galaxies belonging to the cluster according to their  spectroscopic redshift, as explained by \citet{Durret+16}.
We adapted this red sequence to the F606W and F814W filters using the transformations given by \citet{Fukugita+95} and the result is shown in Fig.~\ref{fig:macs0717_coulmag} (red lines). The data points refer to the jellyfish candidates found in the cluster.
We can see that most of them are blue and lie below the red sequence. For the four galaxies located notably above the red sequence, the fit of the SED by a stellar population model (see Sect.~\ref{subsubsec:SED_0717}) gives internal extinctions of 0.4 for three galaxies, and 0 for the fourth one. Thus, except for the last galaxy, their red colours may be at least partly explained by internal dust. 

\subsection{Spatial distribution of jellyfish candidates in MACS~J0717.5+3745}

\begin{figure*}[ht]
\begin{center}
   \includegraphics[width=12cm]{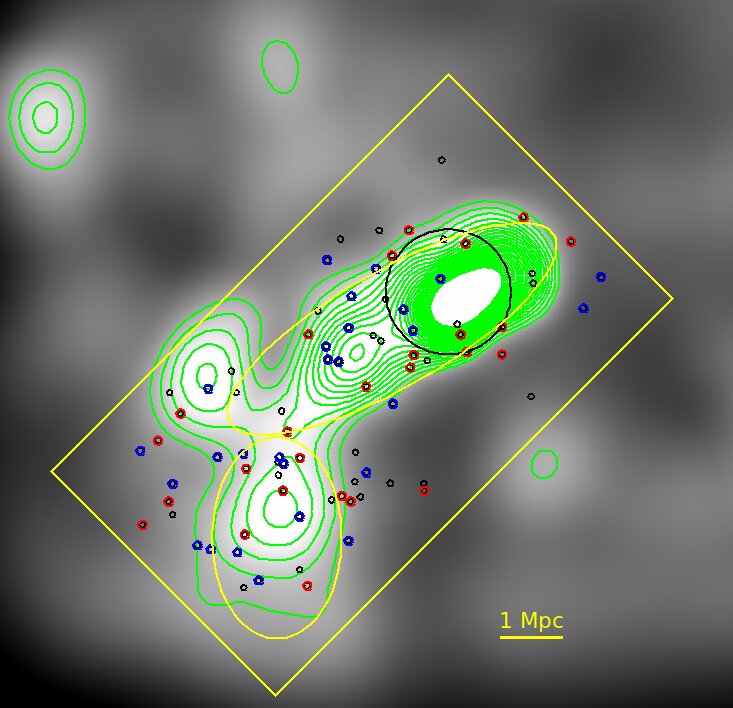} 
\end{center}
\caption{Grey shading and green isocontours show the density map of red sequence galaxies taken from Durret et al. (2016). The
  black circle is centred on the cluster centre, and has a 1~Mpc
  radius. The two yellow ellipses show the 3$\sigma$ contours of
  the density distribution. The positions of the candidate jellyfish galaxies
  are indicated as follows: red circles: galaxies classified as
  jellyfish types 3, 4, and 5 according to the strictest
  classification; blue: additional galaxies classified as types 3, 4, and 5 according
  to the less strict classification; black: all candidate jellyfish galaxies
  from Table~\ref{tab:macs0717}. The yellow rectangle shows the approximate {\it HST} coverage.}
\label{fig:macs0717_posjelly}
\end{figure*}

\begin{figure*}[ht]
\begin{center}
  \includegraphics[width=8cm]{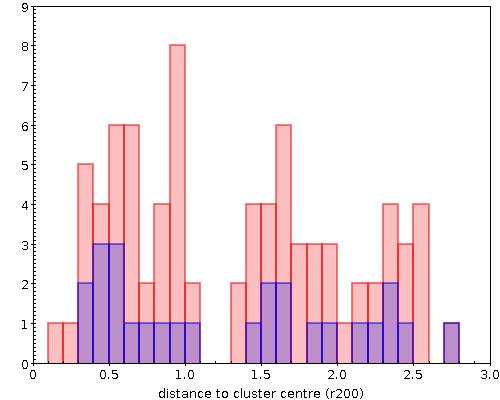}  \kern0.1cm%
  \includegraphics[width=8cm]{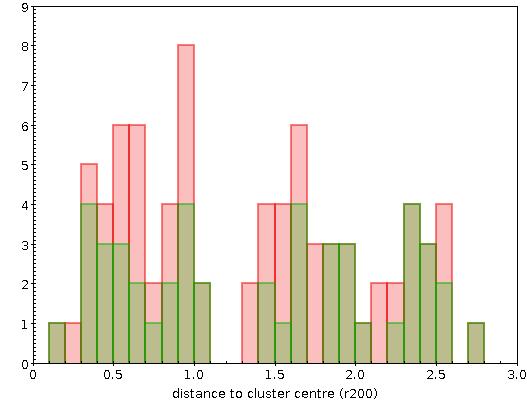}
\end{center}
\caption{Red: histogram of the projected distance to the cluster
  centre of candidate jellyfish galaxies in MACS0717 (81 galaxies). Left: superimposed in blue: candidate jellyfish galaxies of classes 3, 4, and 5 according to the strictest classification (26 galaxies). Right: superimposed in green: candidate jellyfish galaxies of classes 3, 4, and 5 according to the less strict classification (47 galaxies). Distances are in in units of $r_{200}$. }
\label{fig:dist_r_macs0717}
\end{figure*}

The positions of the jellyfish candidates in MACS0717 are shown in
Fig.~\ref{fig:macs0717_posjelly}. We highlight galaxies of types 3, 4 and 5 (considering both classifications),
since they have a high probability of being real jellyfish galaxies.
We can see that a large number is located outside the densest regions 
and about one third lie in regions less dense than the $3\sigma$ contour above the average background density.
Interestingly, about half of the jellyfish candidates are located in the extended filament South region of the cluster (filament C in \citet{Durret+16}, the vertical yellow ellipse in Fig.~\ref{fig:macs0717_posjelly}), 
a low-density zone where only faint X-ray emission is detected \citep{Ma+09}, and therefore where RPS is not expected to be strong but might be gently acting.

Another way to illustrate the spatial distribution of jellyfish galaxies in the large scale MACS0717 structure is to draw the histogram of projected distances to the cluster centre, as shown in Fig.~\ref{fig:dist_r_macs0717}. This figure confirms the paucity of jellyfish galaxies in the innermost cluster regions. The histogram of the galaxy velocities in units of $\sigma _v$ has a similar shape. This lack of jellyfish candidates in the cluster centre agrees with the model by \citet{SafarzadehLoeb19} that predicts no jellyfish galaxies at small clustercentric distances. This result is at odds with the results of the GASP survey, where jellyfish galaxies at low redshift tend to be found in the innermost regions of clusters \citep{Gullieuszik+20}. 

A morphometric analysis of the jellyfish galaxies in MACS0717, comparable to that performed by \citet{Roman+21} with MORFOMETRYKA on a large sample of ram-pressure stripping candidates in the Abell~901/902 multi-cluster system, would be very interesting.

\section{Jellyfish candidates in 22 other clusters}
\label{sec:jelly}

\subsection{The jellyfish catalogue}

Positions and magnitudes of the 97 jellyfish candidates found in 22 clusters are given in
Table~\ref{tab:galaxies}.  For CLASH clusters, galaxy coordinates  
are those of the CLASH catalogue, which always match very well those
measured in the images.

For the DAFT/FADA clusters, galaxy coordinates are those measured by SExtractor on the HST images, as they are more accurate than some of the
coordinates extracted from NED. We checked by superimposing galaxies from the SDSS catalogue that the
astrometry of our {\it HST} images was correct.

\begin{figure}[ht]
\begin{center}
  \includegraphics[width=8cm]{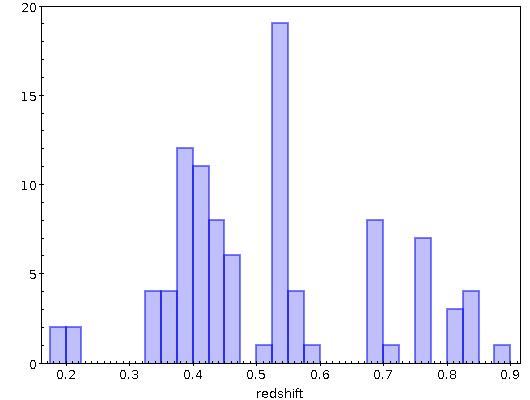} 
\end{center}
\caption{Redshift histogram of the 97 candidate jellyfish galaxies in 22 clusters (excluding MACS0717).}
\label{fig:histoz_103gal}
\end{figure}

\begin{table*}[htbp]
\setcounter{table}{2}
  \caption{List of the 97 candidate jellyfish galaxies in 22 clusters. Columns are: cluster name,
    galaxy identification, galaxy coordinates, redshift, magnitudes
    in the bands where images used for classification are available, jellyfish classes F and S,  projected distance to the cluster centre in kpc and in units of r$_{200}$, and ratio of the difference between the galaxy velocity and that of the cluster divided by the cluster velocity dispersion.}
\label{tab:galaxies}
\resizebox{\textwidth}{!}{
\begin{tabular}{llrrlllllllrrr}
\hline
\hline
  Cluster & G. & RA & DEC & z & F555W & F606W & F702W  & F814W & F & S & Dist. & Dist. & v/$\sigma _v$\\
          &      &    &     &   &       &       & F775W  &       &  &   & (kpc)  & (r$_{200}$) &\\
 \hline
 Cl~0016+16 & a & 4.62890 & 16.42794 & 0.5561 & & & & 19.824 & 4   & 4 & 361 & 0.198 & 2.35\\ 
           & b & 4.63388 & 16.42250 & 0.5498 & & & & 20.742 & 3 & 3 & 391 & 0.215 & 1.37 \\
           & c & 4.63674 & 16.43428 & 0.5382 & & & & 21.770 & 2 & 1 & 129 & 0.071 & $-0.44$\\
           & d & 4.64164 & 16.44986 & 0.56   & & & & 22.033 & 5   & 5 & 274 & 0.151 & 2.95 \\
           & e$^*$ & 4.65058 & 16.44316 & 0.5469 & & & & 22.291 & 2   & 1 & 251 & 0.138 & 0.92 \\
           & f & 4.65223 & 16.42086 & 0.5555 & & & & 21.628 & 2  & 3 & 471 & 0.259 & 2.26 \\
           & g & 4.65749 & 16.44081 & 0.54   & & & & 24.100 & 2   & 2 & 385 & 0.212 & $-0.16$\\
         & h & 4.66072 & 16.45040 & 0.5642 & & & & 19.872 & 3 & 5 & 536 & 0.295 & 3.59 \\
  \hline
  Abell 209         & a & 22.95776 & -13.60326 & 0.2123 & & 19.967 & & 19.283 & 2 & 2 & 176 & 0.073 & 1.25  \\ 
               & b$^*$ & 22.98058 & -13.60446 & 0.2169 & & 19.103 & & 18.821 & 4 & 4 & 133 & 0.055 & 2.16 \\
  \hline
  Cl~0152.7-1357& a$^*$ & 28.12674 & -13.95393 & 0.8474 & & & 21.457 & & 2 & 4 & 1229 & 0.736 & 1.74 \\          
               & b & 28.15464 & -13.95295 & 0.8458 & & & 22.763 & & 2 & 4 & 514 & 0.308 & 1.57 \\
               & c & 28.15784 & -13.93587 & 0.8456 & & & 21.338 & & 2 & 4 & 811 & 0.486 & 1.55 \\
               & d & 28.15924 & -13.92740 & 0.8360 & & & 21.583 & & 1 & 2 & 1012 & 0.606 & $-1.02$ \\
               & e & 28.17714 & -13.93879 & 0.8215 & & & 21.321 & & 2 & 1 & 672 & 0.402 & $-1.02$\\
               & f & 28.17842 & -13.96492 & 0.8371 & & & 23.681 & & 2 & 1 & 218 & 0.131 & 0.65 \\
  \hline
  Abell 383         & a & 42.01002 & -3.55678 & 0.1944 & & 19.487 & & 18.675 & 1 & 3 & 218 & 0.111 & 1.83 \\ 
               & b & 42.03447 & -3.52755 & 0.1914 & & 19.017 & & 18.311 & 2 & 2 & 315 & 0.161 & 1.08 \\
  \hline
  MACS~J0416.1-2403 & a$^*$ & 64.01709 & -24.08955 & 0.396 & & 20.756 & & 19.761 & 1 & 4 & 647 & 0.267 & 0.00  \\ 
                   & b$^*$ & 64.01919 & -24.09614 & 0.3844 & & 22.579 & & 21.676 & 1 & 1 & 716 & 0.296 & $-0.32$ \\ 
                   & c & 64.02498 & -24.09166 & 0.3944 & & 21.533 & & 20.844 & 0 & 1 & 582 & 0.241 & $-0.32$ \\ 
                   & d & 64.02779 & -24.06101 & 0.3918 & & 22.618 & & 21.893 & 1 & 2 & 277 & 0.114 & $-0.62$ \\
                   & e & 64.03268 & -24.07015 & 0.3973 & & 21.410 & & 20.506 & 2 & 2 & 182 & 0.075 & 0.19 \\ 
                   & f & 64.04131 & -24.07134 & 0.3990 & & 20.676 & & 19.707 & 2 & 3 & 101 & 0.042 & 0.44 \\
                   & g$^*$ & 64.04419 & -24.06872 & 0.4071 & & 21.319 & & 20.478 & 2 & 4 & 76 & 0.031 & 1.62  \\ 
                   & h & 64.05088 & -24.06027 & 0.3952 & & 22.508 & & 21.874 & 3 & 3 & 217 & 0.089 & $-0.12$ \\
  \hline
  MACS~J0429.6-0253  & a & 67.38860 & -2.88379 & 0.4    & & 20.798 & & 20.505 & 4 & 4 & 223 & 0.129 & 0.21 \\ 
                    & b & 67.41844 & -2.88832 & 0.4049 & & 20.920 & & 20.665 & 4 & 3 & 360 & 0.208 & 1.22\\
  \hline
 MACS~J0454.1-0300& a & 73.51869 & -3.00396 & 0.5323 & & & & 19.910 & 5 & 5 & 679 & 0.322 & $-2.38$ \\ 
                 & b & 73.53999	& -3.01686 & 0.5483 & & & & 20.118 & 1 & 2 & 152 & 0.072 & $-0.23$ \\
                 & c$^*$ & 73.55029	& -3.00036 & 0.5469  & & & & 20.338 & 1 & 2 & 345 & 0.163 & $-0.41$\\
                 & d$^*$ & 73.56299 & -3.00639 & 0.524   & & & & 22.475 & 5 & 3 & 427 & 0.202 & $-3.52$ \\
                 & e & 73.56342	& -2.99819 & 0.528   & & & & 20.116 & 1 & 2 & 546 & 0.259 & $-2.97$\\
  \hline
  Abell 851   & a & 145.68898 & 46.98263 & 0.407  & & & & 20.668 & 2 & 2 & 1033 & 0.670 & 0.02\\          & b & 145.69411 & 47.00624  & 0.4083 & & & & 20.092 & 2 & 1 & 1005 & 0.652 & 0.32 \\
         & c & 145.71105 & 47.01366 & 0.3958 & & & & 19.165 & 5 & 5 & 798 & 0.517 & $-2.52$\\
         & d & 145.73241 & 47.01542 & 0.3972 & & & & 22.378 & 3 & 3 & 590 & 0.383 & $-2.20$ \\
         & e$^*$ & 145.73256 & 46.99261 & 0.4076 & & & & 19.485 & 3 & 4 & 213 & 0.138 & 0.16 \\
         & f & 145.73862 & 46.96882  & 0.4084 & & & & 20.286 & 2 & 4 & 354 & 0.230 & 0.34 \\
         & g & 145.74384 & 46.99646 & 0.4200 & & & & 21.317 & 1 & 2 & 196 & 0.127 & 2.93 \\
         & h & 145.74773 & 47.01602 & 0.3937 & & & & 21.002 & 4 & 3 & 586 & 0.380 & $-3.00$ \\
         & i & 145.75700 & 47.00907 & 0.4061 & & & & 19.160 & 2 & 3 & 531 & 0.344 & $-0.18$\\ 
         & j & 145.76435 & 46.98741 & 0.41   & & & & 20.839 & 2 & 3 & 444 & 0.288 & 0.70 \\
         & k & 145.76440 & 47.00949  & 0.39   & & & & 21.286 & 1 & 2 & 630 & 0.408 & $-3.85$\\
   \hline
\  LCDCS~0172 & a & 163.57599 & -11.78999 & 0.6965 & & & & 22.109 & 2 & 3 & 793 & 0.911 & $-0.18$\\  
            & b & 163.58247 & -11.77602 & 0.6972 & & & & 21.651 & 3 & 3 & 485 & 0.557 & 0.00 \\  & c & 163.58701  & -11.74937 & 0.6968 & & & & 22.337 & 4 & 3 & 674 & 0.774 & $-0.10$ \\
            & d & 163.58728 & -11.75391 & 0.702  & & & & 20.863 & 3 & 3 & 574 & 0.659 & 1.23 \\
            & e & 163.60573 & -11.76516 & 0.6977 & & & & 20.148 & 1 & 2 & 209 & 0.241 & 0.13 \\
            & f & 163.60592 & -11.79784 & 0.6977 & & & & 22.885 & 4 & 5 & 685 & 0.787 & 0.13  \\
            & g$^*$ & 163.62696 & -11.78158 & 0.6986 & & & & 19.654 & 2 & 3 & 718 & 0.825 & 0.36 \\
            & h & 163.62964 & -11.82359 & 0.6966 & & & & 22.199 & 1 & 2 & 1525 & 1.753 & $-0.15$\\
            & i & 163.64143 & -11.82147 & 0.699  & & & & 21.098 & 2 & 2 & 1650 & 1.897 & 0.46 \\
  \hline
  MACS~J1149.5+2223 & a   & 177.39015 & 22.40389 & 0.543 & & 23.162 & & 21.867 & 2 & 2 & 141 & 0.048 & $-0.14$ \\ 
                   & b$^*$   & 177.39855 & 22.38979 & 0.536 & & 23.149 & & 22.090 & 2 & 3 & 309 & 0.106 & $-0.82$ \\
                   & c$^*$   & 177.39863 & 22.39850 & 0.540 & & 19.258 & & 18.122 & 5 & 4 & 118 & 0.040 & $-0.33$ \\
                   & d   & 177.39977 & 22.39728 & 0.541 & & 21.197 & & 20.223 & 5 & 4 & 155 & 0.053 & $-0.33$ \\
\hline
MACS~J1206.2-0847 & a & 181.53955 & -8.81678 & 0.4356 & & 22.308 & & 21.427 & 2 & 2 & 409 & 0.202 & $-0.73$ \\ 
                   & b$^*$ & 181.56187 & -8.80434 & 0.4265 & & 20.214 & & 19.616 & 3 & 3 & 241 & 0.119 & $-2.24$  \\
                   & c & 181.57174 & -8.80643 & 0.4450 & & 20.156 & & 19.135 & 1 & 1 & 446 & 0.220 & 0.82 \\
\hline
\hline
\end{tabular}}
\end{table*}

\begin{table*}[htbp]
%\centering
\setcounter{table}{2}
  \caption{Continued.}
\resizebox{\textwidth}{!}{%
\begin{tabular}{lcrrlllllllrrr}
  \hline
  \hline
 LCDCS~0541 & a & 188.09897 & -12.86970 & 0.5499 & & & & 21.022 & 2 & 3 & 867 & 1.008 & 2.80 \\
           & b & 188.10412 & -12.86525 & 0.5399 & & & & 20.615 & 2 & 3 & 712 & 0.827 & $-0.50$ \\
           & c & 188.11454 & -12.83518 & 0.5367 & & & & 19.587 & 4 & 1 & 328 & 0.382 & $-1.56$\\
           & d & 188.12779 & -12.83268 & 0.5492 & & & & 21.795 & 2 & 2 & 249 & 0.289 & 2.57\\
           & e & 188.13455 & -12.85739 & 0.5498 & & & & 20.053 & 1 & 2 & 372 & 0.432 & 2.76 \\
           & f & 188.14242 & -12.81575 & 0.5323 & & & & 22.953 & 1 & 2 & 734 & 0.854 & $-3.02$\\
           & g & 188.16354 & -12.89634 & 0.5364 & & & & 21.641 & 2 & 3 & 1482 & 1.723 & $-1.66$\\
  \hline
[MJM98]\_034  & a$^*$ & 203.75482 & 37.82928 & 0.3841 & & & 23.213 & & 2 & 2 & 1303 & 1.300 & 0.22\\ 
  \hline
  LCDCS~0829 & a & 206.87102 & -11.76679 & 0.4534 & & 21.106 & & 20.397 & 4 & 4 & 320 & 0.196 & 0.29 \\ 
  \hline
  LCDCS~0853 & a$^*$ & 208.54225 & -12.51746 & 0.7565 & & & & 23.885 & 3 & 1 & 39 & 0.025 & $-0.79$\\ 
          & b$^*$ & 208.54246 & -12.51466 & 0.7593 & & & & 21.036 & 2 & 3 & 78 & 0.049 & $-0.43$\\
          & c & 208.55306  & -12.55668 & 0.7627 & & & & 22.491 & 2 & 2 & 1102 & 0.693 & 0.00\\
          & d$^*$ & 208.55823 & -12.52258 & 0.7642 & & & & 21.076 & 1 & 3 & 485 & 0.305 & 0.19 \\
          & e & 208.56222 & -12.52241 & 0.7609 & & & & 21.117 & 3 & 2 & 586 & 0.368 & $-0.23$\\
          & f & 208.57370 & -12.51199  & 0.7642 & & & & 24.061 & 2 & 1 & 884 & 0.556 & 0.19 \\
          & g & 208.57877 & -12.51216 & 0.7634 & & & & 21.426 & 2 & 3 & 1017 & 0.640 & 0.09\\
 \hline
3C~295  & a & 212.79865 & 52.20013 & 0.454 & 23.593 & & & 21.908 & 1 & 3 & 1134 & 0.634 & $-1.09$ \\ 
       & b & 212.80549 & 52.16989 & 0.43  & & & & 22.916 & 1 & 2 & 1292 & 0.722 & $-5.54$ \\
       & c & 212.80823 & 52.19388 & 0.4485 & 22.968 & & & 21.459 & 2 & 4& 977 & 0.546 & $-2.10$\\
       & d & 212.82234 & 52.18692 & 0.44  & 23.455 & & & 22.452 & 2 & 2 & 791 & 0.442 & $-3.67$\\
       & e & 212.82708 & 52.20480 & 0.4703 & 23.794 & & & & 4 & 2 & 530 & 0.296 & 1.86\\       
       & f & 212.83317 & 52.19815 & 0.4659  & 23.033 & & & & 3 & 3 & 468 & 0.261 & 1.07\\       
       & g & 212.83551 & 52.20273 & 0.464 & 22.0453 & & & & 3 & 4 & 377 & 0.211 & 0.72\\
       & h & 212.85448 & 52.20978 & 0.47 & 25.132 & & & & 1 & 2 & 61 & 0.034 & 1.80 \\
\hline
  RX~J1532.9+3021 & a & 233.22410 & 30.34982 & 0.3611 & & 17.819 &  & 17.108 & 5 & 5 & 7 & 0.004 & 3.79\\ 
  \hline
MS~1621.5+2640  & a & 245.89661 & 26.57446 & 0.4269 & 23.106 & & & 21.371 & 3 & 4 & 88 & 0.051 & 0.18 \\
             & b & 245.90057 &	26.57651 & 0.4405 & 22.796 & & & 21.103 & 3 & 3 & 136 & 0.079 & 2.84 \\	
              & c & 245.92174 &	26.53319 & 0.4071  & & & & 21.700 & 2 & 5 & 885 & 0.515 & $-3.78$\\	
\hline
  RX~J1716.4+6708 & a & 259.15717 & 67.12481 & 0.8044 & & & & 23.235 & 3 & 2 & 1421 & 0.843  & $-0.94$\\ 
  \hline
  MACS~J1931.8-2634 & a & 292.94157 & -26.59913  & 0.3494 & & 20.890 & & 20.158 & 1 & 3 & 491 & 0.255 & $-0.52$ \\ 
                   & b & 292.9506	 & -26.57826 & 0.3652 & & 20.399 & & 20.131 & 3 & 4 & 115 & 0.060 & 2.61 \\
\hline
MS~2053.7-0449 & a & 314.09464 & -4.59867 & 0.5880 & & 22.293 & & 21.687 & 2 & 2 & 715 & 0.441 & 0.81 \\
\hline
RX~J2248.7-4431 & a & 342.14918 & -44.52740 & 0.3356 & & 21.658 & & 21.043 & 1 & 2 & 569 & 0.247 & $-2.01$\\
               & b & 342.15716 & -44.54514 & 0.3312 & & 23.971 & & 22.954 & 3 & 2 & 515 & 0.224 & $-2.76$\\
	       & c & 342.16731 & -44.51396 & 0.3517 & & 20.443  & & 19.991 & 5 & 5 & 362 & 0.157 & 0.70 \\
	       & d$^*$ & 342.17550 & -44.53546 & 0.3362 & & 19.814 & & 19.106 & 4 & 4 & 155 & 0.067 & $-1.91$ \\
	       & e & 342.20375 & -44.54226 & 0.3552 & & 21.698 & & 21.293 & 2 & 2 & 464 & 0.202 & 1.29 \\
  \hline
  \hline
\end{tabular}}
\end{table*}

The redshift histogram of the 97 jellyfish candidates found in 22 clusters (other than MACS0717) is shown in Fig.~\ref{fig:histoz_103gal}. 

\subsection{Images and notes on individual objects}

The images of the 97 jellyfish candidates are shown in Appendix~\ref{appendix2}.
In some cases, we give below a few indications on specific galaxies when we think it
is useful and we indicate if clusters are merging whenever this information is available. In particular, we mention if the positions of the jellyfish candidates lie in the direction of the general elongation of the cluster, defined both from the position angle of the brightest cluster galaxy, from the alignment of the brightest cluster galaxies, and from the red-sequence galaxy density maps drawn by \citet{Durret+16} and 
 \citet{Durret+19} when available. This direction should trace the orientation of the filamentary regions, at large scale, where each cluster is embedded, and along which one would expect the largest galaxy infall to happen  \citep[e.g.][]{West94,West+17}. Thus, these regions are the privileged areas for infalling late-type galaxies to become jellyfish as they enter with high speed and move across the cluster denser environment. 

When two galaxies are located in the same frame, the galaxy with a spectroscopic redshift at the cluster redshift is indicated with a circle in images of Appendix~\ref{appendix2}.

\subsubsection {Cl~0016+16 (z=0.5455)} %4.638875,16.4432778

Images of the eight jellyfish galaxy candidates in Cl0016+16 are shown in
Fig.~\ref{fig:cl0016}.1.  Out of the eight jellyfish galaxies (out of 103 galaxies at the cluster redshift), four are well aligned with the general cluster elongation (see \citet{Durret+19}, Fig.~B.1), and three others are not far from this region/direction. 

\subsubsection {Abell 209 (z=0.206)} %22.9708333,-13.609444

Images of the two jellyfish galaxy candidates (out of 39 galaxies at the cluster redshift) are shown in Fig.~\ref{fig:a209}. One of them is aligned along the cluster main elongation region (\citep{Durret+19}, fig.~B.2).

\subsubsection {Cl~0152.7-1357 (z=0.831)} %28.1707917,-13.9625

Images of the six jellyfish galaxy candidates (out of 66 galaxies at the cluster redshift) are shown in Fig.~\ref{fig:cl0152}. All of them are in the Northern part of this merging cluster (see \citet{Guennou+14} and references therein), which is the main zone covered by the HST image. Based on X-ray data and on a large number of galaxy spectroscopic redshifts, \citet{Girardi+05} showed that Cl0152.7-1357 consists of three galaxy clumps of different mean velocities: a low velocity clump in the central-South-West region, a high velocity clump in the Eastern region, and a weaker Eastern clump. 

\subsubsection {Abell 383 (z=0.1871)} %42.0083333,-3.5375

Images of the two jellyfish galaxy candidates (out of 32 galaxies at the cluster redshift) are shown in Fig.~\ref{fig:a383}.  One is located in the cluster elongation area (\citet{Durret+19}, Fig.~B.5), one is close to this region, and the third one is further out.

\subsubsection {MACS~J0416.1-2403 (z=0.396)} %64.04125,-24.066111

Images of the eight jellyfish galaxy candidates (out of 205 galaxies at the cluster redshift) are shown in
Fig.~\ref{fig:macs0416}.  Galaxy~f may be superimposed on a gravitational arc.
Out of the eight jellyfish galaxies, seven seem to be roughly spatially aligned with the main cluster merging axis.

\subsubsection {MACS~J0429.6-0253 (z=0.399)} %67.4,-2.8855556

Images of the two jellyfish galaxy candidates (out of only two galaxies at the cluster redshift) are shown in Fig.~\ref{fig:macs0429}, none of them being located along the main cluster elongation area.

\subsubsection {MACS~J0454.1-0300 (z=0.5377)} %73.5455,-3.0186389

Images of the five jellyfish galaxy candidates (out of 92 galaxies at the cluster redshift) are shown in
Fig.~\ref{fig:macs0454}.  Galaxy~a may not be a jellyfish galaxy, but
we keep it in the sample because of its two nuclei. Two galaxies follow the main cluster elongation, and a third one is close-by.

\subsubsection {Abell~851 (z=0.4069)} %145.736,46.9894167

Images of the eleven  jellyfish galaxy candidates (out of 102 galaxies at the cluster redshift) are shown in Fig.~\ref{fig:a851}. All but one are located in the Northern half of the cluster, but with no specific alignment. This is a merging cluster  \citep{Durret+16}.

\subsubsection {LCDCS~0172 (z=0.6972)} %163.600792,-11.771667

The images of the nine candidate jellyfish galaxies (out of 45 galaxies at the cluster redshift) are displayed in Fig.~\ref{fig:lcdcs0172}, and show no particular spatial distribution.

\subsubsection {MACS~J1149.5+2223 (z=0.544)} %176.396208,22.4030278

Images of the four jellyfish galaxy candidates (out of 106 galaxies at the cluster redshift) are shown in
Fig.~\ref{fig:macs1149}. Three of them follow the main cluster elongation. Galaxies c and d are interacting, with many
filaments in their neighbourhood. Thus the image also shows their environment.

\subsubsection {MACS~J1206.2-0847 (z=0.44)} %181.550833,-8.8002778

Images of the three jellyfish galaxy candidates (out of 64 galaxies at the cluster redshift) are shown in Fig.~\ref{fig:macs1206}.
Two of these galaxies are located along the main cluster elongation.

\subsubsection {LCDCS~0541 (z=0.5414)} %188.127042,-12.842472

Images of the seven jellyfish galaxy candidates (out of 80 galaxies at the cluster redshift) are shown in Fig.~\ref{fig:lcdcs0541}. Their spatial distribution shows no particular trend. The bright arc North and East of galaxy~e is a gravitational arc.

\subsubsection {[MJM98]\_034 (z=0.383)} %203.807417,37.8156389

Image of the single jellyfish galaxy candidate (out of eight galaxies at the cluster redshift) is shown in Fig.~\ref{fig:mjm98}. As noted by \citet{Guennou+14} this cluster is at redshift z=0.383 and not at z=0.5950 as found in NED.

\subsubsection {LCDCS~0829=RXJ1347 (z=0.451)} %206.883292,-11.761667

Image of the single jellyfish galaxy candidate (out of 35 galaxies at the cluster redshift) is shown in Fig.~\ref{fig:lcdcs0829}.
Its position is roughly aligned with the cluster elongation.

\subsubsection {LCDCS~0853 (z=0.7627)} %206.883292,-11.761667

Images of the seven jellyfish galaxy candidates (out of 20 galaxies at the cluster redshift) are shown in Fig.~\ref{fig:lcdcs0853}.
All seven galaxies are located in the South-East quarter of the cluster.

\subsubsection {3C~295 (z=0.4600)} %212.833958,52.2025

Images of the eight jellyfish galaxy candidates (out of 30 galaxies at the cluster redshift) are shown in
Fig.~\ref{fig:3c295}. They show no particular distribution throughout the cluster. Due to different spatial coverage, some appear in both filters, others only in
one.

\subsubsection {RX~J1532.9+3021 (z=0.345)} %233.224167,30.3494444

Image of the single jellyfish galaxy candidate (this galaxy is in fact the brightest cluster galaxy, and there is only one other galaxy with a measured redshift in the cluster) is shown in Fig.~\ref{fig:rx1532}, as well as a zoomed image showing a disturbed structure. 

\subsubsection {MS~1621.5+2640 (z=0.426)} %245.898625,26.5637778

Images of the three jellyfish galaxy candidates (out of 31 galaxies at the cluster redshift) are shown in Fig.~\ref{fig:ms1621}. They show no particular distribution in the cluster. The two filters do not cover exactly the same region. Galaxy a may be interacting with one or two galaxies, but spectroscopic redshifts are not available for these objects. Galaxy b
has a plume of emission to the South, and seems to be surrounded by gravitational arcs in the North. Spectroscopy is also needed to confirm the jet-like feature East of galaxy c.

\subsubsection {RX~J1716.4+6708 (z=0.813)} %259.206667,67.1416667

Image of the single jellyfish galaxy candidate (out of 22 galaxies at the cluster redshift) is shown in Fig.~\ref{fig:rx1716}.

\subsubsection {MACS~J1931.8-2634 (z=0.352)} %292.956667,-26.576111

Images of the two jellyfish galaxy candidates (out of only three galaxies at the cluster redshift) are shown in Fig.~\ref{fig:macs1931}.

\subsubsection {MS~2053.7-0449 (z=0.583)} %314.09321,-4.62873

Image of the single jellyfish galaxy candidate (out of 32 galaxies at the cluster redshift) is shown in Fig.~\ref{fig:ms2053}.
It lies exactly North of the cluster centre, positioned along the direction of the cluster elongation.

\subsubsection {RX~J2248.7-4431 (z=0.348)} %342.18125,-44.528889

Images of the five jellyfish galaxy candidates (out of 42 galaxies at the cluster redshift) are shown in Fig.~\ref{fig:rx2248}.
Two out of five galaxies are located within the cluster elongation area/direction.

\section{Jellyfish galaxy candidate properties}
\label{sec:properties}

\subsection{Spatial distribution relative to the cluster centre}
\label{subsec:spatial_distribution}

\begin{figure*}[ht]
\begin{center}
  \includegraphics[width=8cm]{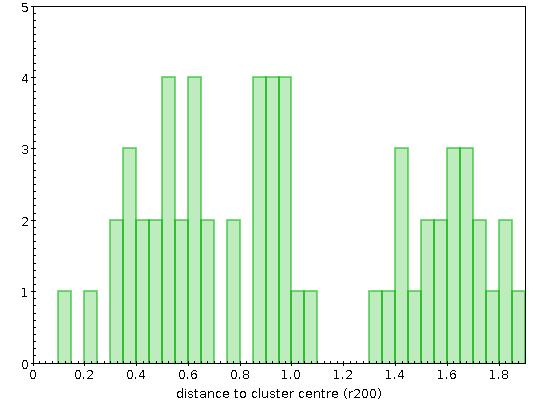}  \kern0.1cm%
  \includegraphics[width=7.5cm]{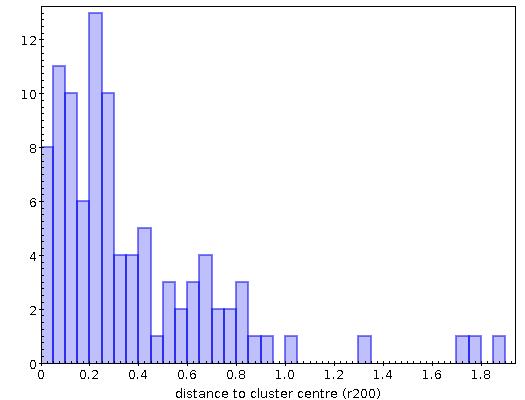}
\end{center}
\caption{Left: histogram of the projected distance of 61 jellyfish galaxy candidates to the cluster centre for MACS0717, in units of $r_{200}$, with the same limit as the figure on the right.
 Right:  same histogram for the 97 jellyfish galaxy candidates of the other clusters. }
\label{fig:dist_r}
\end{figure*}

The histogram of the projected distances of jellyfish galaxy candidates to the centre of the respective cluster is shown in Fig.~\ref{fig:dist_r}, where we compare the result already shown for MAC0717 but limited to within 1800 kpc (on the left) with the results obtained for all the other clusters taken together (right panel). This figure shows that the number of jellyfish galaxy candidates in the central region of MACS0717 is rather small, whereas it is large in the ensemble of all the other clusters. Unlike for MACS0717, our coverage of the remaining clusters is far from complete; nonetheless, this result indicates that in MACS0717 there seems to be a real absence of jellyfish galaxy candidates in the innermost cluster region.

\subsection{Spectral energy distribution }

\subsubsection{SED fitting for jellyfish galaxy candidates in MACS~J0717.4+3745}
\label{subsubsec:SED_0717}

\begin{figure*}[ht]
\begin{center}
\includegraphics[width=9cm]{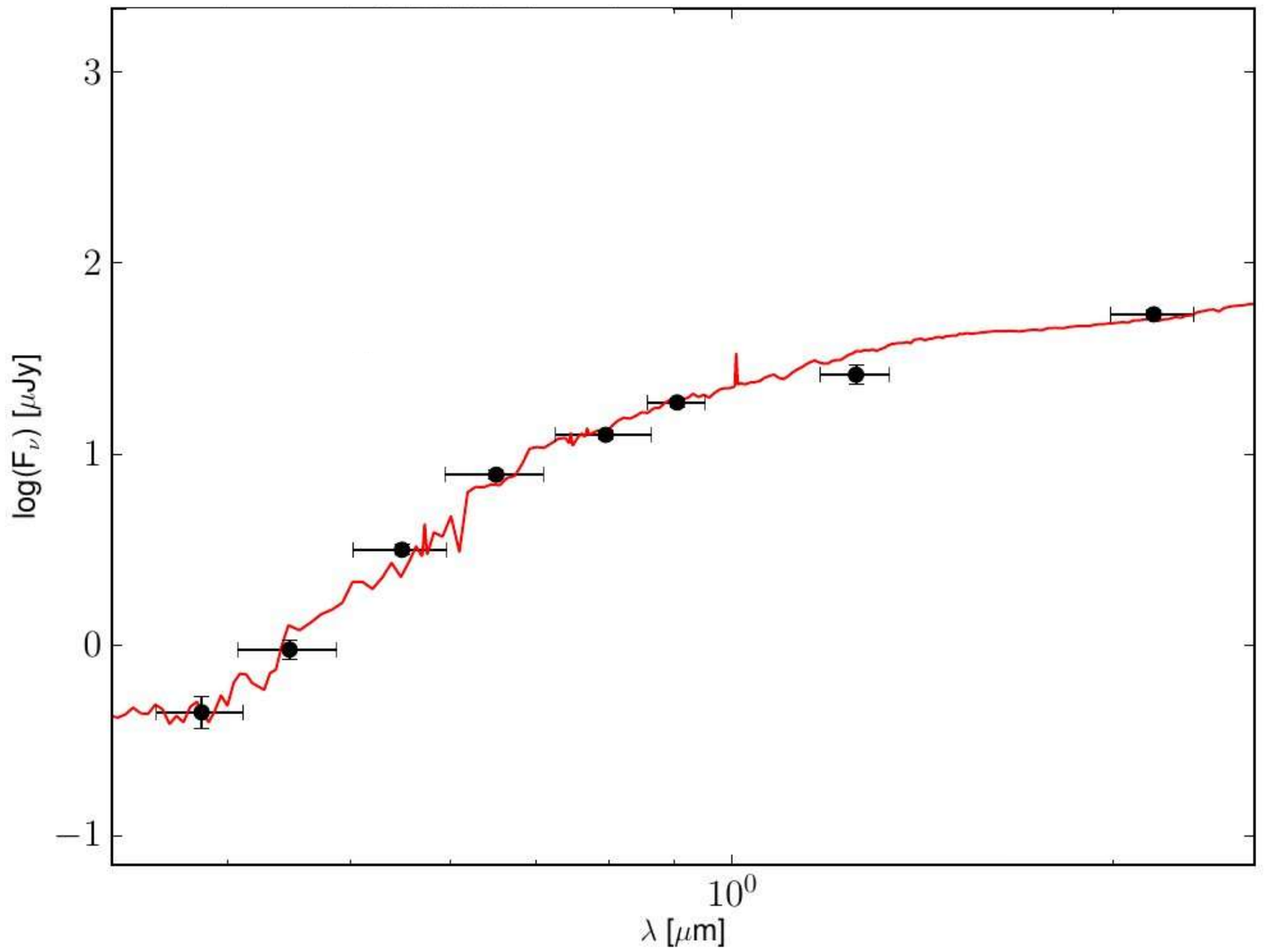} \kern0.1cm%
\includegraphics[width=9cm]{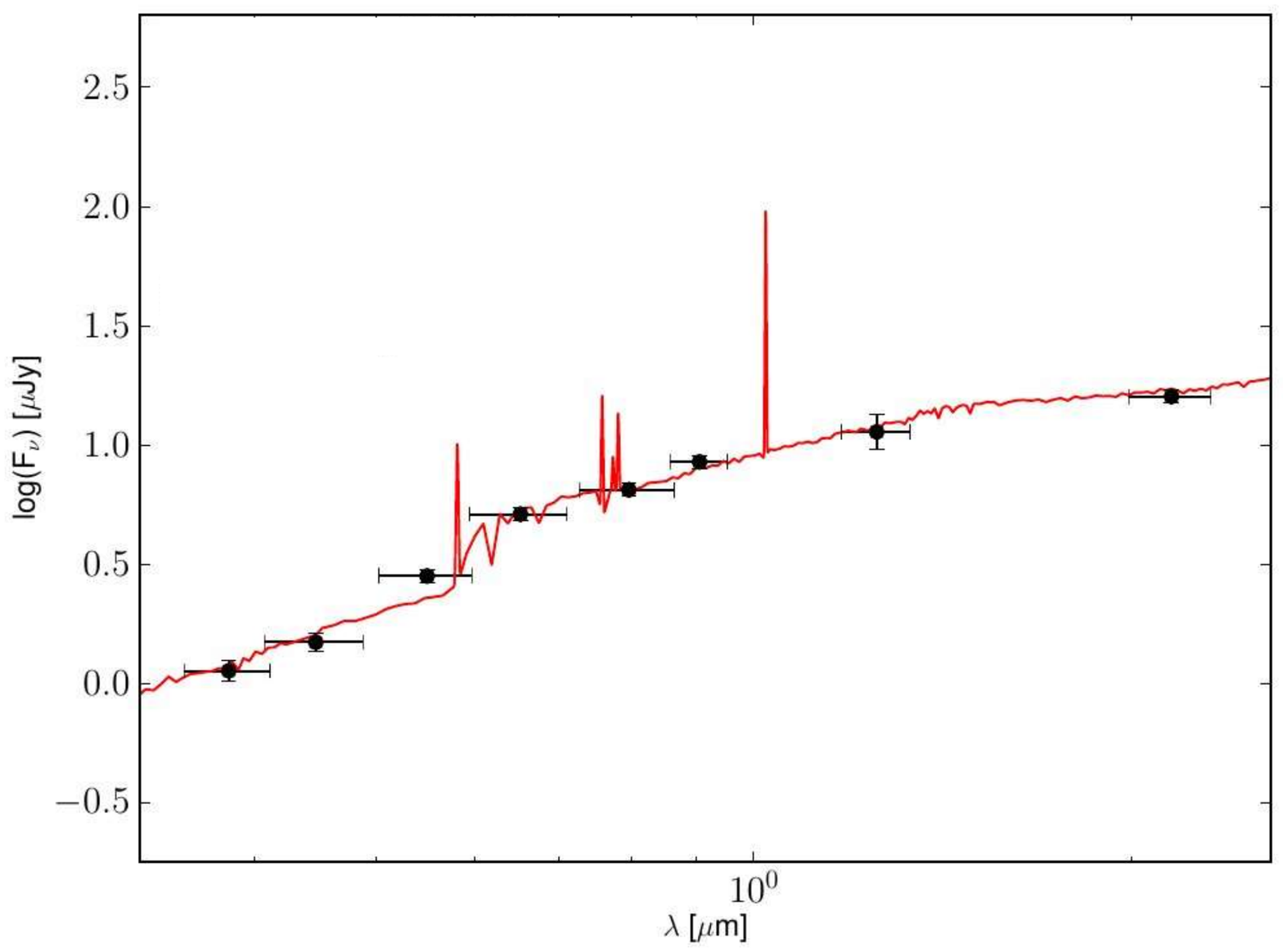} 
\end{center}
\caption{For two of the jellyfish galaxy candidates in MACS0717, in black: 8 magnitudes available from the \citet{Jauzac+12} catalogue, in red: best stellar population fit superimposed, one with a weak H$\alpha$ emission line (left) and one with several strong emission lines (right).}
\label{fig:spectres}
\end{figure*}

\begin{figure*}[ht]
\begin{center}
\includegraphics[width=8cm]{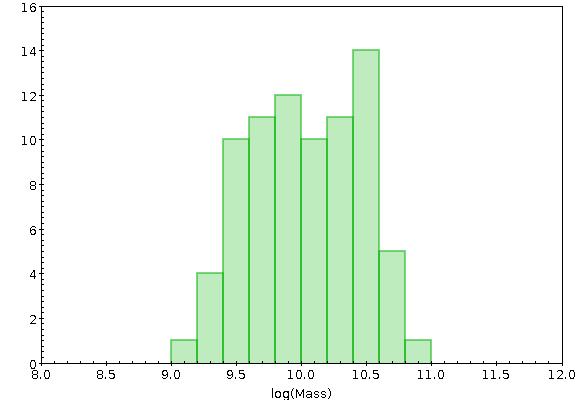} \kern0.1cm%
 \includegraphics[width=8cm]{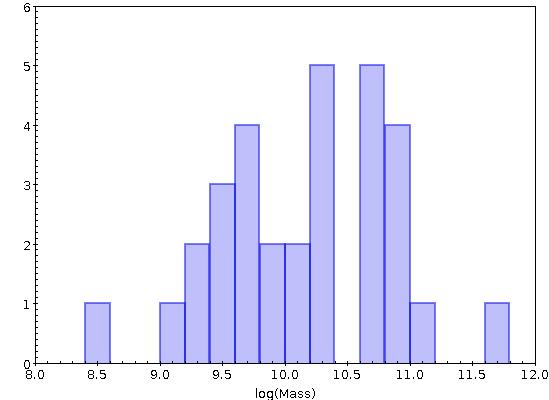} 
\end{center}
\caption{Histograms of the stellar masses for the 79 jellyfish galaxy candidates in MACS0717 (left), and 31 galaxies of other clusters (right).}
\label{fig:m0717_histo_mass}
\end{figure*}

\begin{figure*}[ht]
\begin{center}
\includegraphics[width=8.5cm]{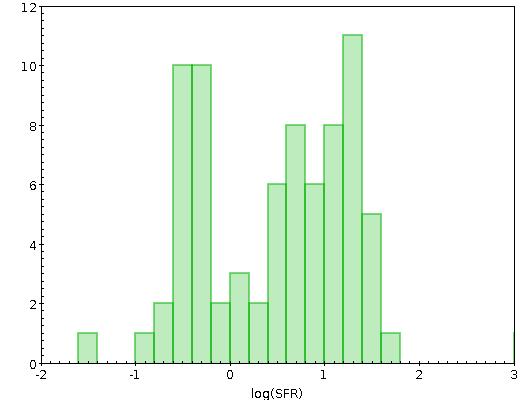} \kern0.1cm%
 \includegraphics[width=8cm]{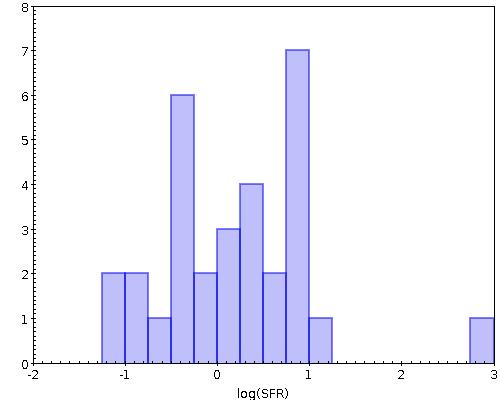} 
\end{center}
\caption{Histograms of the star formation rates for the 79 jellyfish galaxy candidates in MACS0717 (left), and 31 galaxies of other clusters (right).}
\label{fig:m0717_histo_SFR}
\end{figure*}

\begin{figure*}[ht]
\begin{center}
\includegraphics[width=8.5cm]{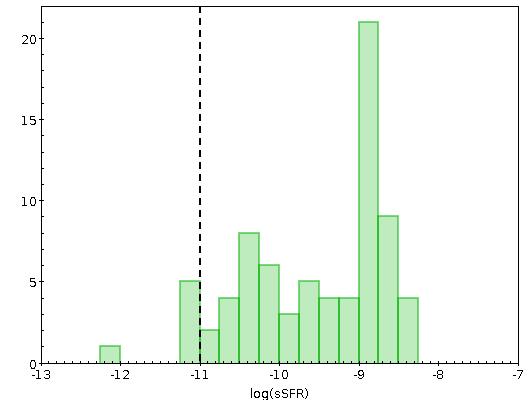} \kern0.1cm%
 \includegraphics[width=8cm]{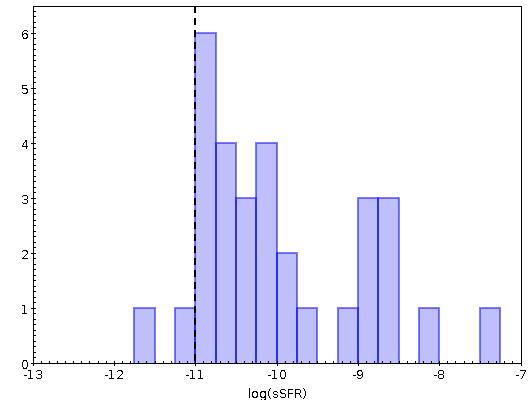} 
\end{center}
\caption{Histograms of the specific star formation rates of the 79 jellyfish galaxy candidates in MACS0717 (left), and 31 galaxies in the other clusters (right). The black vertical dashed line shows the value of -11 below which galaxies are considered as quiescent.}
\label{fig:m0717_histo_sSFR}
\end{figure*}

\begin{figure*}[ht]
\begin{center}
\includegraphics[width=8.5cm]{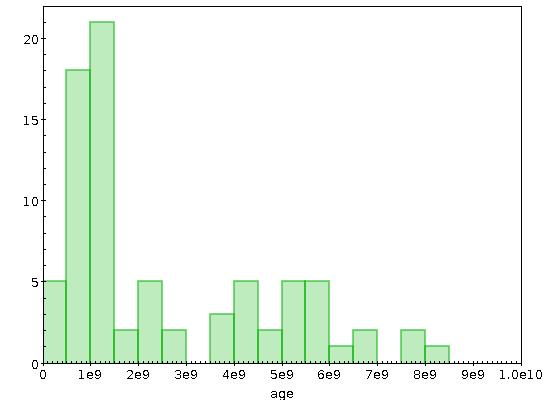} \kern0.1cm%
 \includegraphics[width=8cm]{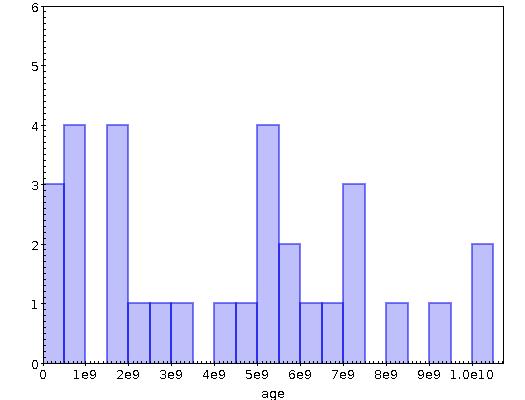} 
\end{center}
\caption{Histograms of the stellar population ages of the 79 jellyfish galaxy candidates in MACS0717 (left), and 31 galaxies in the other clusters (right).}
\label{fig:m0717_histo_age}
\end{figure*}

\begin{figure*}[ht]
\begin{center}
\includegraphics[width=8.5cm]{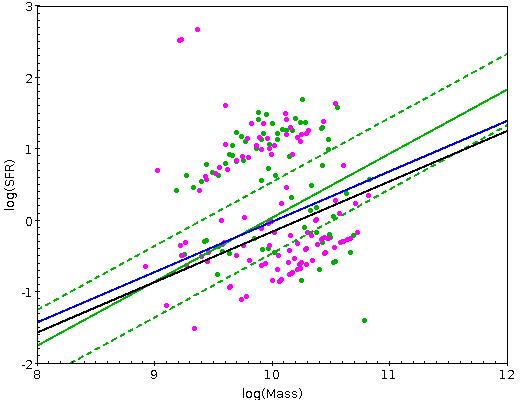} \kern0.1cm%
 \includegraphics[width=8cm]{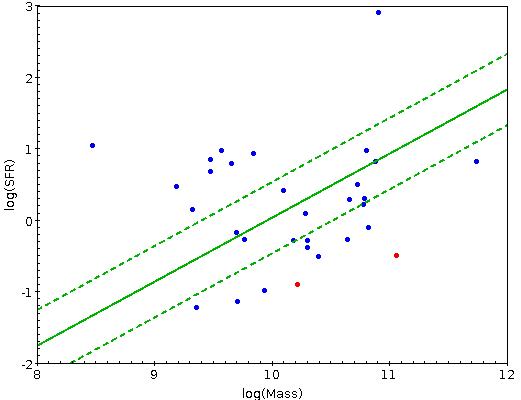} 
\end{center}
\caption{SFR as a function of stellar mass for the 79 jellyfish galaxy candidates in MACS0717 (left) and 31 galaxies of other clusters (right). On both figures, the three green lines indicate the relation found by \citet{Peng+10} and its approximate dispersion of $\pm 0.5$ (dashed lines). On the left figure, pink dots correspond to ``normal'' cluster galaxies with log(sSFR) $\geq -11$ (see text). Blue and black lines show the relations found by \citet{Vulcani+18} for the disk SFR–mass relation for stripping and control sample galaxies respectively (see their Fig.~1). On the right panel, the two red points highlight the two galaxies with
log(sSFR) $<-11$ }
\label{fig:m0717_mass_SFR}
\end{figure*}

We matched our catalogue of 81 jellyfish galaxy candidates detected in MACS0717 with the eight band optical and infrared ground-based catalogue covering the entire region of MACS0717 described in Section~2. We found 79 galaxies in common using a match radius of 1.5~arcsec.

We then used LePhare \citep{Ilbert+06}, through the GAZPAR interface\footnote{https://gazpar.lam.fr/home}, to fit the spectral energy distributions (SED) of these 79 jellyfish galaxy candidates with the \citet{BC03} models and the \citet{Chabrier+03} initial mass function: based on an input catalogue with positions, magnitudes and corresponding errors, and in our case spectroscopic redshift, LePhare fits the galaxy SED, computes absolute magnitudes in the input bands, and infers, from the best fit template in each case, the stellar mass, star formation rate, specific star formation rate, mean stellar population age, 
as well as other quantities that we will not consider here. 
The input parameter space was carefully selected so as to cover the expected characteristics of late-type galaxies with probable SF activity.

We can note that for all 79 galaxies but a few
the best fit template spectrum includes an H$\alpha$ line, and among these 
about 80\% of the template spectra include all the main emission lines in the optical ([OII]3727, [OIII]4959, 5007, H$\beta$ and H$\alpha$), thus implying that the majority of our jellyfish candidates are forming stars. 
As an illustration of the obtained fits, we show in Fig.~\ref{fig:spectres} the SED and best fit templates for two galaxies, one with only a weak H$\alpha$ emission line and one with several emission lines in the best fit template. We see that these fits are quite good, and this was indeed the case for all galaxies that GAZPAR/LePhare analysed.

The histogram of jellyfish galaxy candidate stellar masses in MACS0717 is shown in the left-hand panel of Fig.~\ref{fig:m0717_histo_mass}. The galaxies cover the range of stellar masses between $10^9$ and $10^{11}$~M$_\odot$. 
We divided the sample into high-mass (log M $\geq 10$) and low-mass (log M $< 10$) galaxies and checked their spatial distribution but found no difference between these two samples.

The histogram of jellyfish galaxy candidate star formation rates for MACS0717 is shown in the left hand panel of Fig.~\ref{fig:m0717_histo_SFR}. The SFRs cover a large range, essentially between 0.1 and 60~M$_\odot$~yr$^{-1}$.
We also show in Fig.~\ref{fig:m0717_histo_sSFR} the histogram of the specific star formation rates in MACS0717:  values go from $10^{-12}$ to $10^{-8}$~yr$^{-1}$ (except for two galaxies that have very low sSFRs and do not appear on the figure), and more than 30 galaxies have sSFR $> 10^{-9}$~yr$^{-1}$. We mark on the sSFR histograms the indicative value of log(sSFR)$ = -11$ below which galaxies are commonly considered to be quiescent.

The histogram of the stellar population 
age for the jellyfish galaxy candidates in MACS0717 is shown in Fig.~\ref{fig:m0717_histo_age}. We can see that more than half of the galaxies
are, on average, younger than $1.5\times 10^9$~yrs, so the stellar population is globally quite young.

The relation between the SFR and galaxy stellar masses is shown in Fig.~\ref{fig:m0717_mass_SFR}. We superimposed on this plot the main sequence of SF galaxies as determined by \citet{Peng+10}, based on a very large sample of galaxies from the SDSS and zCOSMOS surveys, and its dispersion that we estimated to be $\pm 0.5$ around the relation (see their Fig.~1). We see that 12 (15\%) jellyfish galaxies are below this sequence, 23 (29\%) are in the interval of \citet{Peng+10}, and the 44 other ones (56\%) lie above the SF main sequence. 

In agreement with the previously noted fact that SEDs of the jellyfish galaxies in MACS0717 are in majority best fitted by a spectrum including one or several emission lines, this confirms that, in average, jellyfish galaxy candidates seem to have a higher SFR than ``normal'' star forming galaxies of the same stellar mass. Figure~\ref{fig:m0717_mass_SFR} suggests that the majority seems to form a sequence parallel to that of \citet{Peng+10}, but with a SFR about 10 times higher. There are, however, a few cases of very low specific star formation rates: these galaxies are apparently quenched, as indicated by the log(sSFR) $ < -11$ criterion (see Fig.~\ref{fig:m0717_histo_sSFR}). 

As a comparison, we also obtain an SED fit for 442 galaxies in the same redshift interval but not classified as jellyfish candidates. Out of these, 113 galaxies can be considered as non-quiescent (specific star formation rate  log(sSFR)$\geq -11$). We also include these galaxies in Fig.~\ref{fig:m0717_mass_SFR}. One can see that they cover more or less the same region as the jellyfish candidates. However, if we calculate the average sSFR for the jellyfish candidates and for this control sample, both limited to logsSFR$\geq -11$ (respectively: 71 and 113 galaxies), we find respective values of  $-9.36$ and $-9.87$, suggesting that the sSFR is about 3 times larger for jellyfish candidates. In view of the errors on these quantities, this value of 3 should not be taken at face value, but it does suggest that jellyfish galaxies have a sSFR higher than that of ``normal'' galaxies.

We also overplot on Fig.~\ref{fig:m0717_mass_SFR} (left panel) the relations found by \citet{Vulcani+18} for disks of ram pressure stripped galaxies (in blue), and undisturbed galaxies (in black). One can note that at least half of our jellyfish candidates are located above both their sequences.

As a final check, and since jellyfish galaxies are very often systems with an increased star formation rate (see the Introduction), we verified if our results were affected by classification errors in the following manner. We divided our sample in two sub-samples based on the classF column of Table 2:  types 1 and 2 on one side, types 3, 4 and 5 on the other. We then checked the numbers of each type in Fig.~\ref{fig:m0717_mass_SFR} (left), to see if the galaxies with a SFR lower than the main sequence were in majority of types 1 or 2 (i.e. galaxies that may not be jellyfish galaxies after all). 
The percentages of jellyfish galaxy candidates of types 1 and 2 with a SFR below, in, and above the Peng sequence are comparable to those given above for all the jellyfish galaxy candidates. We therefore consider that galaxies with a low SFR are not necessarily doubtful jellyfish galaxies. This means that a number of our jellyfish candidates are indeed undergoing a phase of low 
star formation activity.

%------------------------------------------------------------------------------
\subsubsection{AGN activity in MACS0717 jellyfish candidates}
\label{subsubsec:AGN_0717}

An interesting link between the presence of jellyfish features and AGN activity was explored by \cite{PoggiantiNatur+17}, who found a strong correlation between these parameters for the most extreme examples of jellyfish galaxies in their GASP sample. Motivated by this, we checked if any of our jellyfish candidates in MACS0717 showed signs of hosting an AGN. This is a hard task with the limited data available so, considering the possibilities, we opted for using the MIR criteria, based on WISE colors, developed by \cite{Mateos+12} and \cite{Stern+12}. Both are optimized to select luminous AGN so we will likely just uncover the tip of the iceberg. 

We used the Table Access Protocol (TAP) Query service of TOPCAT \citep{Taylor+05} to access and download the DR8 tractor catalogue limited to the area covered by the HST observations. This catalogue contains magnitudes in all four WISE bands, extracted by the DESI team in preparation for their Legacy Survey, that go about 1 magnitude deeper than the original AllWISE ones (D. Schlegel, private communication). We converted these de-redenned AB magnitudes to the Vega system, following the DESI webpage information\footnote{https://www.legacysurvey.org/dr8/description/}, to apply directly the above mentioned criteria. We then matched this catalogue with our own: for a search radius of 1.5 arcsec, 79 of our 81 galaxies had data in the DR8 tractor catalogue, allowing us to check their W1-W2 color index. Out of these, only two barely pass the \cite{Stern+12} threshold for identifying AGN, i.e. a color index W1-W2>0.8, imposing as well a S/N>5 for the WISE individual magnitudes.
These are galaxies \#9 and \#59 of Table~\ref{tab:macs0717}.
Applying the stricter \cite{Mateos+12} criterion, in the W1-W2 versus W2-W3 color plane, results in no AGN candidates.

Our marginal AGN candidates are located in regions of high density. One lies on top of the 1 Mpc central radius (black circle in Fig.~\ref{fig:macs0717_posjelly}), to the right, has v/$\sigma _v ~\sim -1$, W1-W2=0.90 and J class 4 attributed by both FD and SC: it is galaxy \#9 of Table \ref{tab:macs0717}, marked with a red symbol in Fig.~\ref{fig:macs0717_posjelly}. The other one lies close to the northern edge of the yellow vertical ellipse of 
Fig.~\ref{fig:macs0717_posjelly}, where it is marked by a blue symbol (galaxy \#63; J class 2 and 3 by FD and SC, respectively) and has v/$\sigma _v ~\sim -0.5$ and W1-W2=0.84.
The respective stellar masses of galaxies \#9 and \#59 are $5.7\times 10^9$ and $5.5\times 10^9$~M$_\odot$,
and their respective SFRs are 6.9 and 14.7~M$_\odot$~yr$^{-1}$. However, if these galaxies indeed host an AGN, then the LePhare output parameters can only be taken as indicative since they are likely affected by larger uncertainties (the SED fits did not take into account any AGN contribution).

\subsubsection{SED fitting results for other CLASH clusters}
\label{subsubsec:SED_32}

\begin{table}[h]
\centering
\caption{CLASH clusters in which the spectral energy distribution (SED) of candidate 
  jellyfish galaxies was analysed. Columns are: cluster name and number of jellyfish candidates for
  which the SED was analysed.}
  \label{tab:32gal}
\begin{tabular}{lc}
\hline
\hline
  Cluster              & \# jellyfish \\
                       &   candidates \\
\hline
  Abell~209            &  2       \\ 
  Abell~383            &  2        \\
  MACS~J0416.1-2403    &  8       \\
  MACS~J0429.6-0253    &  2        \\
  MACS~J1149.5+2223    &   4        \\
  MACS~J1206.2-0847    &   3        \\
  LCDCS~0829           &   1        \\
  RX~J1532.9+3021      &   1        \\
  MACS~J1931.8-2634    &   2        \\
  RXC~J2248.7-4431     &   5        \\
  \hline
\end{tabular}
\end{table}

In our cluster sample, there are 11 CLASH clusters listed in Table~\ref{tab:32gal}, in which we find a total of 31 jellyfish galaxy candidates. These galaxies have optical and infrared magnitudes available (up to 17 bands between 225~nm and 1.6~$\mu$m, from the CLASH program, see section \ref{sec:sample}).

We fit SEDs of these 31 galaxies in the same way as those in MACS0717 with GAZPAR/LePhare. 
The best fit template spectrum includes an H$\alpha$ line for all but two galaxies and, except for these two, more than half of the
best fit template spectra include all main emission lines in the optical ([OII]3727, [OIII]4959, 5007, H$\beta$ and H$\alpha$), here also suggesting that the majority of these jellyfish galaxy candidates are forming stars. 
 
Similar plots to those shown for MACS0717 are given in the right panels of Figs.~\ref{fig:m0717_histo_mass} to 
\ref{fig:m0717_mass_SFR}. 
We can see that stellar masses of these 31 galaxies cover a range comparable to that covered by the jellyfish galaxy candidates in MACS0717. Their SFRs are also comparable to those in MACS0717, but we can note that only two galaxies out of 31 (6\%) have SFR $ > 10$~M$_\odot$~yr$^{-1}$ (though one is possibly overestimated), whereas there are 25 (32\%) in MACS0717. Fig.~\ref{fig:m0717_mass_SFR} also shows that while MACS0717 seems to be quite rich in jellyfish galaxy candidates having a high SFR, for the other clusters a clear assessment for comparison is difficult due to their incomplete coverage. The distribution of the specific star formation rate is quite different in the two samples (Fig.~\ref{fig:m0717_histo_sSFR}): 35 (44\%) of the galaxies in MACS0717 have sSFR $\geq 10^{-9}$~yr$^{-1}$ while the distribution of sSFRs for the 31 other jellyfish galaxy candidates is ``smoother", with 12 (39\%) galaxies having sSFR $\geq 10^{-10}$~yr$^{-1}$, though very few are quiescent (sSFR $<10^{-11}$~yr$^{-1}$). Consistently, the distributions of the mean ages of the stellar populations of the two samples are also quite different, as seen in Fig.~\ref{fig:m0717_histo_age}: half of the galaxies in MACS0717 are on average younger than $1.5\times 10^9$~yrs, while the age distribution in the sample of 31 galaxies is flatter.

Again, we underline that these comparisons are merely indicative, since the sample of 31 jellyfish galaxy candidates results from a spectroscopic and spatial coverage that is quite incomplete, and therefore by no means identical to that of MACS0717.

We also looked for variations of the stellar mass, SFR and sSFR with redshift for these 31 galaxies, but the dispersion is large, specially for the first two quantities, so we cannot claim there are clear correlations.

Finally, we checked if there was a correlation between the jellyfish classification and the stellar mass, both in the 79 galaxies of MACS0717 and in the 31 galaxies belonging to other clusters, and indeed found none, in agreement with the results of  \citet{Poggianti+16}.

\subsection{Clusters with no jellyfish galaxy candidate identified}

There are 17 clusters in our sample in which no jellyfish galaxy candidate was detected (see Table~\ref{tab:clusters}).
For 13 of them, there are only between one and three galaxies with a spectroscopic redshift available at the cluster redshift, and located in the zone covered by {\it HST} images. For the remaining four clusters, BMW-HRI\_J122657.3+333253, ZwCl~1332.8+5043, MACS J1423+24, and Abell~2261, there are respectively 23, 6, 7, and 14 galaxies with a spectroscopic redshift available at the cluster redshift, and located in the zone covered by {\it HST}. So obviously, at least in the last three of these four clusters, the absence of jellyfish galaxies in these clusters can simply be due to the small number of available redshifts.

\section{Discussion and conclusions}
\label{sec:conclusion}

We searched for jellyfish galaxy candidates in an initial sample of 40 clusters in the redshift range $0.2<z<0.9$ from the DAFT/FADA and CLASH surveys with HST optical images available. To this purpose, two of us examined the shapes of all galaxies with a spectroscopic redshift in the approximate cluster range. This approach led us to find one or several jellyfish galaxy candidates in 23 clusters (from the original set of 40), that were classified from J=1 to J=5, following the classification scheme proposed by \citet{Ebeling+14}. In the remaining 17 clusters we found no jellyfish candidate. We  analysed cluster MACS0717 separately, because it has a large {\it HST} coverage and spectroscopic catalogue: in this system, we found 81 jellyfish candidates. In the remaining 22 clusters, we detect a total of 97 jellyfish galaxy candidates.

For all these jellyfish candidates, we give positions, redshifts, magnitudes in one or two optical filters (usually F606W and F814W), and show images in the Appendix. Whenever images are available for the galaxies in these two wavebands, we provide both: the comparison of galaxy images in the F606W and F814W filters shows that our candidates are morphologically quite similar in both, but with more evidence for star formation in the bluer filter. as expected. 

A colour-magnitude diagram for MACS0717 shows that most of the 81 jellyfish candidates are blue and located below the cluster red sequence. This is reinforced by the UVJ diagram \citep{Williams+09}, where only two galaxies appear to be quenched; all remaining ones seem to be SF galaxies, even if several of them lie in the region of dust obscured objects (thus can be red sequence objects). 
For the 79 jellyfish candidates in this cluster having multiwavelength data available, the SED fitting that was carried out with LePhare finds that almost all are best fit by template spectra that have one or several of the main optical emission lines usually associated with ongoing star formation. As a consequence, the stellar mass – SFR plane (where these quantities were also obtained by LePhare directly from the best fit template spectra), shows that at least 80\% of the jellyfish candidates are star-forming galaxies - and among these SF systems, about 70\% have increased SFR relatively to the main sequence galaxies. The SED fit results thus provide another indication that the majority of jellyfish galaxies in this cluster have notably high SFR for their stellar masses (about 60\% have sSFR $> 10^{-10}$~yr$^{-1}$, stellar masses ranging from $10^{9.15}$ to $10^{10.6}$~M$_\odot$). Though affected by the usual uncertainties associated with any SED fitting method, this result is similar to what was obtained at low redshift by \citet{Poggianti+16}.

If we now look at the location of these galaxies inside the cluster, their redshift histogram does not hint for any particular placement along the line-of-sight: jellyfish candidates share the global redshift distribution of all galaxies within the structure (i.e. all galaxies within the adopted redshift interval for this system), so it is impossible to infer any particular kinematical behavior. As for their spatial distribution on the plane of the sky, jellyfish candidates spread throughout the cluster and its extended filament but avoid the cluster central, densest region. Since MACS0717 filaments are well detected in projection, and the redshift histogram of the whole structure (Fig.~\ref{fig:macs0717_histoz}) is, rather surprisingly, Gaussian-like, it does look as if the main infall, at large scale, should essentially take place along the plane of the sky, in the areas where we found our jellyfish candidates. 

In the cluster core, and apparently in compliance with the model of \cite{SafarzadehLoeb19}, jellyfish galaxy candidates are almost absent, which may further be a result of the very rough core environment of this massive merging cluster. A detailed study of the inner 1 Mpc carried out with combined optical and X-ray data by \cite{Ma+09} does point to much more complex dynamics in the cluster core (when compared with what we can infer from the global redshift histogram of Fig.~\ref{fig:macs0717_histoz}). Interestingly enough, \citet{Ellien+19} analyzed the distribution of intra-cluster light (ICL) in this system and detected a large amount of it in the cluster core but no such light in the cosmic filament. As ICL is thought to be made up of disrupted galactic material, the findings of \citet{Ellien+19} corroborate the existence of a harsh environment in the core, contrasting with softer conditions along the filament. This is expected and the distribution of jellyfish galaxy candidates found in this work may reflect that, quite probably, transient phenomena such as jellyfish features cannot survive for long in the tumultuous core of MACS0717.

From all these pieces of evidence, we think that most jellyfish candidates identified here could be a population of rather recent infallers that have felt the first impact and effects of penetrating into a denser environment, which altered their morphology and generally increased their SFR. A tentative MIR analysis singles out two possible AGN hosts among the jellyfish candidates located in the denser cluster regions.

On the other hand, the apparent paucity of jellyfish galaxies in the cluster core could simply be the result of a selection effect, imposed by the wavelengths that we rely upon to identify them - probing, approximately, restframe B and V optical emission. Though these wavebands have been successfully used by \cite{Poggianti+16} to select their GASP jellyfish candidates at redshifts 0.04-0.07, we expect that at the higher redshifts of our sample, the surface brightness of jellyfish structures becomes significantly dimmer. Moreover, tails and other jellyfish characteristics can indeed lie undetected in some filters since the material making up the jellyfish structures emits at selective wavelengths, as documented by several examples mentioned in Section \ref{sec:intro}. Just to highlight a couple of illustrative examples, and regardless of their location within the host cluster, we can mention FGC1287 in Abell~1367 that has a 250 kpc-long HI tail with no optical counterpart \citep{Scott+12}, and D100 in Coma, which presents a remarkably long and narrow ($60\times 1.5$~kpc) H$\alpha$ gas tail, whereas the optical image shows an apparently normal spiral galaxy \citep{Cramer+19}.
Such galaxies with tails are rather extreme examples of jellyfish characteristics but they seem to be located - at least in what concerns the low-redshift universe, i.e. mostly below z=0.05 but reaching up to z=0.2 - within the inner 40\% of their cluster virial radius (i.e. within 1~Mpc) with very few exceptions (T. Scott, private communication). This is the case even for those with stellar tails - that could thus potentially be unveiled in our observations. So, as far as optically detected jellyfish galaxy candidates are concerned, there don’t seem to be many in the inner regions of MACS0717, at least as far as our images’ depth can probe, and none shows a conspicuous tail.\\

Unlike the large spectroscopic coverage we have for MACS0717, that allowed us to detect in a more complete way jellyfish candidates in this cluster, the lack of redshifts in the case of most of the other clusters that we analysed prevents us from drawing major conclusions. Our aim here was simply to detect candidates, characterize them and make, whenever possible, a comparison with what was found for MAC0717. The jellyfish candidates detected in all remaining 22 clusters cover the same stellar masse range but don’t seem to avoid the centre of their host clusters, as seen in Fig.~\ref{fig:dist_r}. This might explain the generally lower SFR values (only reaching about  $10$~M$_\odot$~yr$^{-1}$) when compared with jellyfish candidates that are members of MACS0717. 

Finally, and considering the whole sample together, we next attempt to infer proportions of jellyfish galaxy candidates in the clusters analysed here and the existence of any trend with cluster relaxation states.

We already mentioned the incompleteness of our redshift catalogues for the various clusters. As jellyfish galaxies are most often quite bright, and/or show emission lines in their spectra, they may be easier to observe spectroscopically, and thus their proportion may be overestimated. Besides, since our aim is to detect jellyfish galaxies, we may have classified as such, galaxies which are merely somewhat strange--looking spirals with a distorted morphology. For these reasons, and due to the incompleteness of our sample, estimating the proportion of jellyfish galaxies in clusters remains difficult. We will therefore just give a few numbers. If we consider the 21 clusters for which more than ten spectroscopic redshifts are available, we find an average proportion of jellyfish galaxy candidates of 9.5\% (by using the numbers in Table~\ref{tab:clusters}). If we now consider MACS0717, where statistics are more robust (81 jellyfish candidates detected, with a large spectroscopic and spatial coverage), we find a percentage of 13\%. This seems to mean that jellyfish galaxies are not that rare after all, and it is a clear encouragement to pursue such studies with more complete spectroscopic data.

Since the main mechanisms leading to jellyfish galaxies appear to be RPS and/or harassment \citet[see the Introduction and][]{Poggianti+17}, the proportion of jellyfish galaxies can be expected to vary with the relaxation state of clusters to which they belong. To estimate this relaxation state, we looked at several properties: first, the  histograms of all the redshifts available in a large zone around each cluster (for the DAFT/FADA clusters, these histograms were given by \citet{Guennou+14} and \citet{Durret+16}), for the other clusters we retrieved all the redshifts available in NED and drew their histogram in the approximate cluster redshift range. We also looked at
the matter distribution based on a weak lensing analysis by \citet{Martinet+16} or on the shape of the red sequence density map by \citet{Durret+16} or \citet{Durret+19}. 
For clusters with at least 10 spectroscopic redshifts available within the {\it HST} images analyzed in the present paper, we looked at the proportion of jellyfish galaxy candidates relative to the relaxation state of the cluster. With the available data, we found no relation between the proportion of jellyfish galaxy candidates and the relaxation state of the cluster.

\begin{acknowledgements}
   
It is a pleasure to thank Tom Scott for enlightening discussions about RPS and its effect on cluster galaxies. We are grateful to the referee for her/his prompt report and interesting suggestions that helped to improve the paper.
F. Durret acknowledges continuous financial support from CNES since 2002. C. Lobo acknowledges support by Funda\c cao para a Ci\^encia e a Tecnologia (FCT) through the research grants UIDB/04434/2020 and UIDP/04434/2020. 
M. Jauzac is supported by the United Kingdom Research and Innovation (UKRI) Future Leaders Fellowship `Using Cosmic Beasts to uncover the Nature of Dark Matter' (grant number MR/S017216/1).

This work is partly based on tools and data products produced by GAZPAR operated by CeSAM-LAM and IAP and we further acknowledge the dedicated support of O. Ilbert.
This research has made use of the SVO Filter Profile Service (http://svo2.cab.inta-csic.es/theory/fps/) supported from the Spanish MINECO through grant AYA2017-84089.
This research has also made use of the NASA/IPAC Extragalactic Database (NED), which is funded by the National Aeronautics and Space Administration and operated by the California Institute of Technology.
We further acknowledge M. Taylor for developing TOPCAT, making our life so much easier.

\end{acknowledgements}

\bibliographystyle{aa}
\bibliography{biblio-jelly.bib}

\begin{appendix}

\section{Images of jellyfish galaxies in MACS~J0717.5+3745}
 \label{appendix1}

The images of our 81 jellyfish galaxy candidates in MACS0717
are shown below. For each galaxy, we indicate the classifications estimated by two of us in parentheses (as given in Table~\ref{tab:galaxies}).

\begin{figure*}[h]
\begin{center}
\caption{MACS~J0717.5+3745: galaxies \#1 (type 3-2), \#2 (3-2), \#3 (2-3), \#4 (2-1), \#5 (1-2), \#6 (1-2), \#7 (2-3), \#8 (3-3), \#9 (4-4), \#10 (3-3)
\#8 (4-4), and \#11 (3-3) in the F606W (left) and F814W (right) filters.}
  \includegraphics[width=7cm]{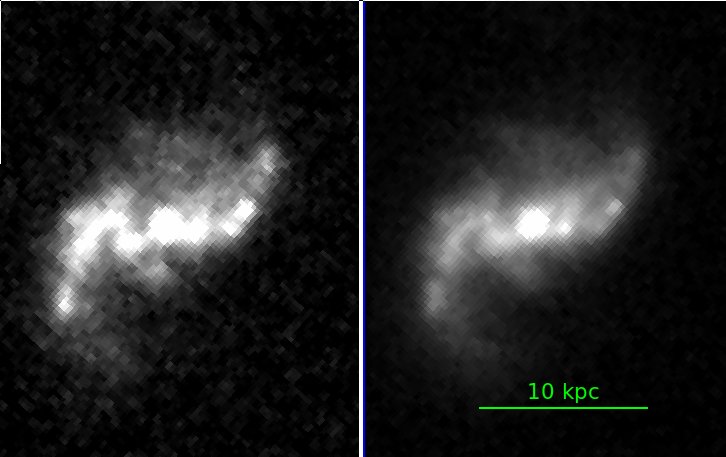}  \kern0.1cm%
  \includegraphics[width=7cm]{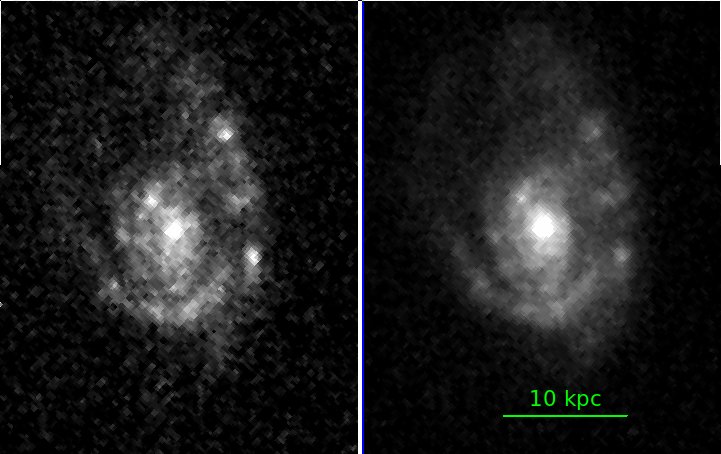} 
  \includegraphics[width=7cm]{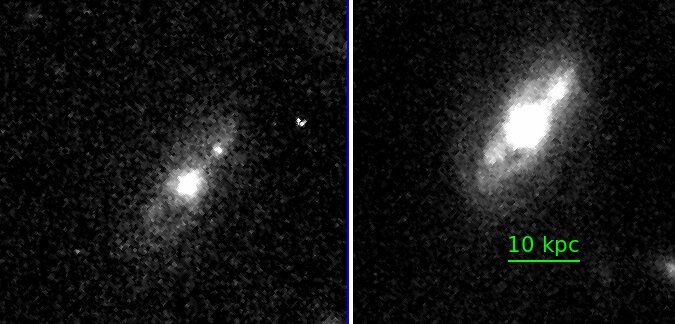}  \kern0.1cm%
  \includegraphics[width=7.5cm]{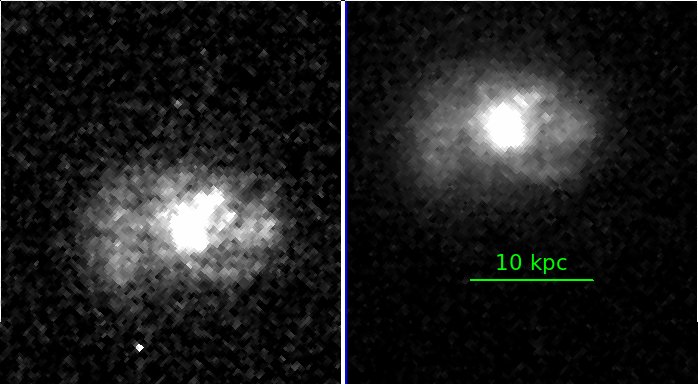} 
  \includegraphics[width=7cm]{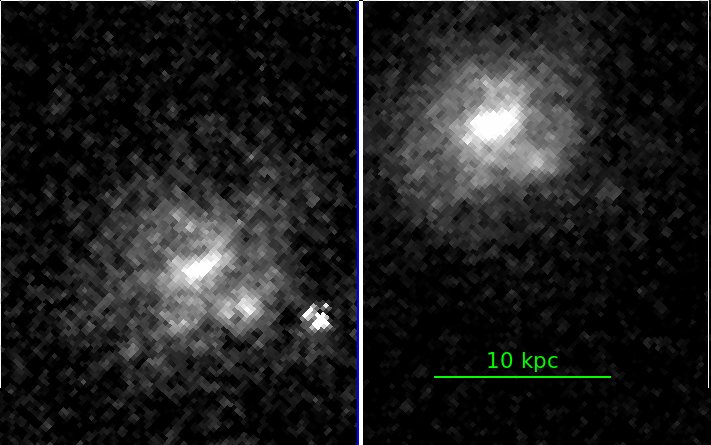}  \kern0.1cm%
  \includegraphics[width=7cm]{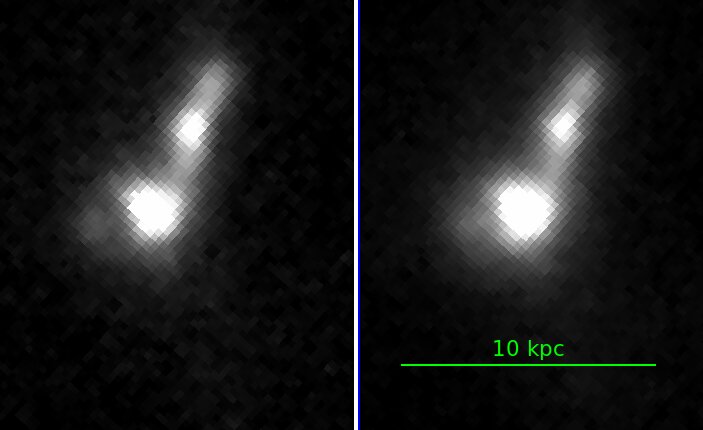}  
  \includegraphics[width=7cm]{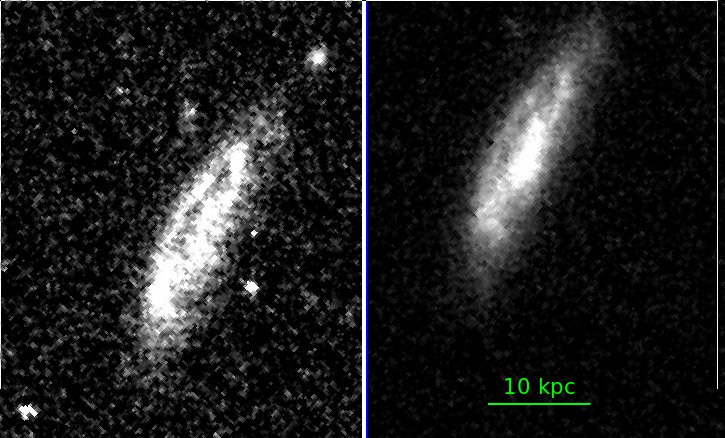} \kern0.1cm%
  \includegraphics[width=7cm]{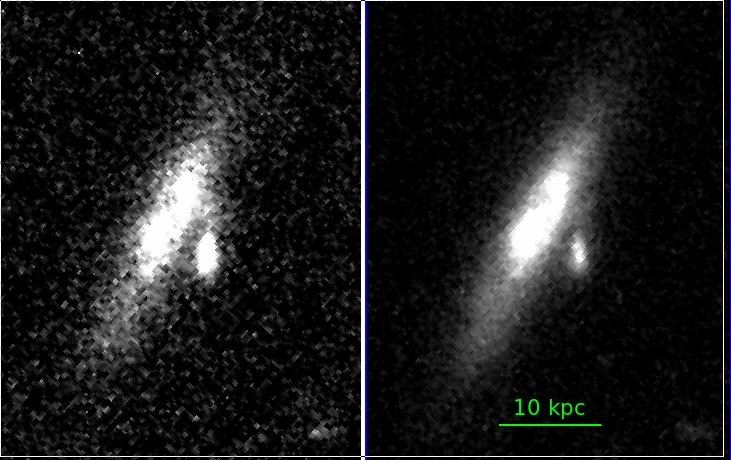}  
  \includegraphics[width=7cm]{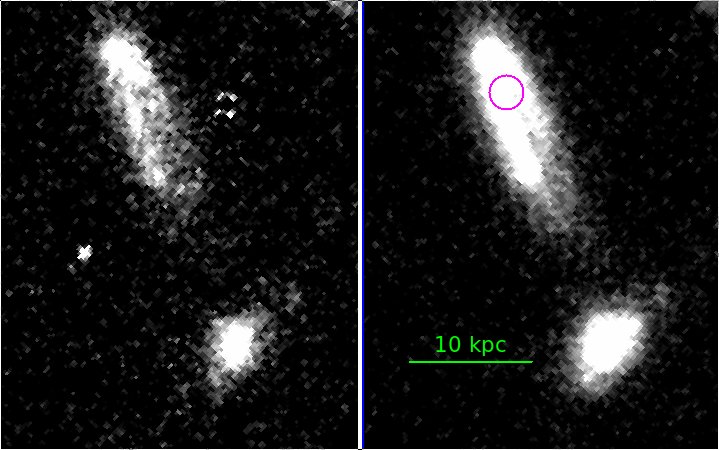} \kern0.1cm%
  \includegraphics[width=7cm]{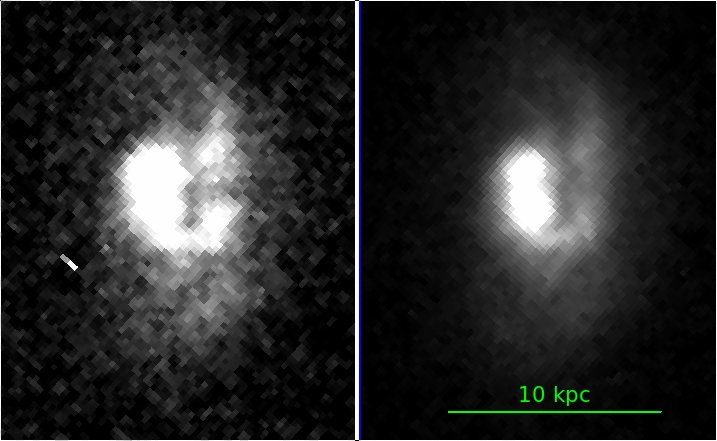} 
\end{center}
\end{figure*}
  
\begin{figure*}[h]
\begin{center}
  \caption{MACS~J0717.5+3745: galaxies \#11 (5-5), \#12 (4-4), \#13 (2-2), \#14 (2-2), \#15 (1-1), 
\#16 (3-2), \#17 (1-1), \#18 (4-4), \#19 (2-2), \#20 (2-4) in the F606W (left) and F814W (right) filters. }
  \includegraphics[width=7cm]{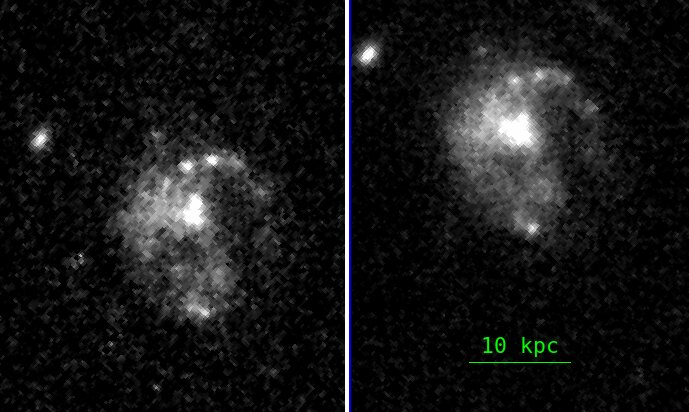}  \kern0.1cm%
  \includegraphics[width=7cm]{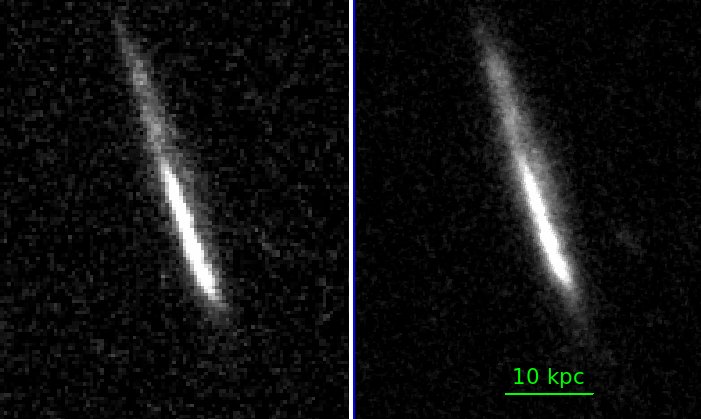}  
  \includegraphics[width=7cm]{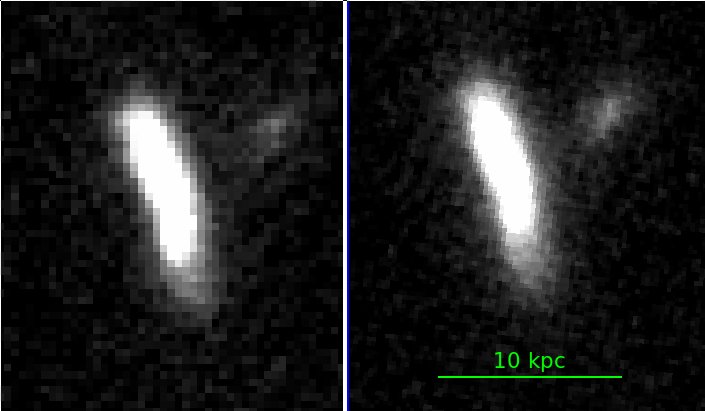} \kern0.1cm%
  \includegraphics[width=7cm]{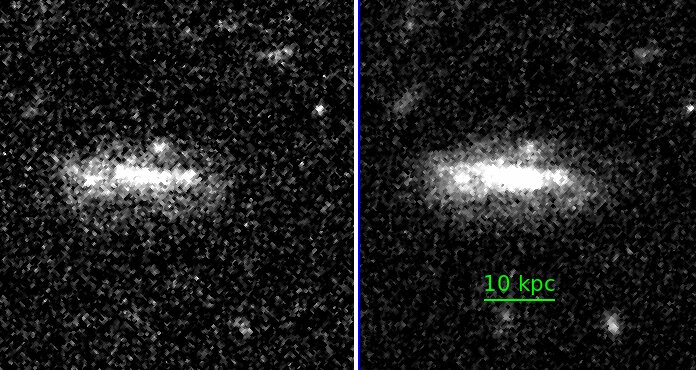}  
  \includegraphics[width=7cm]{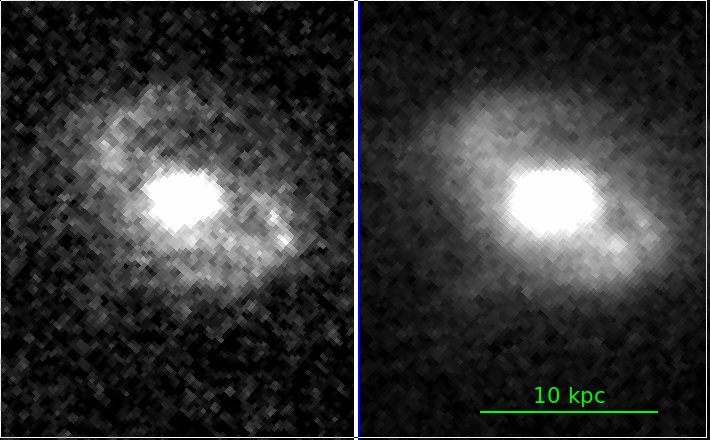} \kern0.1cm%
   \includegraphics[width=7.5cm]{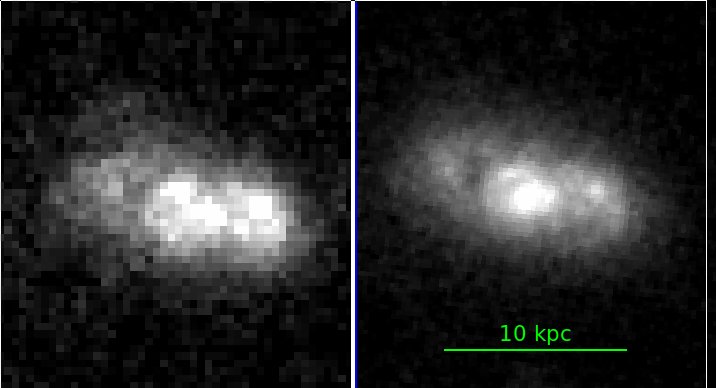} 
  \includegraphics[width=7.5cm]{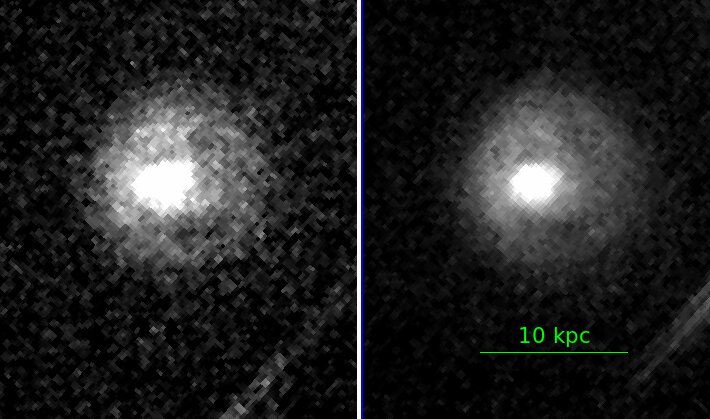}  \kern0.1cm%
\includegraphics[width=7cm]{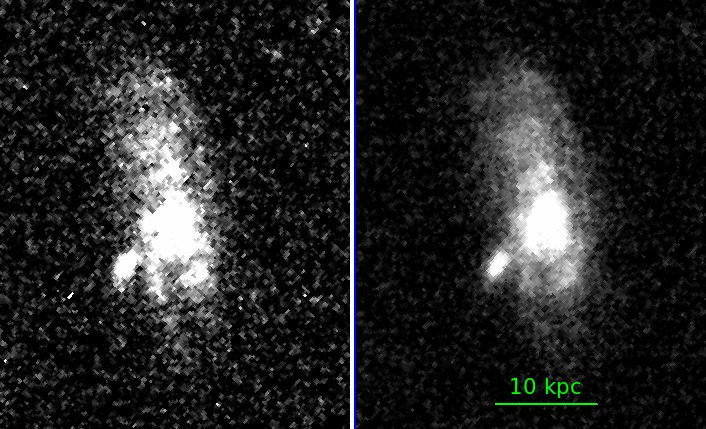}  
\includegraphics[width=7cm]{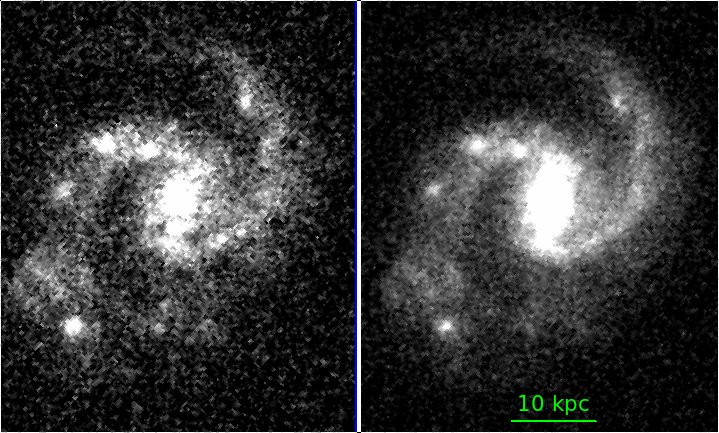} \kern0.1cm%  
  \includegraphics[width=7cm]{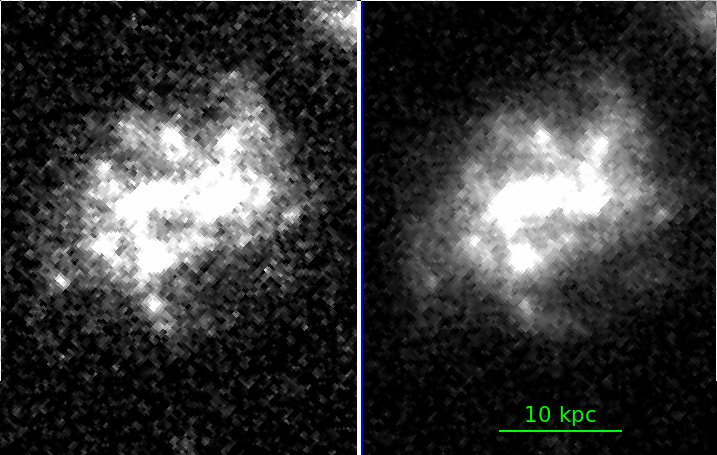} 
\end{center}
\end{figure*}

\begin{figure*}[h]
\begin{center}
  \caption{MACS~J0717.5+3745: galaxies \#21 (3-2), \#22 (4-4), \#23 (4-3), \#24 (3-2), \#25 (3-2), \#26 (3-3), \#27 (2-2), \#28 (1-2), \#29 (2-0), and \#30 (1-0) 
  in the
    F606W (left) and F814W (right) filters. }
  \includegraphics[width=7cm]{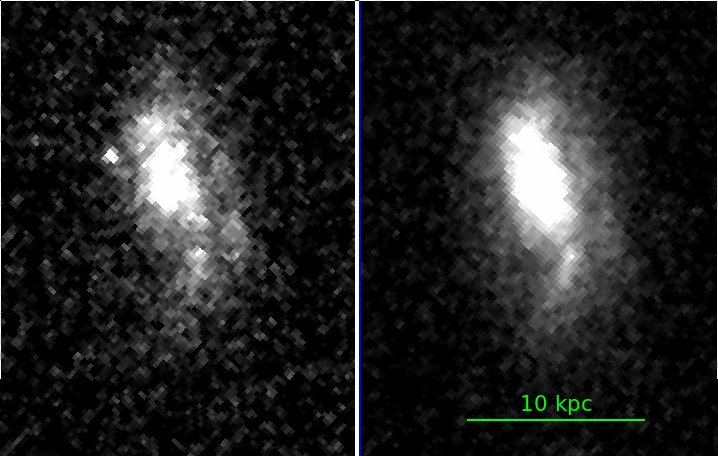} \kern0.1cm%
  \includegraphics[width=7cm]{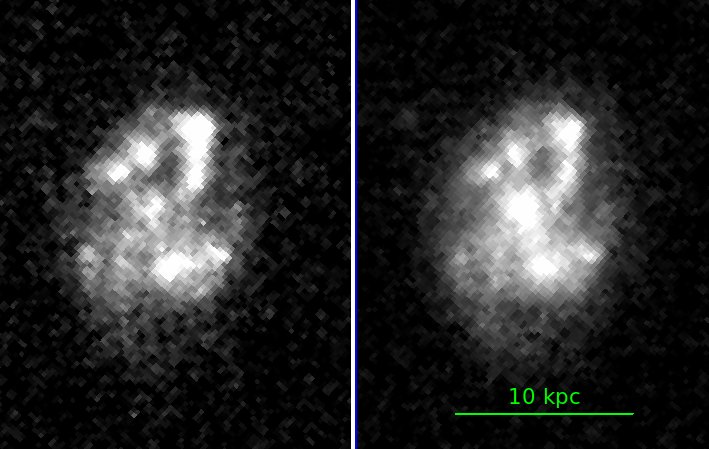} 
  \includegraphics[width=7cm]{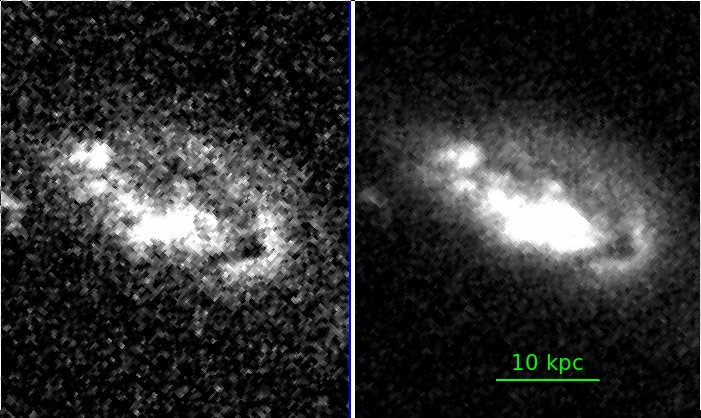} \kern0.1cm%
  \includegraphics[width=8cm]{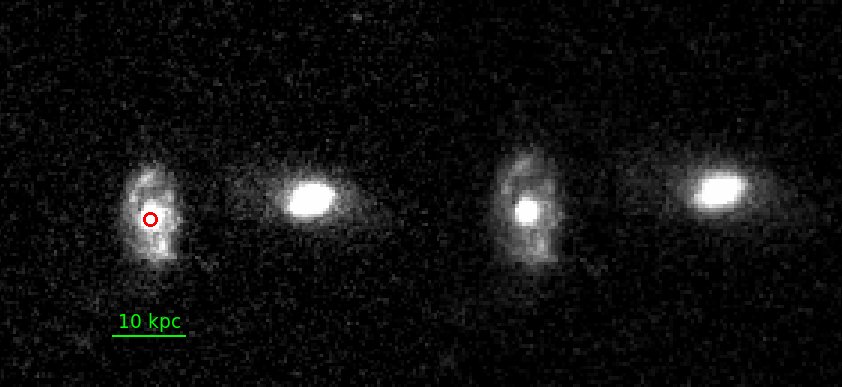} 
  \includegraphics[width=7.5cm]{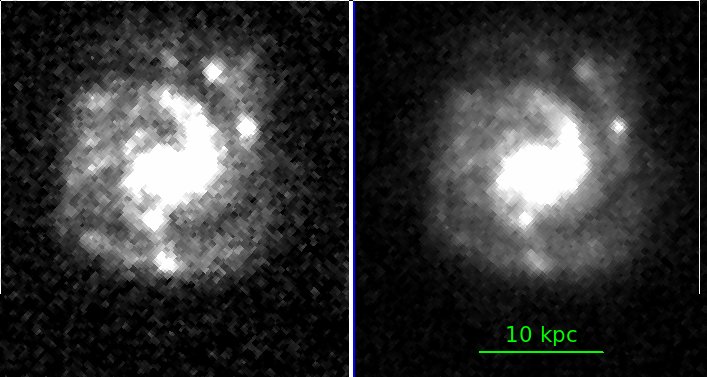} \kern0.1cm%
  \includegraphics[width=7cm]{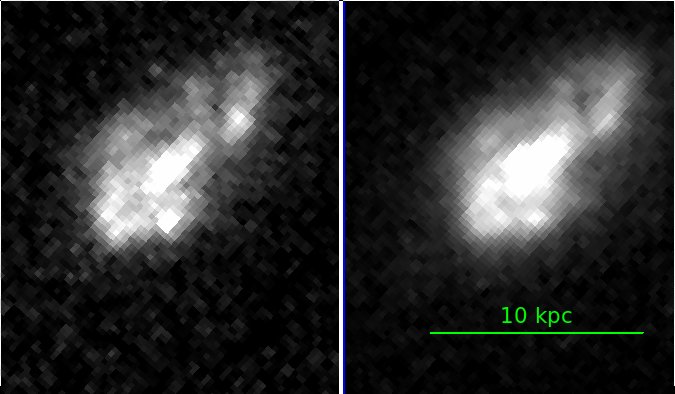} 
  \includegraphics[width=8cm]{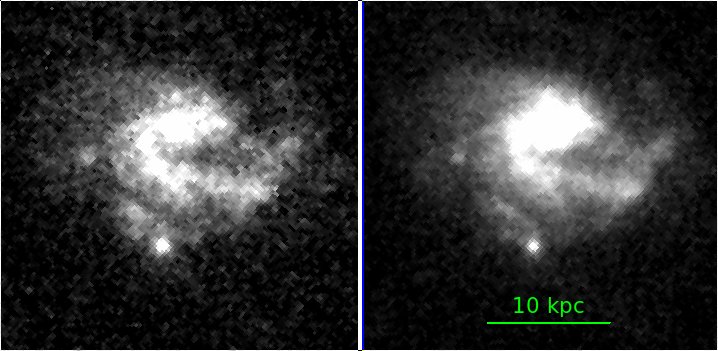} \kern0.1cm%
  \includegraphics[width=7cm]{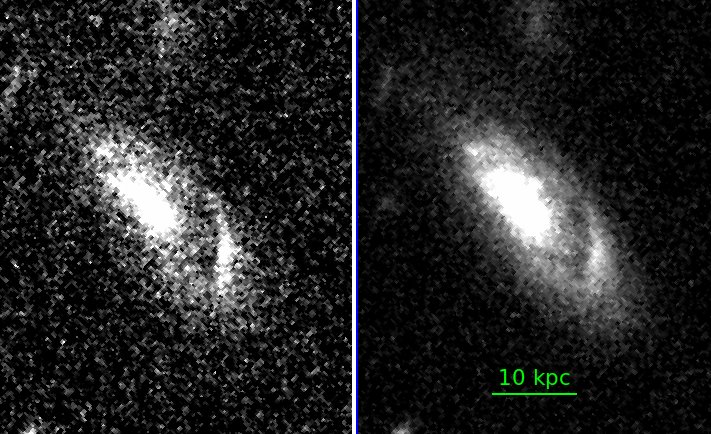} 
  \includegraphics[width=7cm]{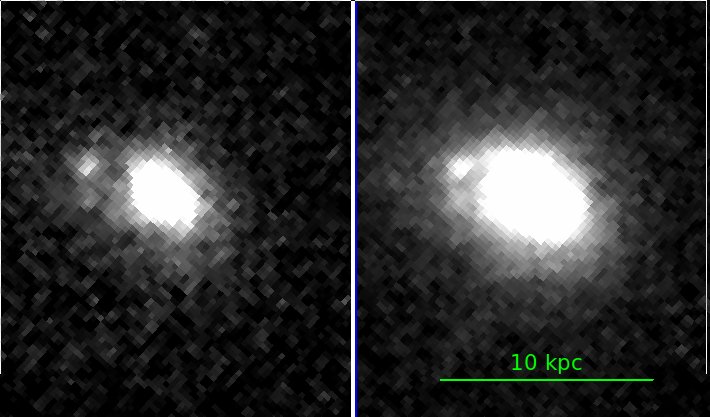} \kern0.1cm%
  \includegraphics[width=7cm]{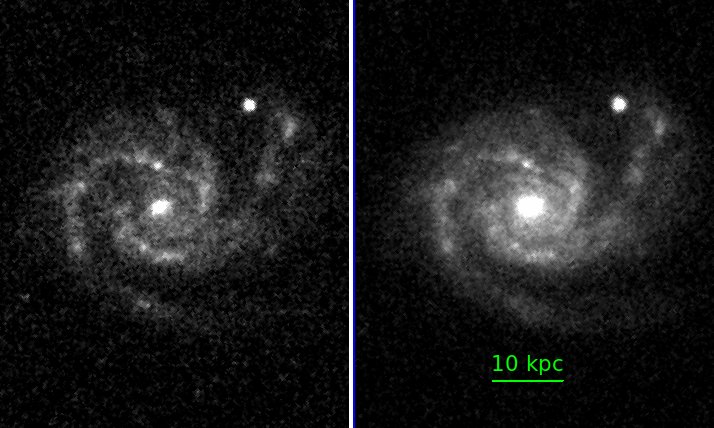} 
\end{center}
\end{figure*}

\begin{figure*}[h]
\begin{center}
  \caption{MACS~J0717.5+3745: galaxies \#31 (3-1), \#32 (1-0), \#33 (3-2), \#34 (2-3), \#35 (2-1), 
  \#36 (2-2), \#37 (2-2), \#38 (4-2),\#39 (5-4), and \#40 (5-2)
  in the F606W (left)
    and F814W (right) filters.}
  \includegraphics[width=7cm]{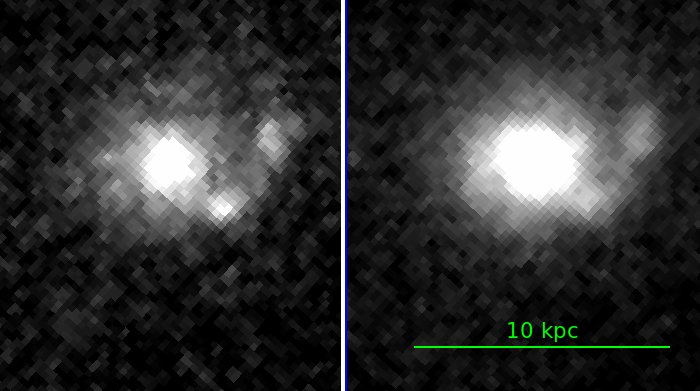} \kern0.1cm%
  \includegraphics[width=7cm]{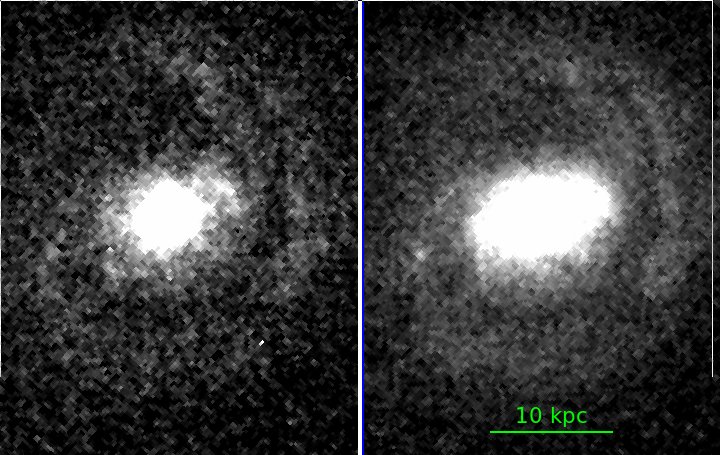} 
  \includegraphics[width=7cm]{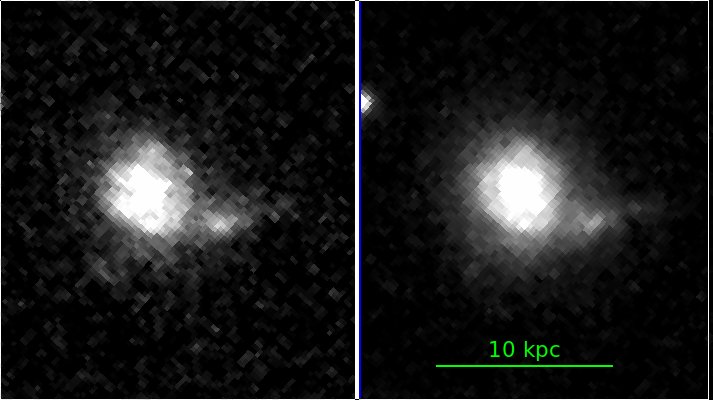} \kern0.1cm%
  \includegraphics[width=7cm]{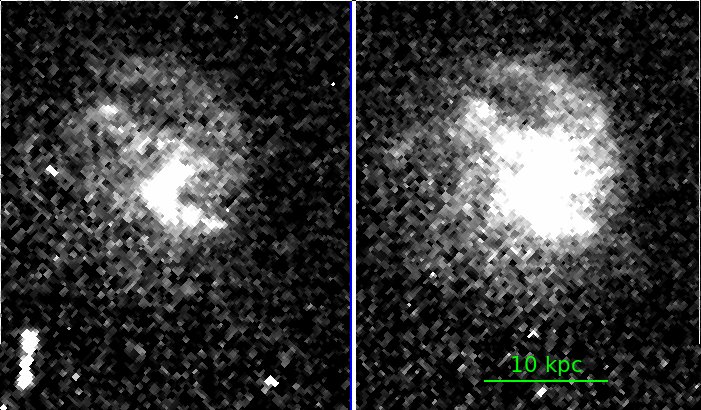}  
  \includegraphics[width=7cm]{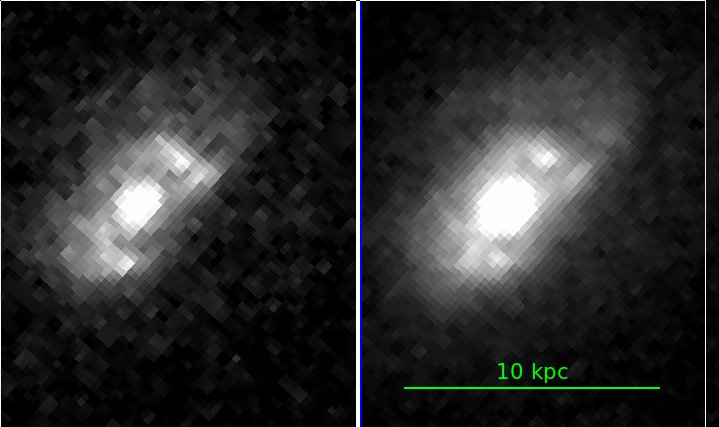} \kern0.1cm%
  \includegraphics[width=7cm]{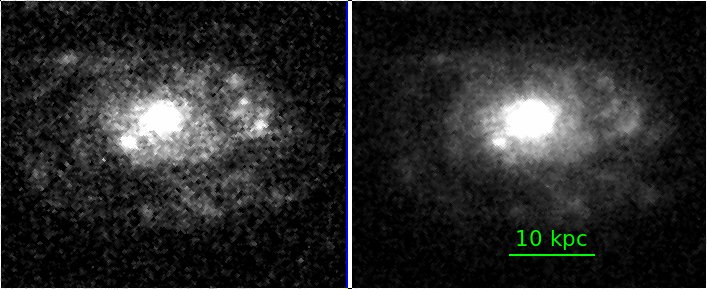}  
  \includegraphics[width=7cm]{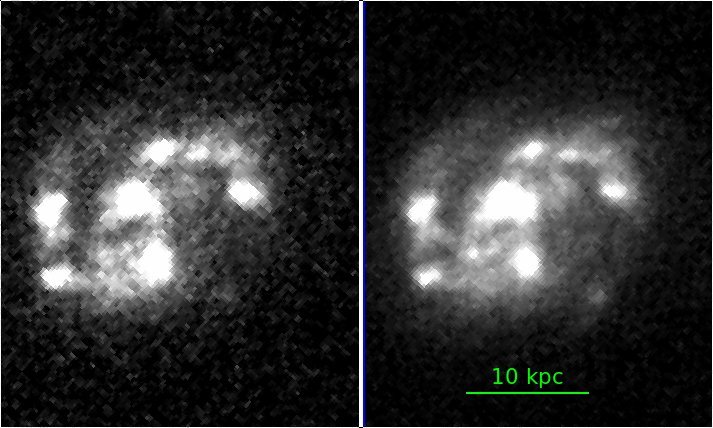} \kern0.1cm%
  \includegraphics[width=7cm]{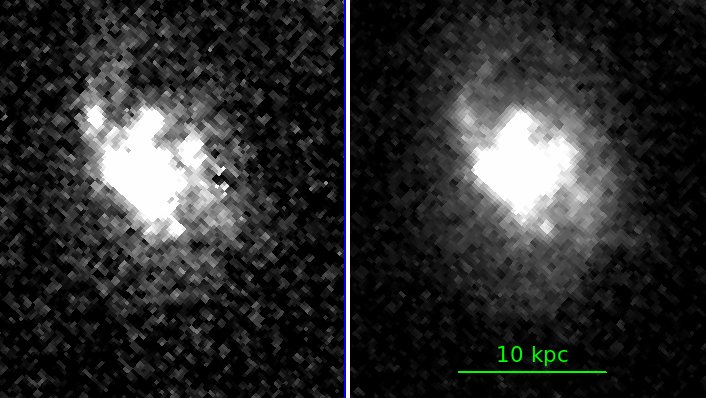}  
  \includegraphics[width=7cm]{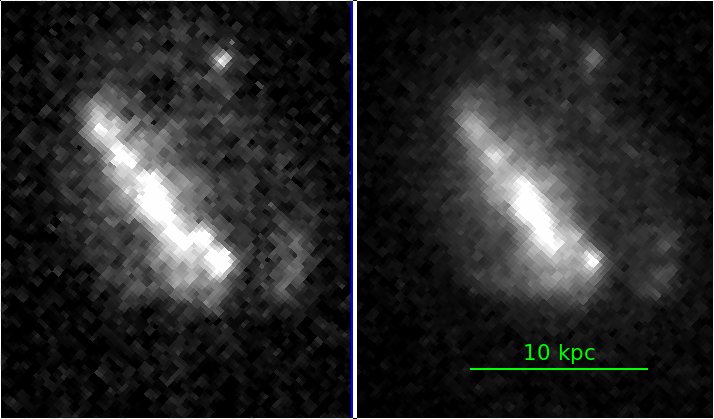} \kern0.1cm%
  \includegraphics[width=7cm]{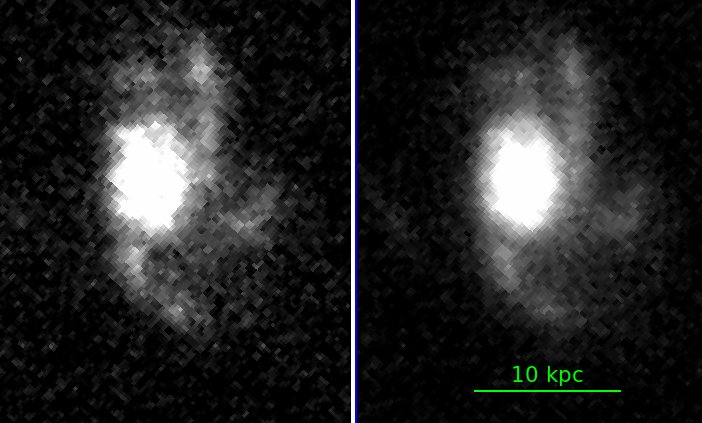}  
\end{center}
\end{figure*}

\begin{figure*}[h]
\begin{center}
  \caption{MACS~J0717.5+3745: galaxies  \#41 (4-1), \#42 (3-3), \#43 (2-2), \#44 (5-2), \#45 (1-0), \#46 (3-2), \#47 (3-0), \#48 (4-2),  \#49 (2-0), and \#50 (4-5) in the F606W (left)
    and F814W (right) filters. }
  \includegraphics[width=7cm]{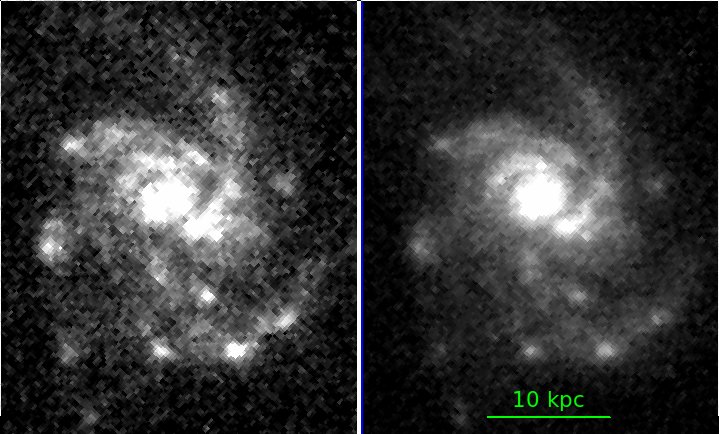} \kern0.1cm%
  \includegraphics[width=7cm]{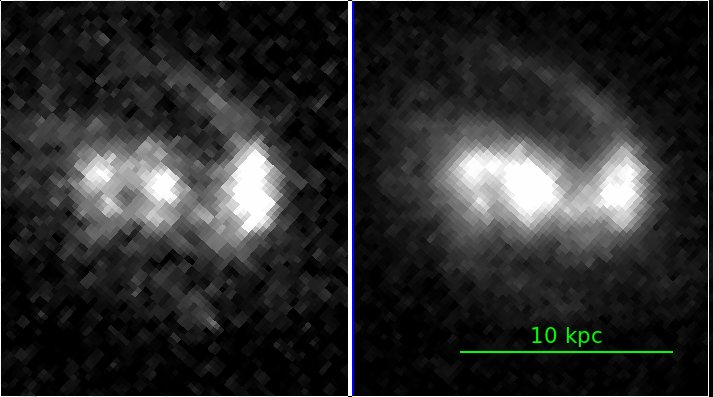} 
  \includegraphics[width=7cm]{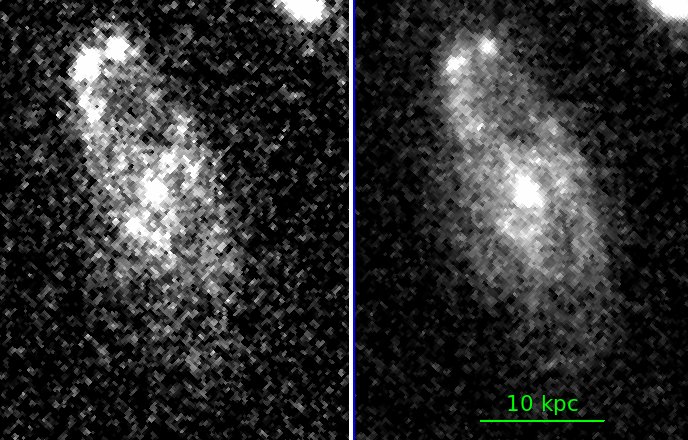} \kern0.1cm%
  \includegraphics[width=7cm]{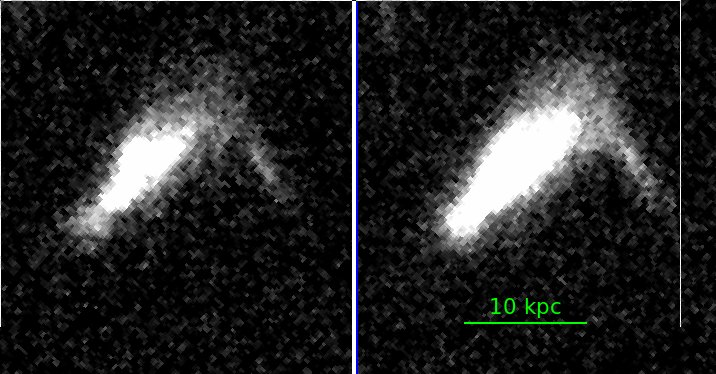} 
  \includegraphics[width=7cm]{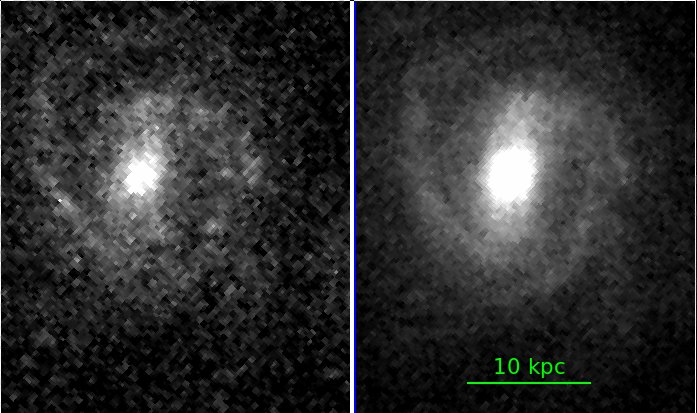} \kern0.1cm%
  \includegraphics[width=7cm]{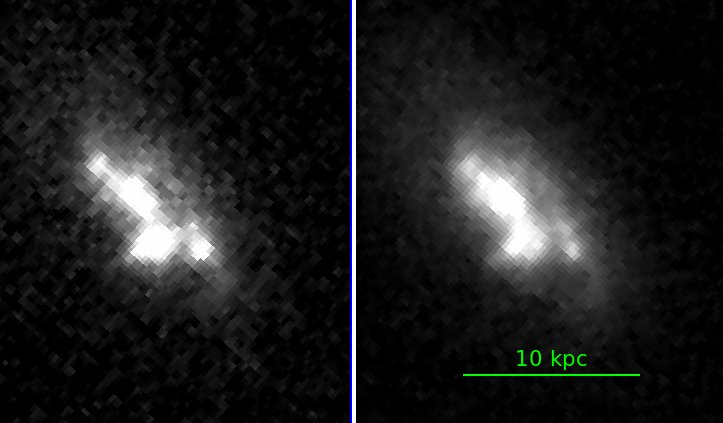} 
  \includegraphics[width=7cm]{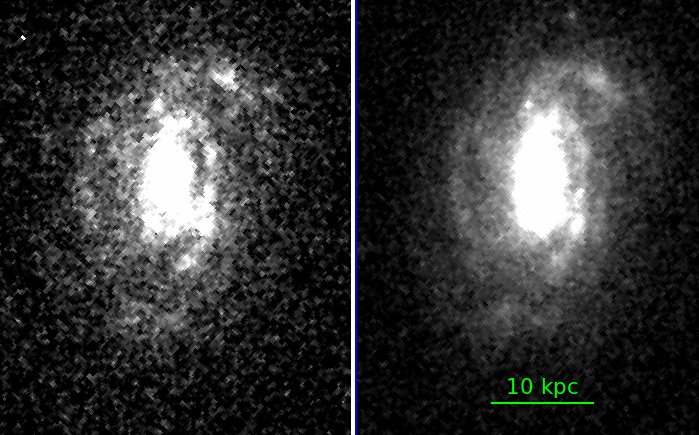} \kern0.1cm%
  \includegraphics[width=7cm]{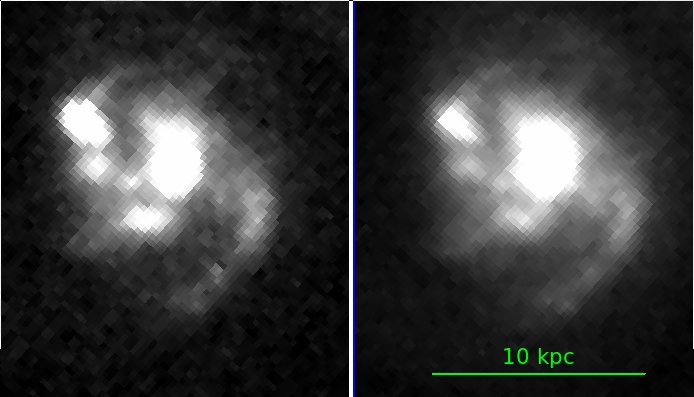}
  \includegraphics[width=7cm]{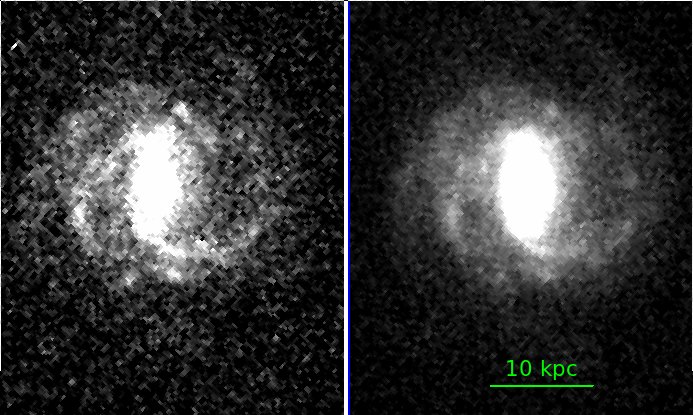} \kern0.1cm%
    \includegraphics[width=7cm]{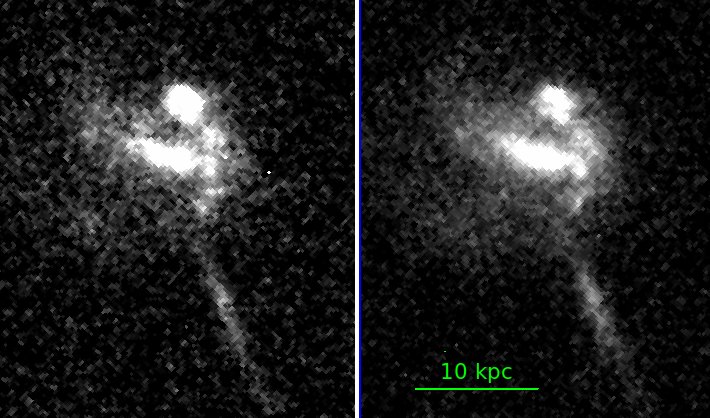}
\end{center}
\end{figure*}

\begin{figure*}[h]
\begin{center}
\caption{MACS~J0717.5+3745: galaxies  \#51 (4-3), \#52 (2-0), \#53 (2-3), \#54 (5-2), \#55 (4-3), 
\#56 (3-1), \#57 (3-3), \#58 (1-0), \#59 (3-2), and \#60 (2-2) in the F606W (left) and F814W (right) filters.}
  \includegraphics[width=7cm]{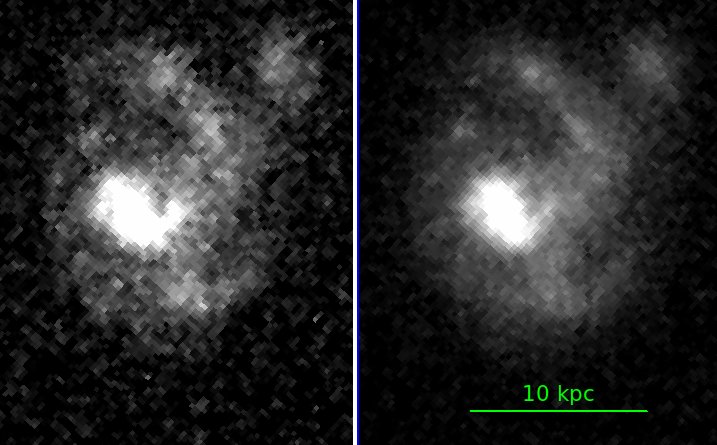}  \kern0.1cm%
  \includegraphics[width=7cm]{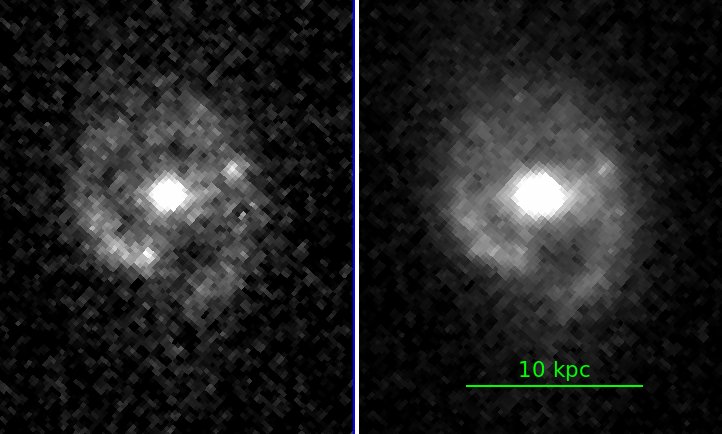} 
  \includegraphics[width=7cm]{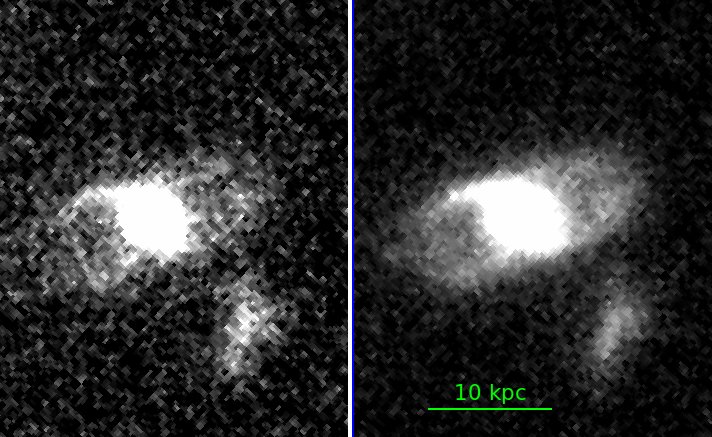}  \kern0.1cm%
  \includegraphics[width=7cm]{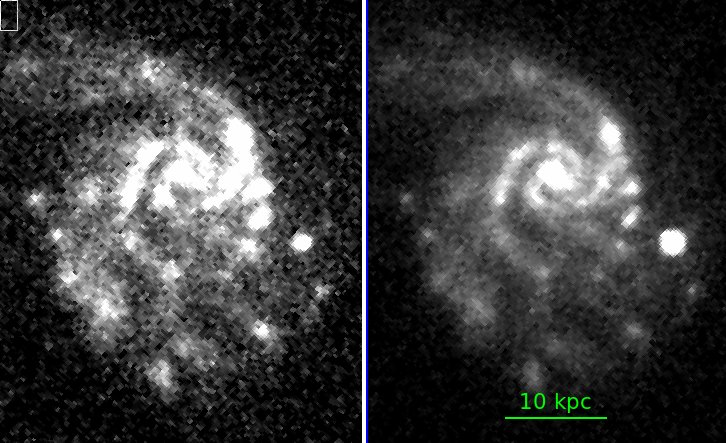} 
  \includegraphics[width=7cm]{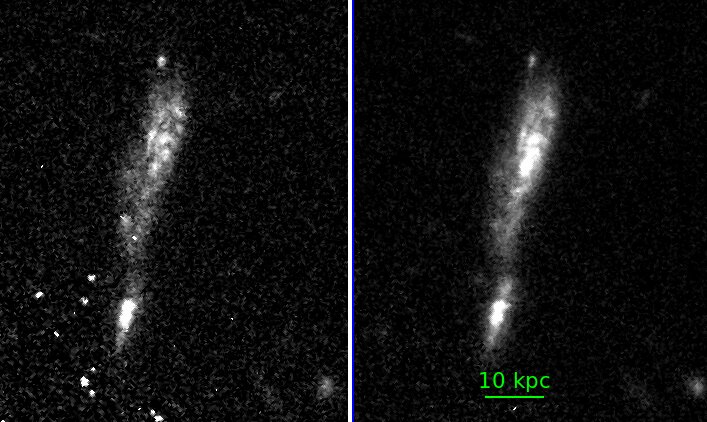} \kern0.1cm%
  \includegraphics[width=7cm]{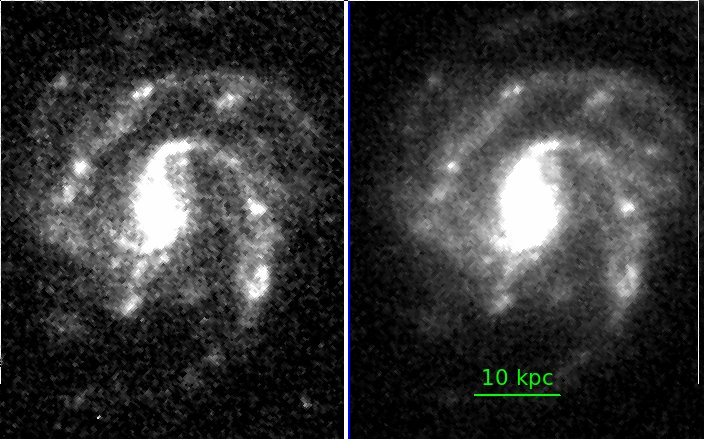}  
  \includegraphics[width=7cm]{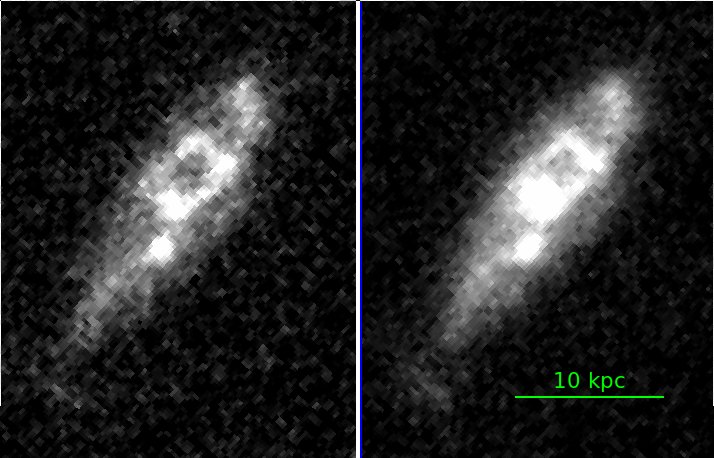} \kern0.1cm%
  \includegraphics[width=7cm]{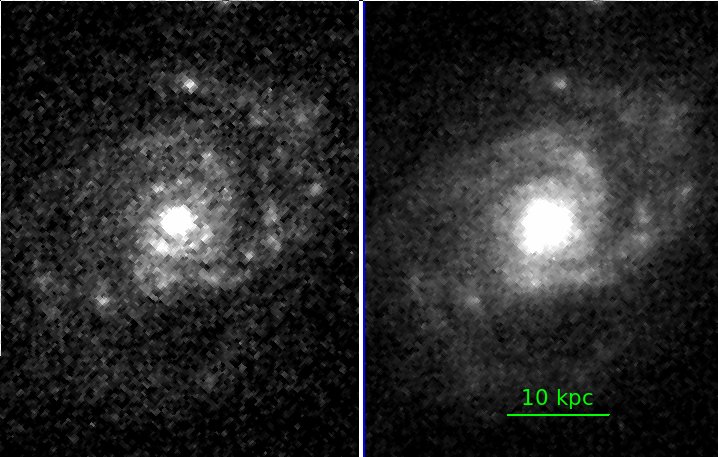}  
   \includegraphics[width=7cm]{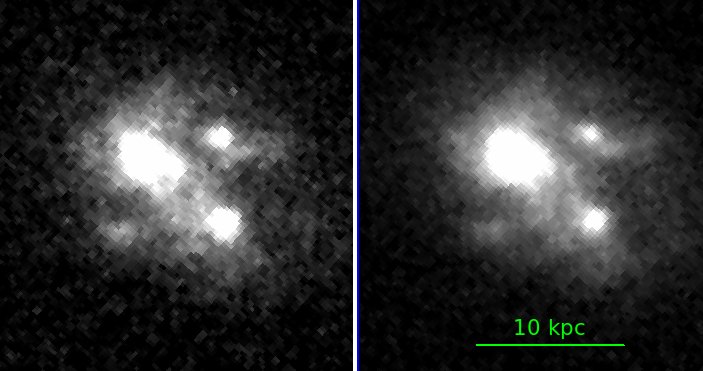} \kern0.1cm%
  \includegraphics[width=7cm]{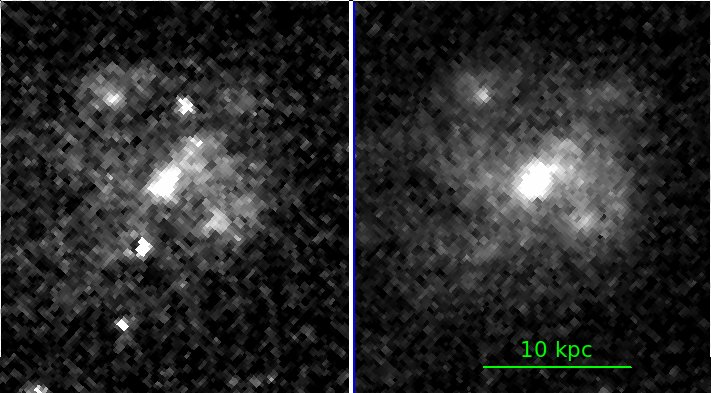} 
\end{center}
\end{figure*}

\begin{figure*}[h]
\begin{center}
\caption{MACS~J0717.5+3745: galaxies  \#61 (2-2), \#62 (4-2), \#63 (4-4), \#64 (5-4), \#65 (1-0), \#66 (3-2), \#67 (5-2), \#68 (4-2), \#69 (2-2), and \#70 (3-2),
in the F606W (left) and F814W (right) filters.}
  \includegraphics[width=7cm]{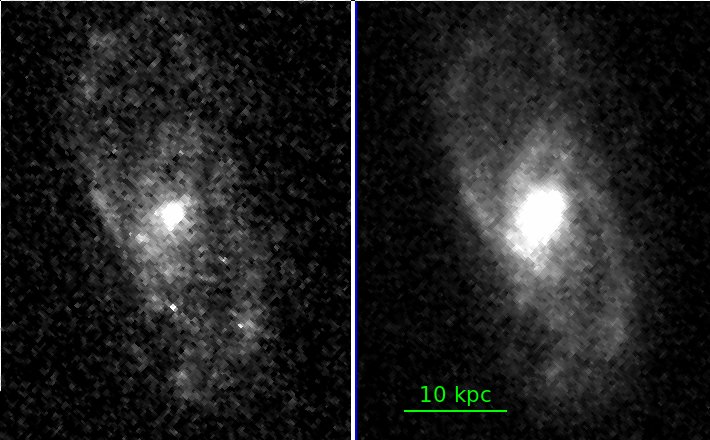} \kern0.1cm%
  \includegraphics[width=7cm]{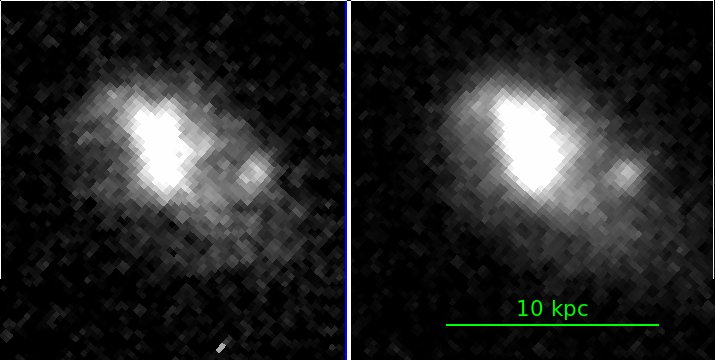} 
  \includegraphics[width=7cm]{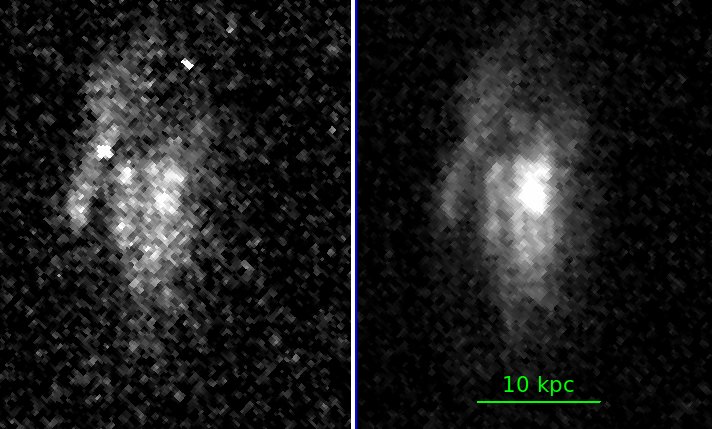} \kern0.1cm%
  \includegraphics[width=7cm]{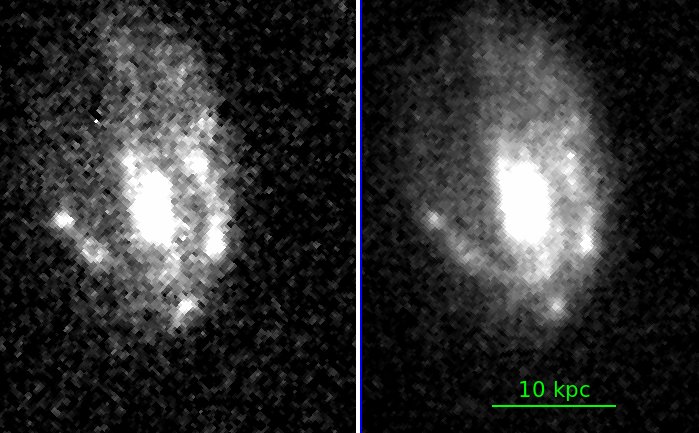} 
  \includegraphics[width=7cm]{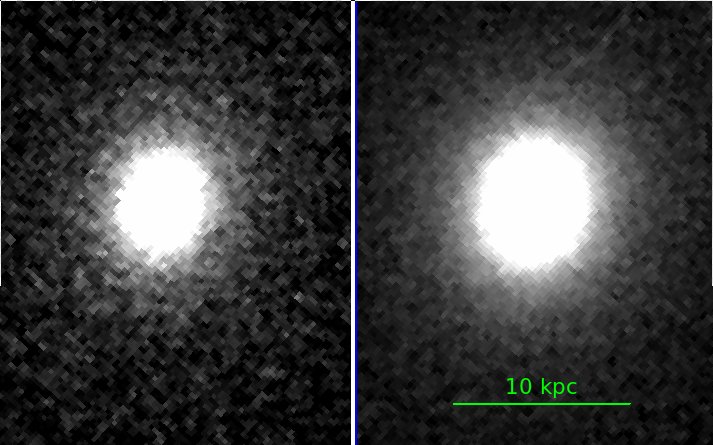}\kern0.1cm%
  \includegraphics[width=7cm]{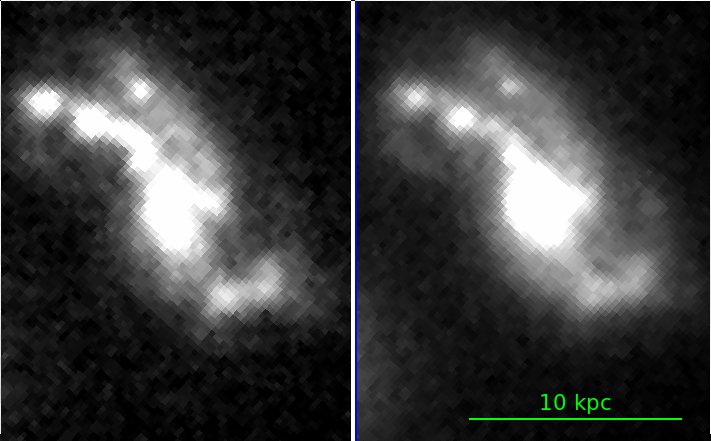}
  \includegraphics[width=7cm]{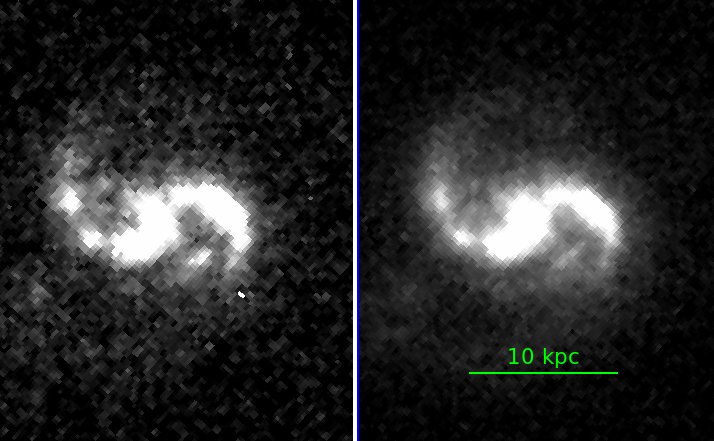} \kern0.1cm%  
  \includegraphics[width=7cm]{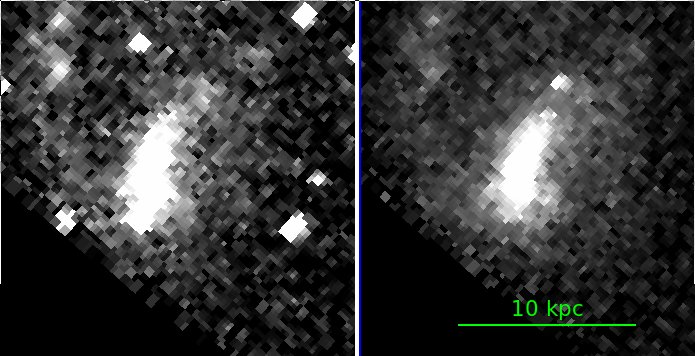} 
  \includegraphics[width=7cm]{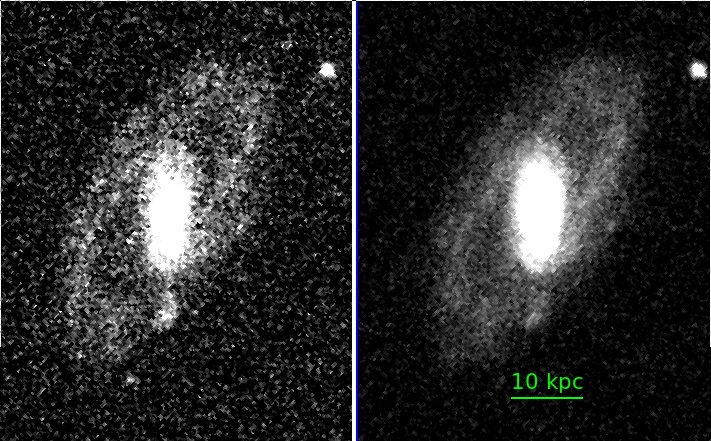} \kern0.1cm%
  \includegraphics[width=7cm]{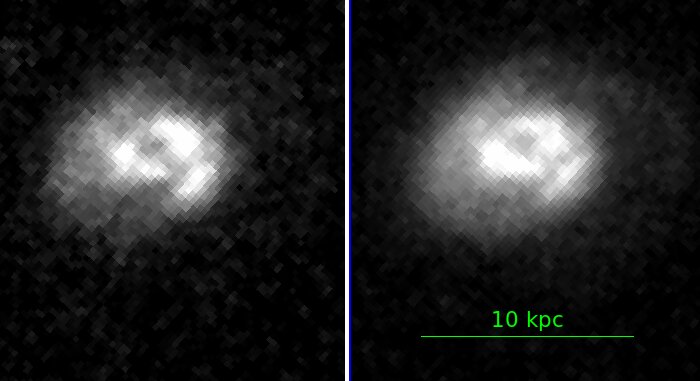} 
\end{center}
\end{figure*}

 \begin{figure*}[h]
\begin{center}
\caption{MACS~J0717.5+3745: galaxies   \#71 (1-0), \#72 (3-2), \#73 (4-2), \#74 (2-3), \#75 (4-0), \#76 (2-2), \#77 (2-0), and \#78 (5-4) 
in the F606W (left) and F814W (right) filters.}
  \includegraphics[width=7cm]{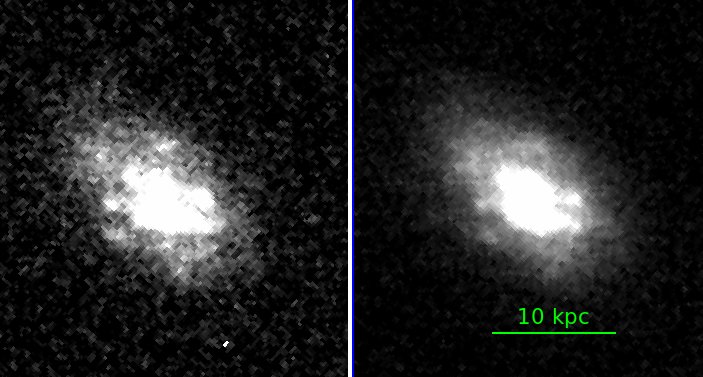} \kern0.1cm%
  \includegraphics[width=7cm]{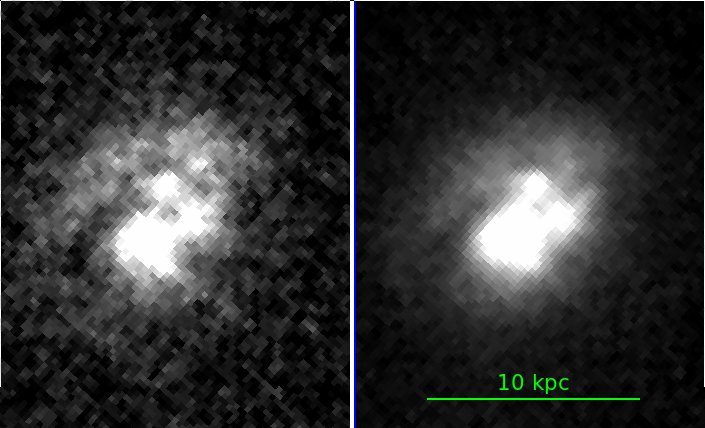} 
  \includegraphics[width=7cm]{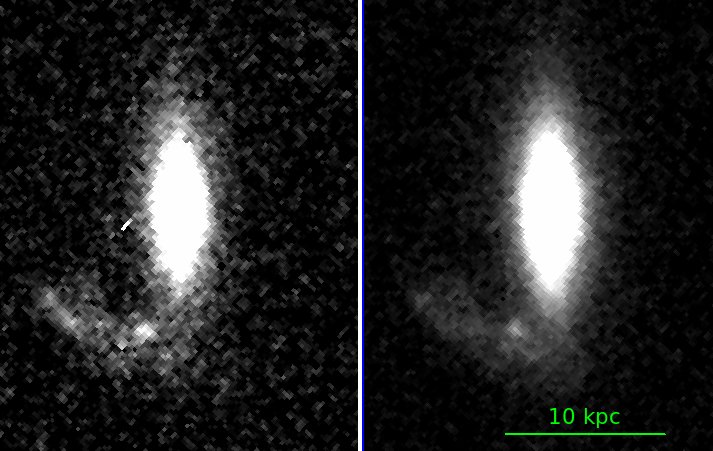} \kern0.1cm%
  \includegraphics[width=7cm]{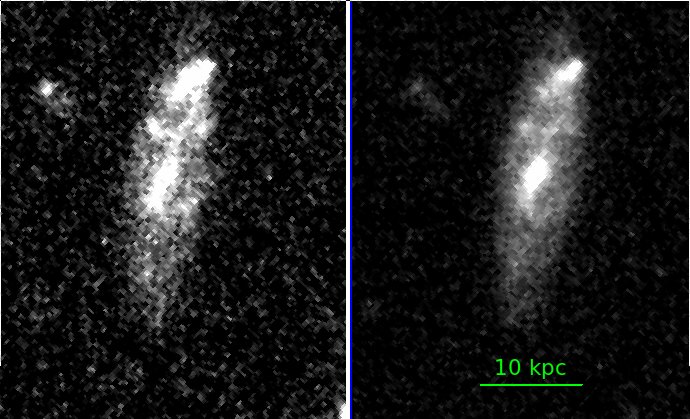} 
  \includegraphics[width=7cm]{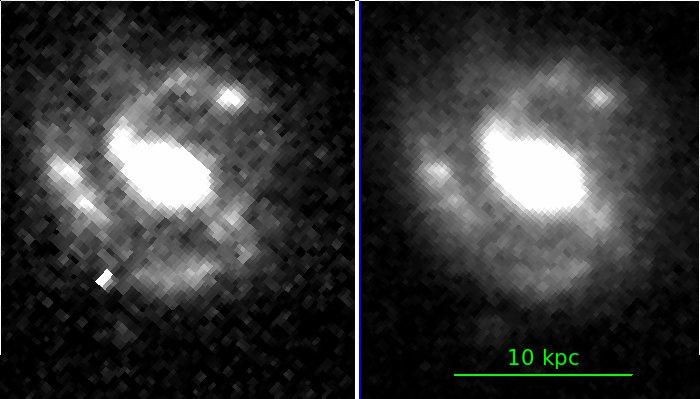} \kern0.1cm% 
  \includegraphics[width=7cm]{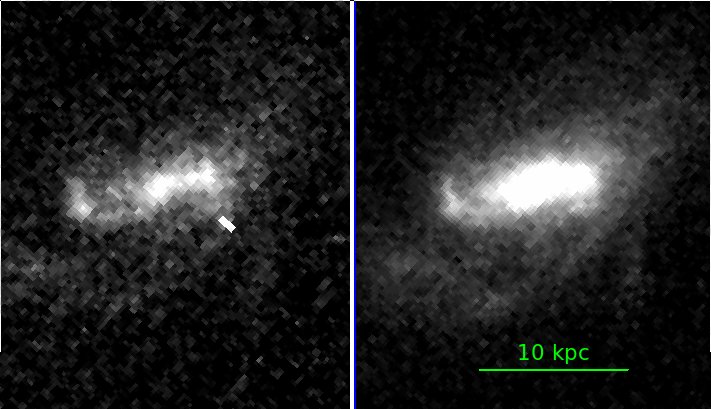} 
  \includegraphics[width=7cm]{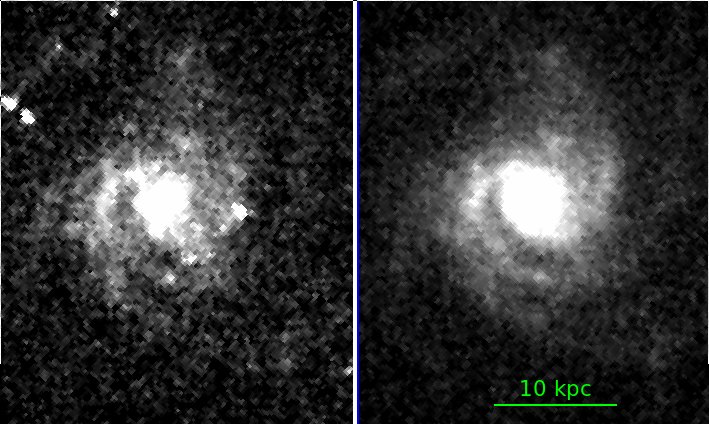} \kern0.1cm%
  \includegraphics[width=7cm]{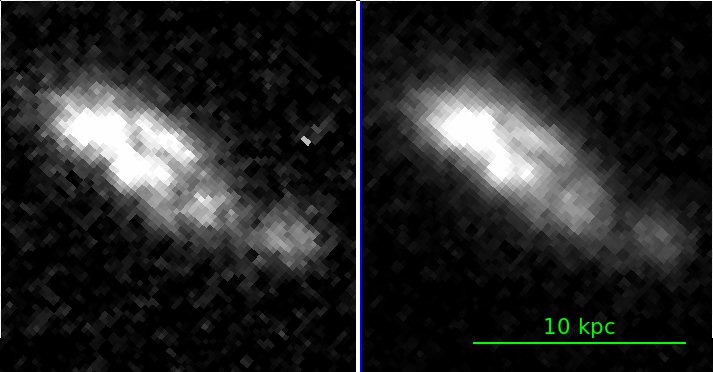} 
\end{center}
\end{figure*}

\begin{figure*}[ht]
\begin{center}
\caption{MACS~J0717.5+3745: galaxies  \#79 (4-3), \#80 (4-4) and \#81 (3-2) in the F606W (left) and F814W (right) filters.}
  \includegraphics[width=7cm]{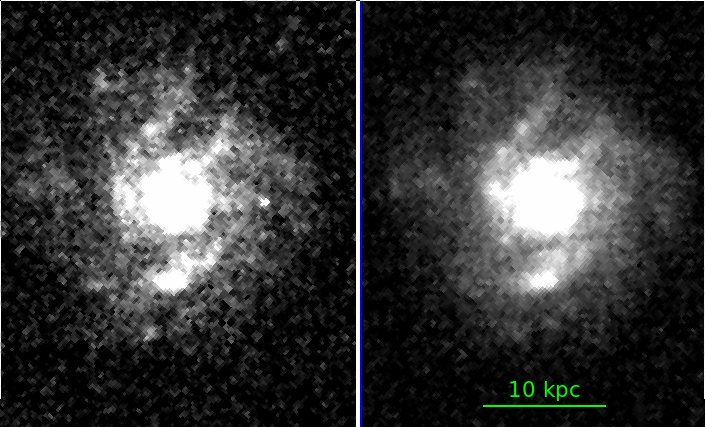}  \kern0.1cm%
  \includegraphics[width=7cm]{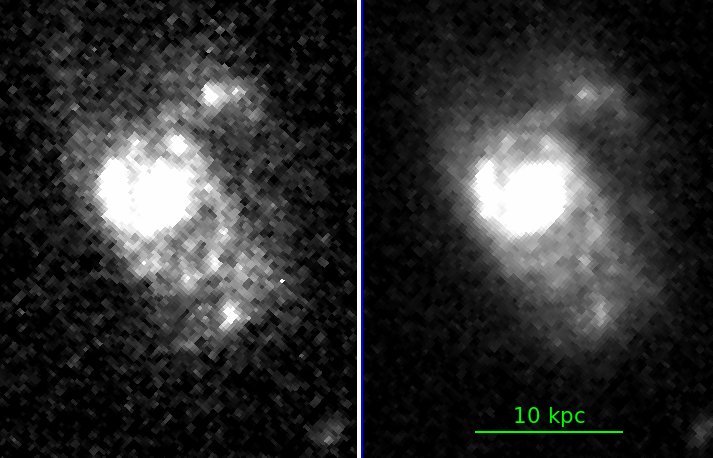} 
  \includegraphics[width=7cm]{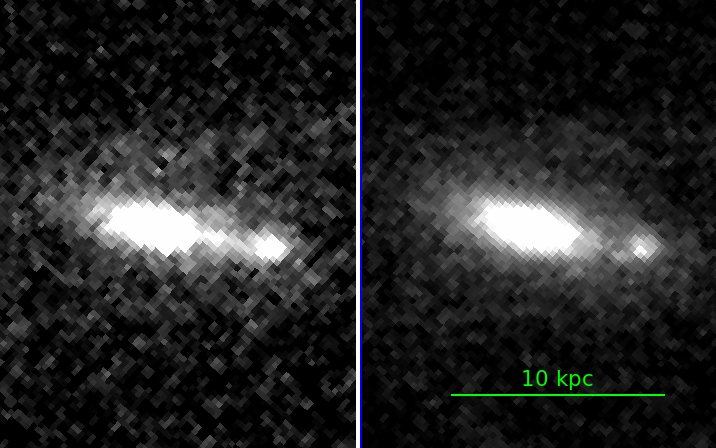} 
\end{center}
\end{figure*}

\section{Images of jellyfish galaxies in all clusters except MACS~J0717.5+3745}
\label{appendix2}

\begin{figure*}[h]
\begin{center}
  \includegraphics[width=5cm]{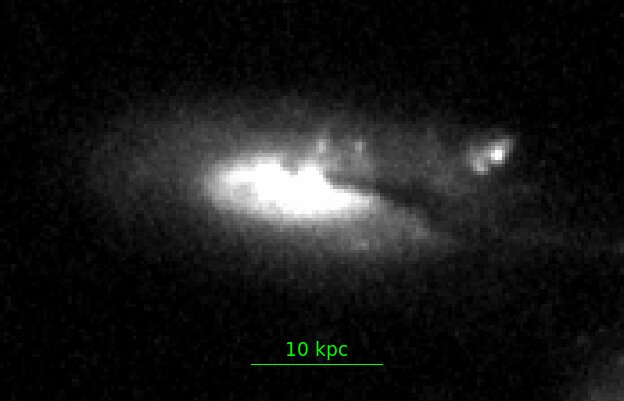}
  \includegraphics[width=5cm]{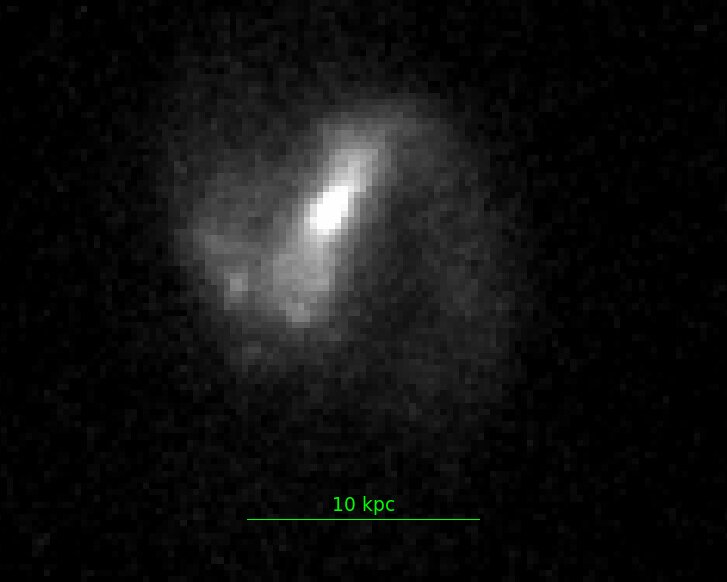}
  \includegraphics[width=5cm]{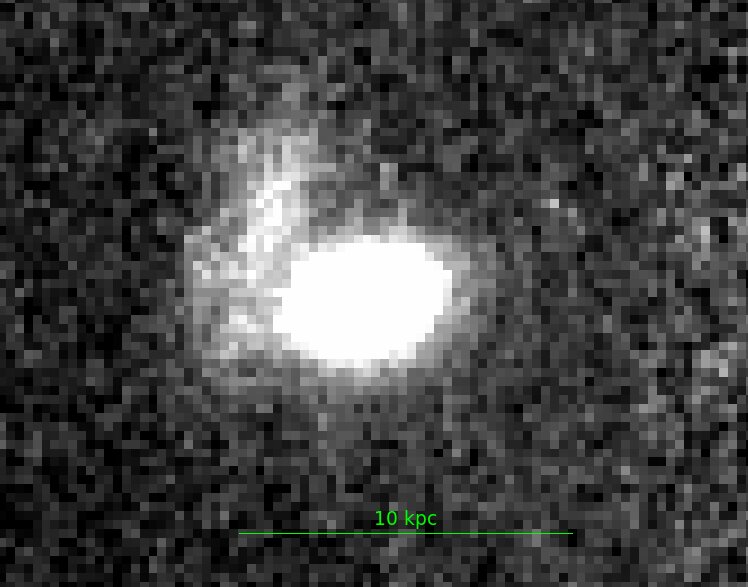}
  \includegraphics[width=5cm]{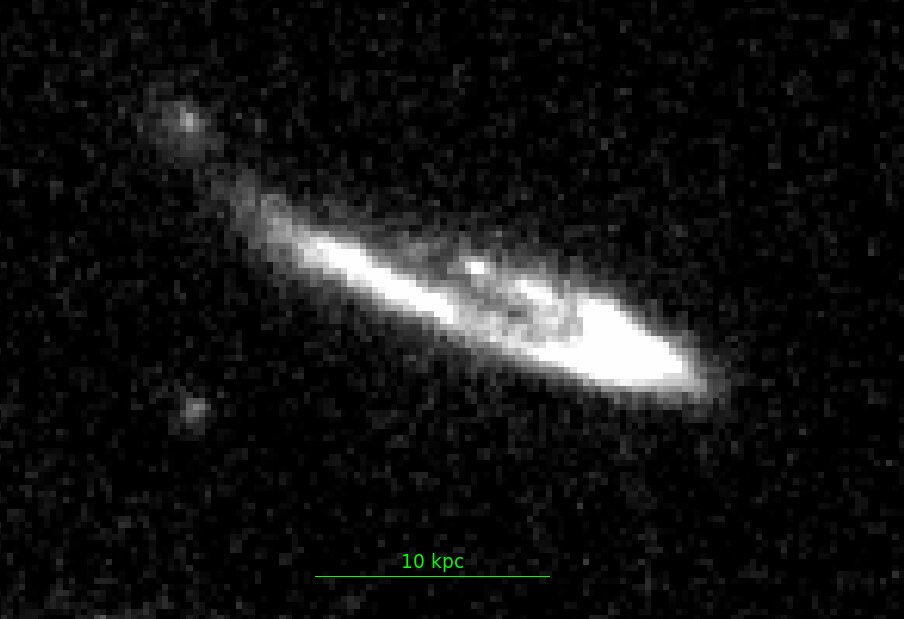}
  \includegraphics[width=5cm]{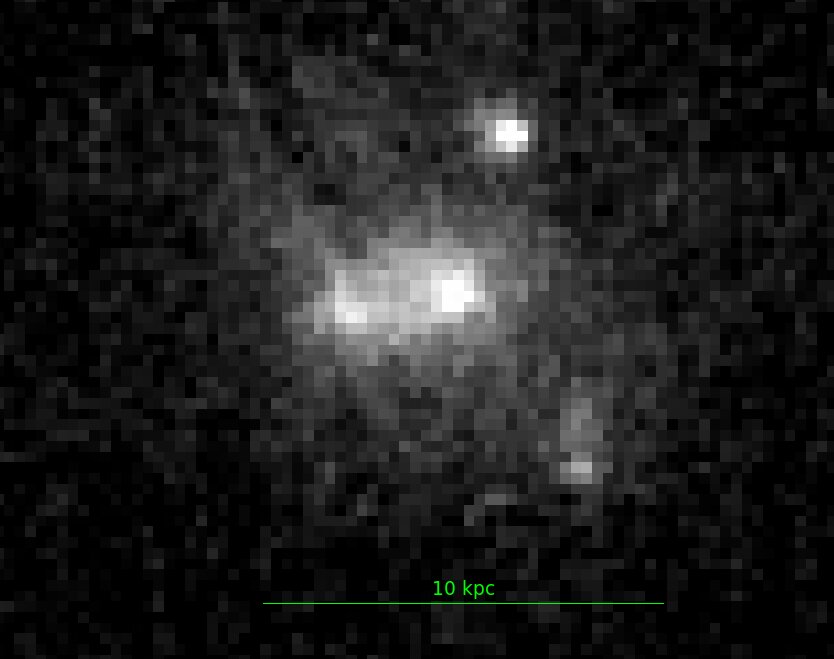}
  \includegraphics[width=5cm]{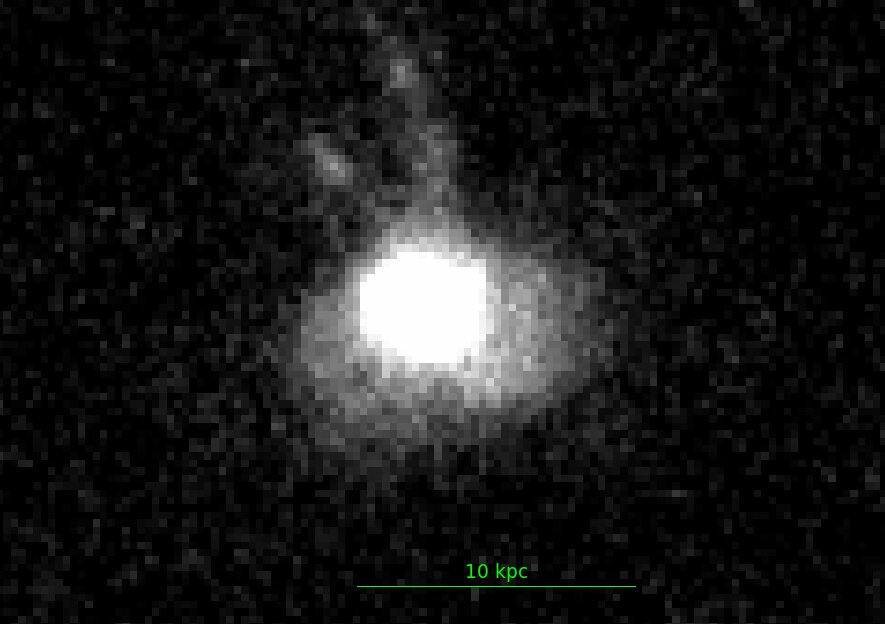}
  \includegraphics[width=5cm]{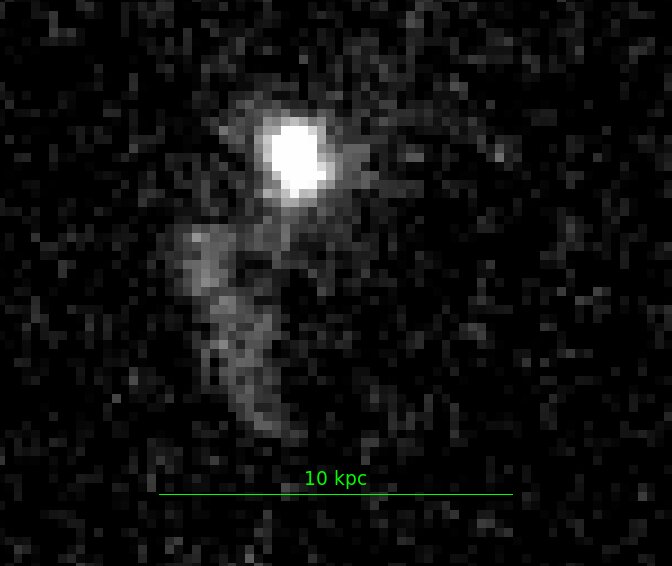}
  \includegraphics[width=5cm]{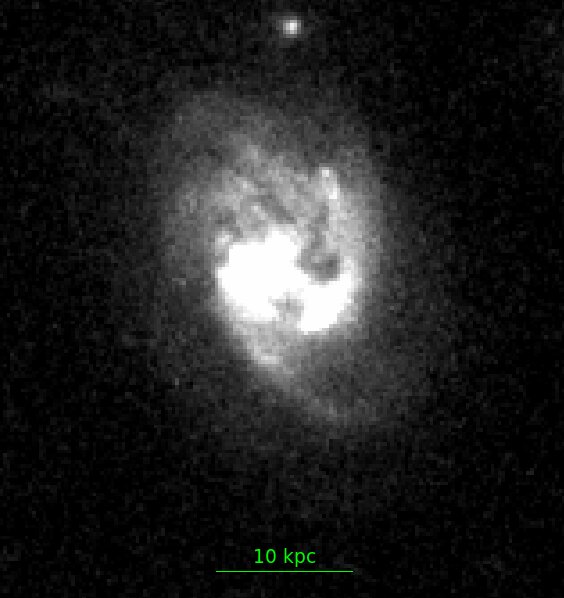}
\end{center}
\caption{Cl0016+16 (z=0.5455). All the images are in F814W. Row 1:
  galaxies a, b and c. Row 2: galaxies d, e and f. Row 3: galaxies g
  and h.}
\label{fig:cl0016}
\end{figure*}

\begin{figure*}[h]
\begin{center}
  \includegraphics[width=10cm]{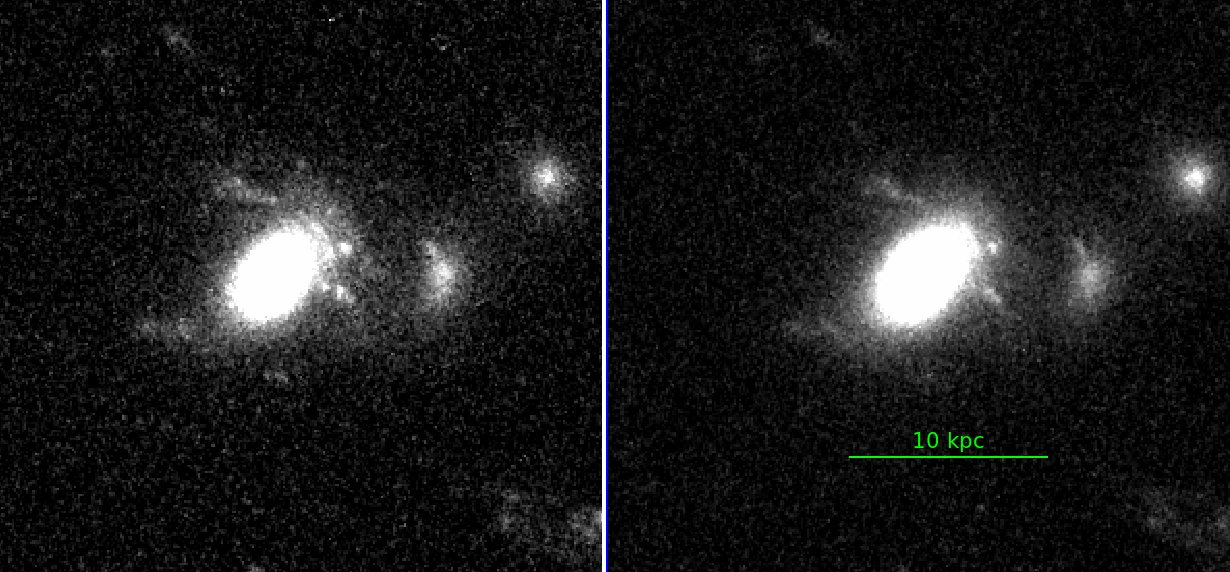}
  \includegraphics[width=10cm]{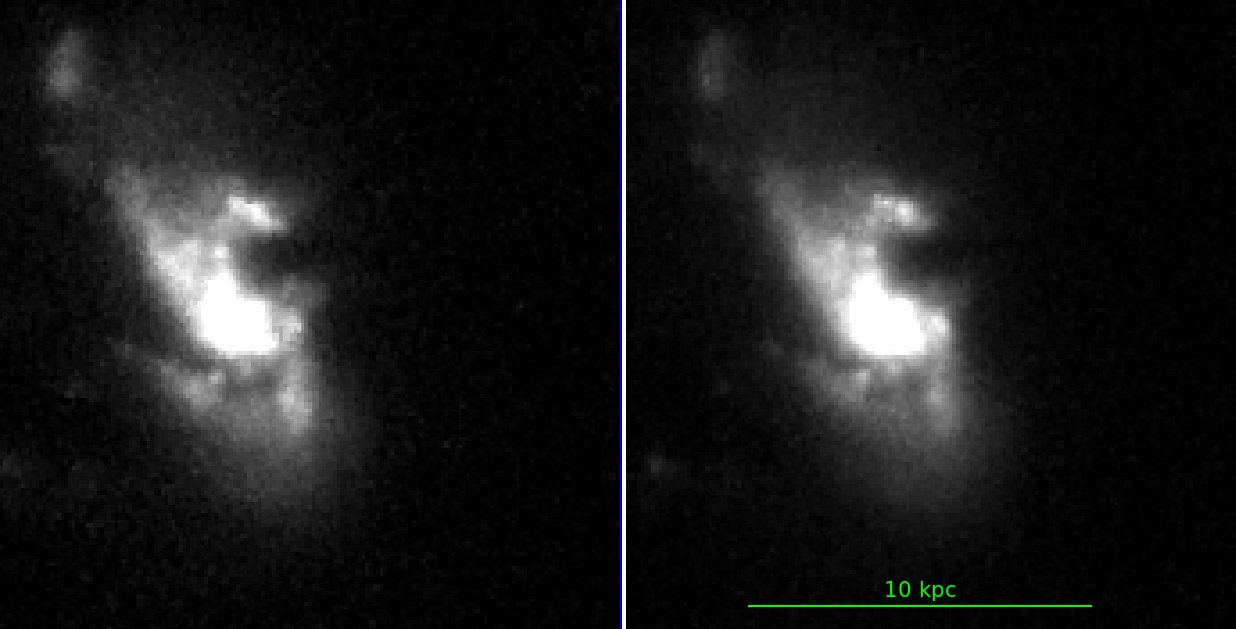}
\end{center}
\caption{A209 (z=0.206). Row 1: galaxy a in F606W and F814W. Row 2: galaxy
  b in F606W and F814W.}
\label{fig:a209}
\end{figure*}

\begin{figure*}[h]
\begin{center}
  \includegraphics[width=5cm]{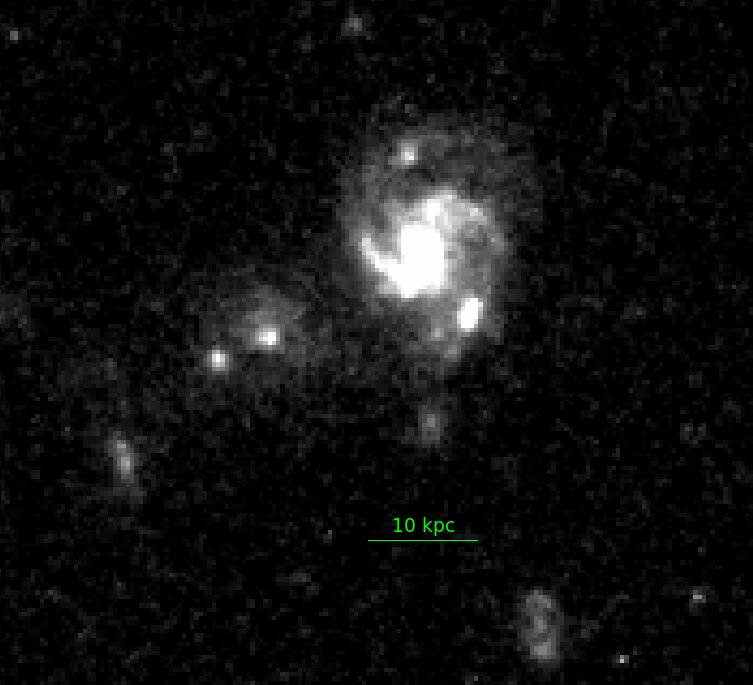}
  \includegraphics[width=5cm]{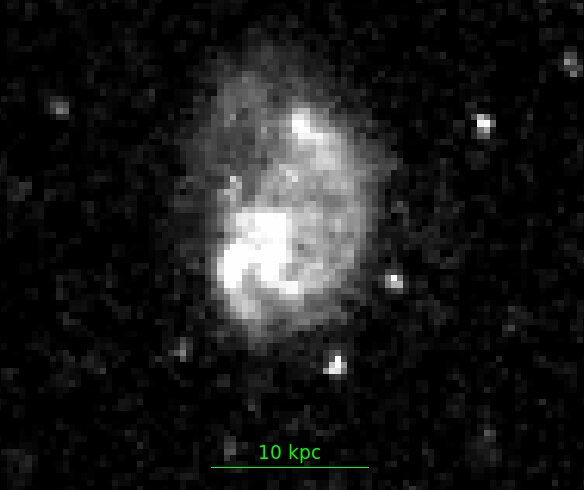}
  \includegraphics[width=5cm]{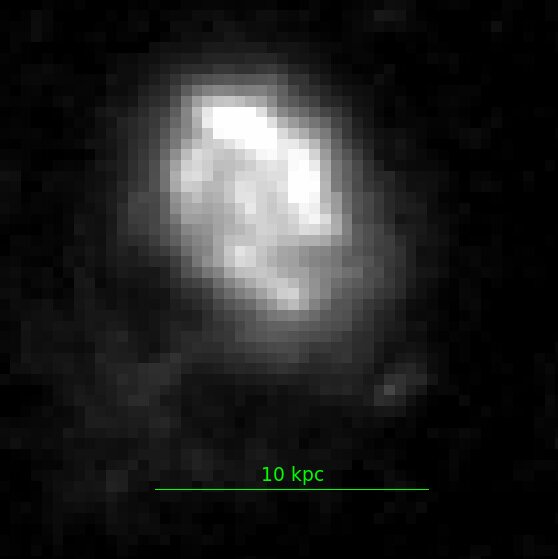}
  \includegraphics[width=5cm]{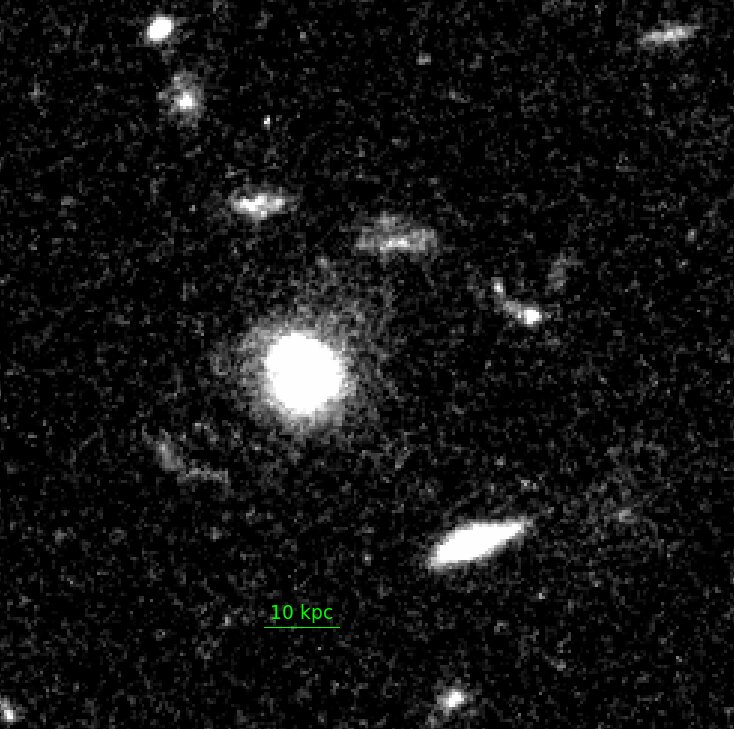}
  \includegraphics[width=5cm]{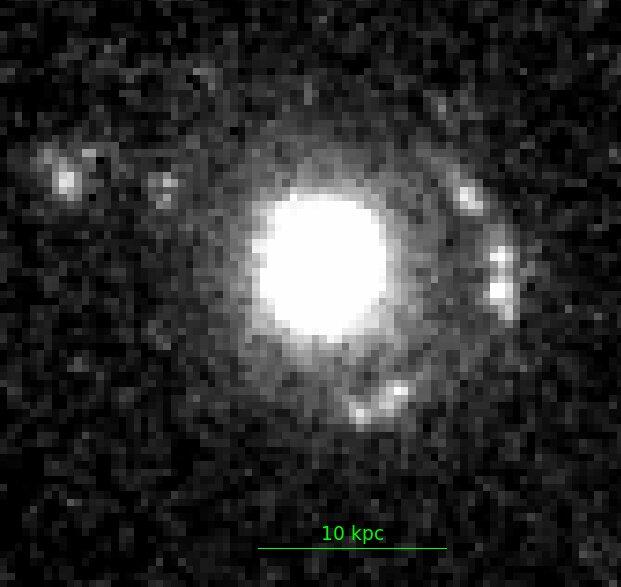}
  \includegraphics[width=5cm]{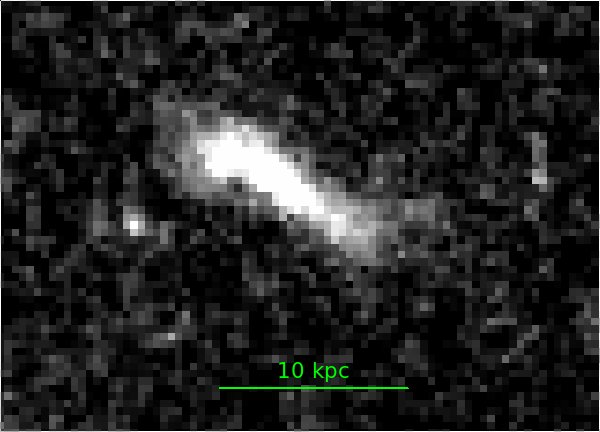}
\end{center}
\caption{Cl0152.7-1357 (z=0.831). All the images are in F775W. Row 1:
  galaxies a, b and c. Row 2: galaxies d, e, and f.}
\label{fig:cl0152}
\end{figure*}

\begin{figure*}[h]
\begin{center}
  \includegraphics[width=10cm]{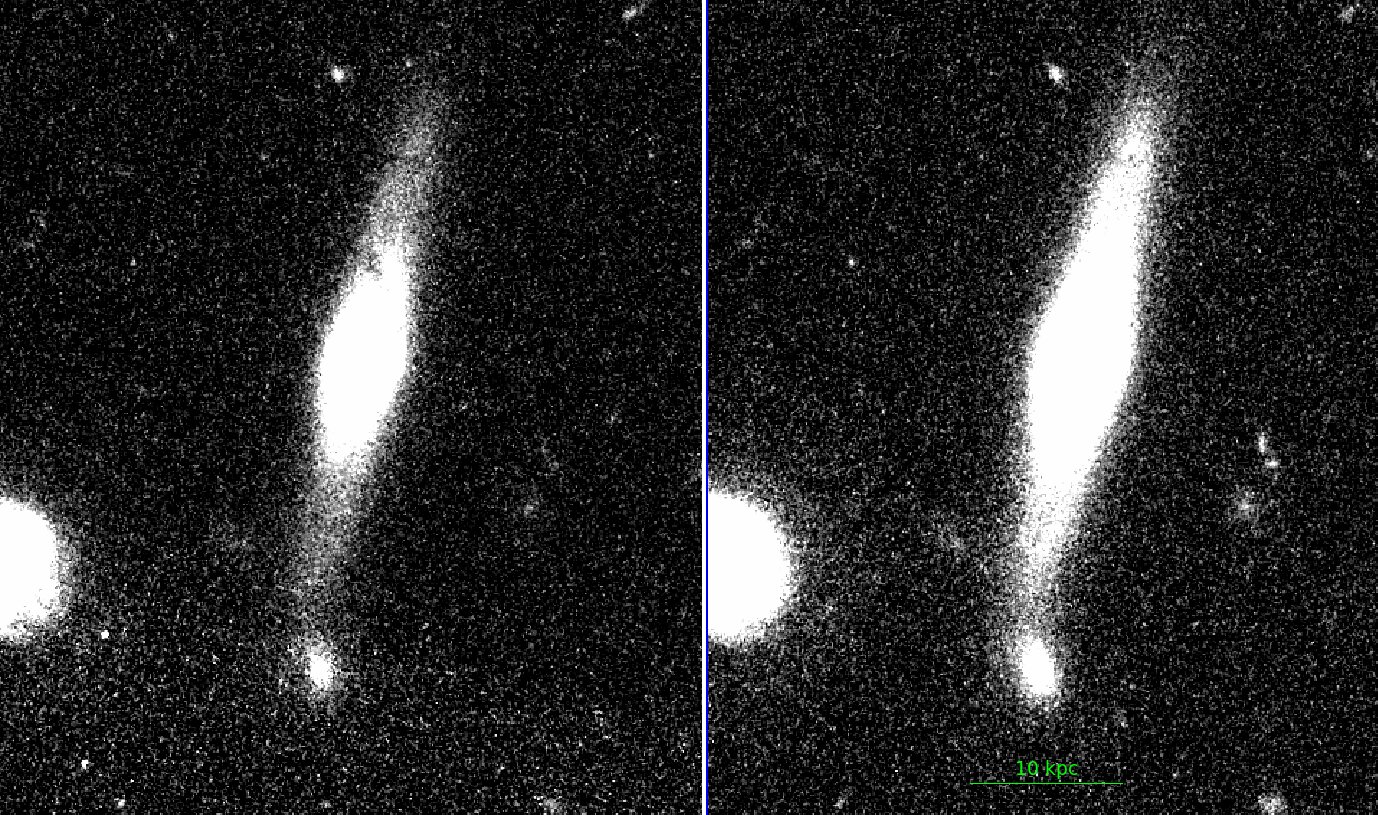}
  \includegraphics[width=10cm]{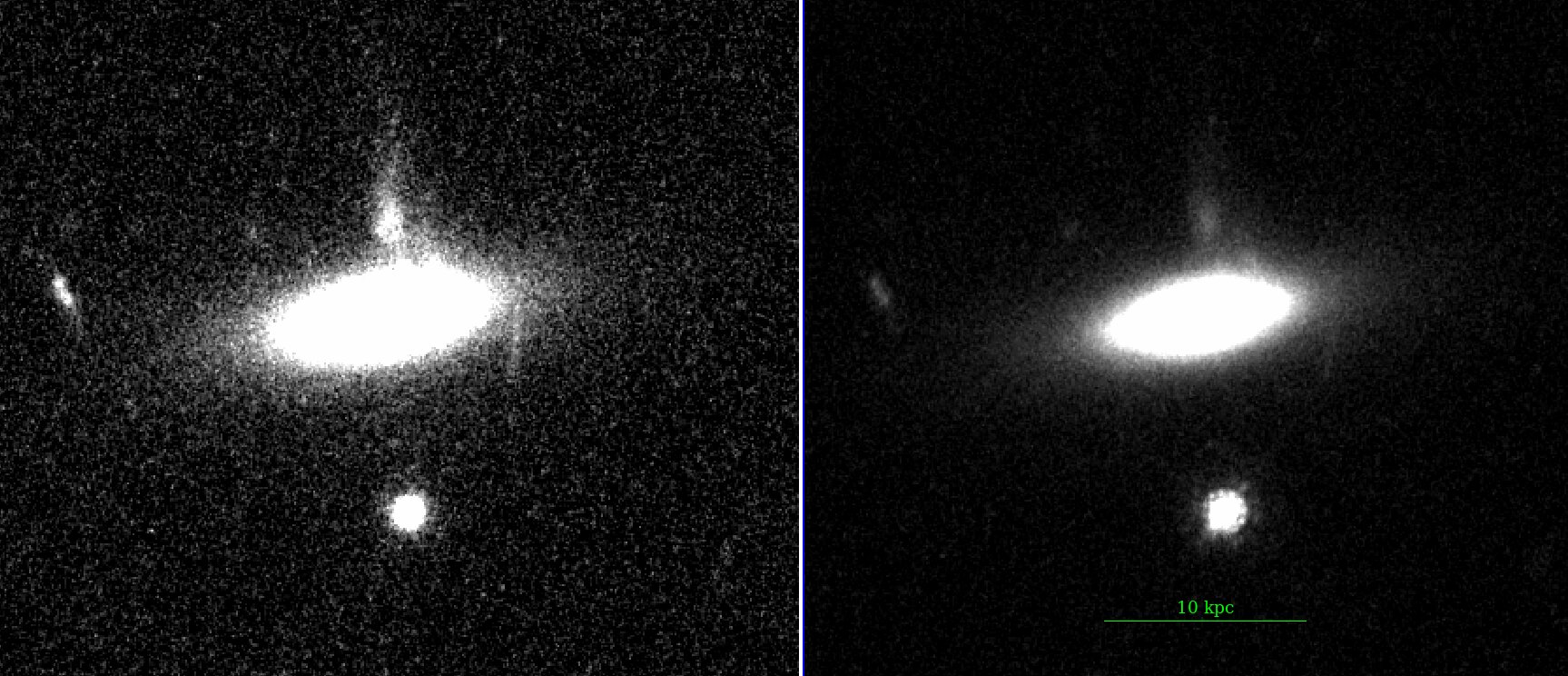}
\end{center}
\caption{A383 (z=0.1871) in F606W (left) and F814W (right).  Row 1:
  galaxy a, row 2: galaxy b.}
\label{fig:a383}
\end{figure*}

\begin{figure*}[h]
\begin{center}
  \includegraphics[width=8cm]{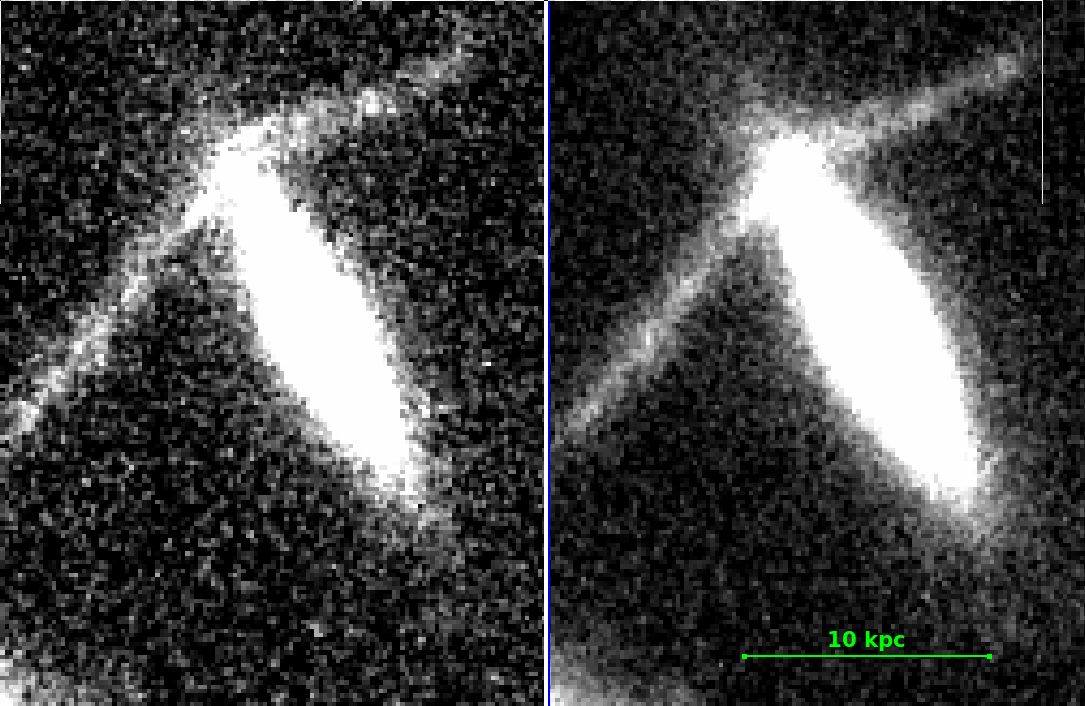}
  \includegraphics[width=8cm]{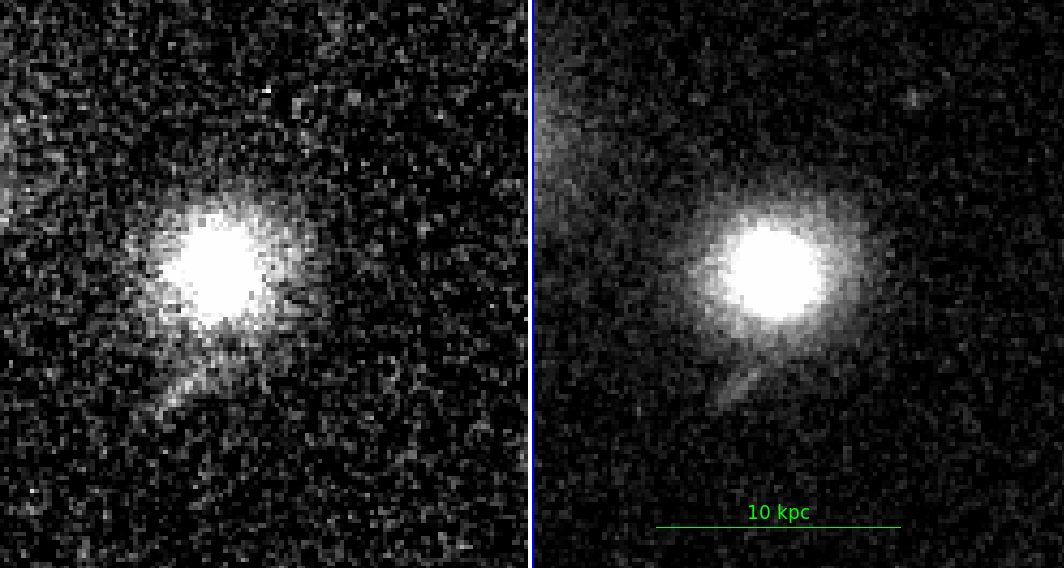}
  \includegraphics[width=8cm]{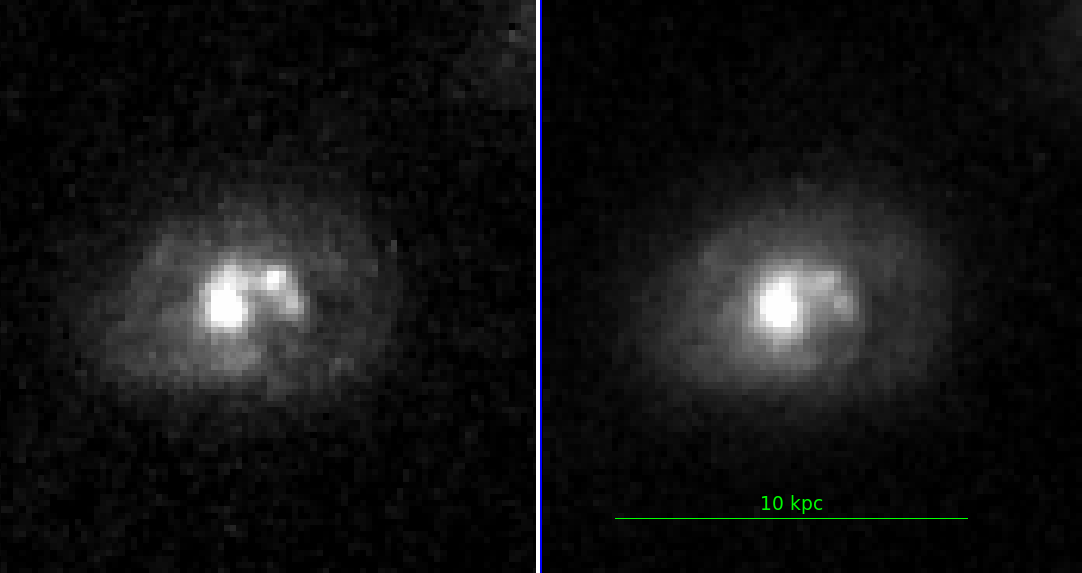}
  \includegraphics[width=8cm]{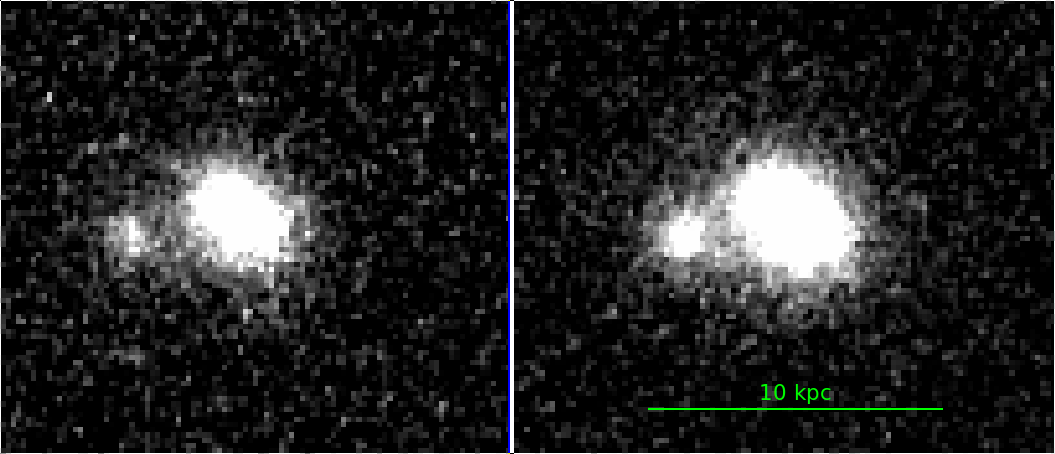}
  \includegraphics[width=8cm]{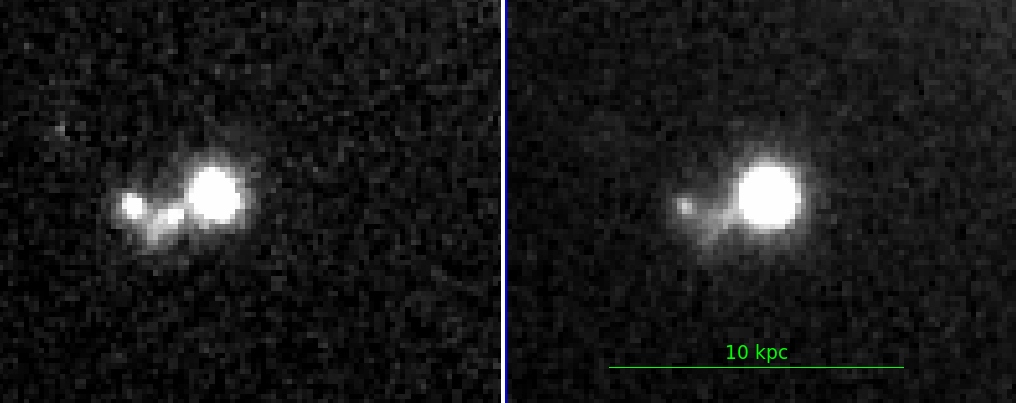}
  \includegraphics[width=8cm]{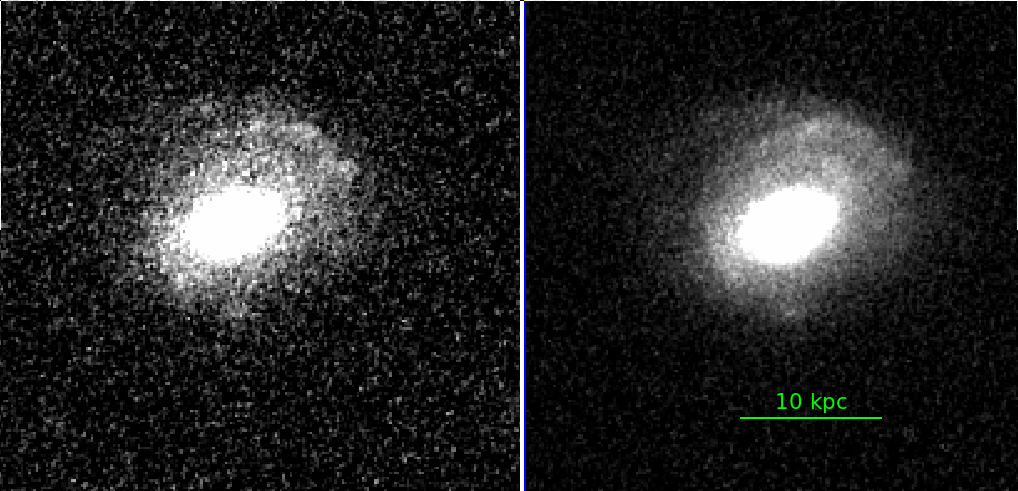}
  \includegraphics[width=8cm]{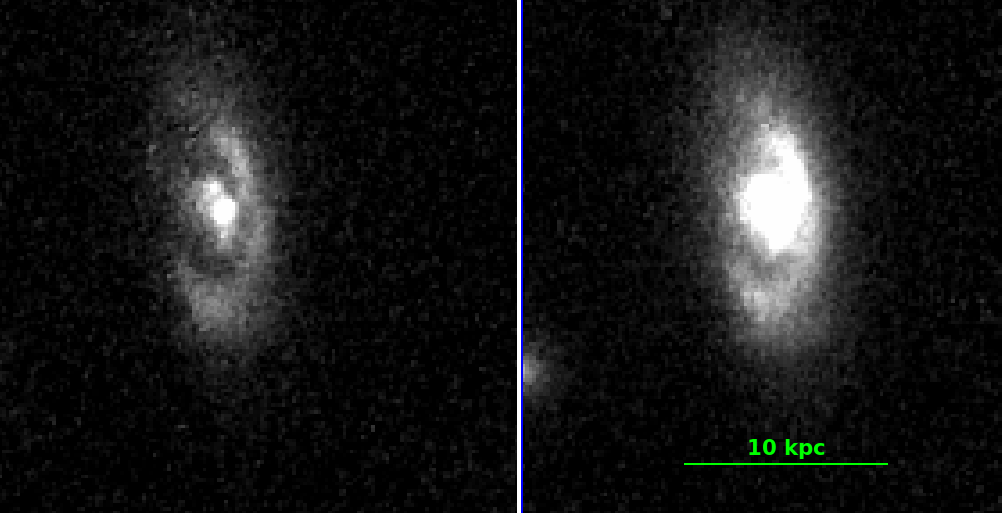}
  \includegraphics[width=8cm]{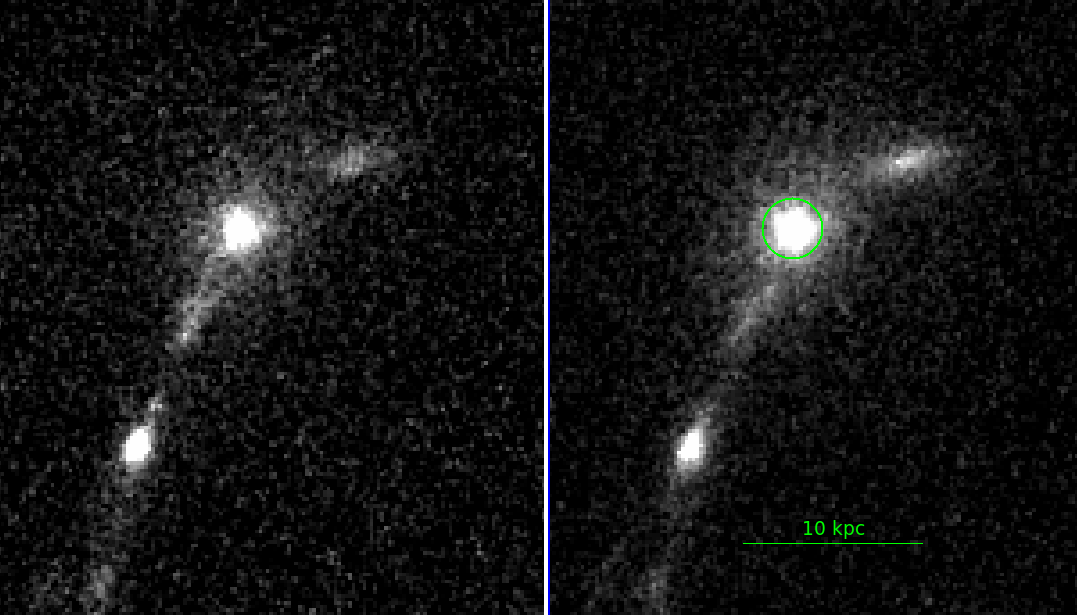}
\end{center}
\caption{MACS0416 (z=0.396) in F606W and F814W.  Row
  1: galaxies a and b. Row 2: galaxies c and d. Row 3: galaxies e and
  f. Row 4: galaxies g and h (green circle).}
\label{fig:macs0416}
\end{figure*}

\begin{figure*}[h]
\begin{center}
  \includegraphics[width=8cm]{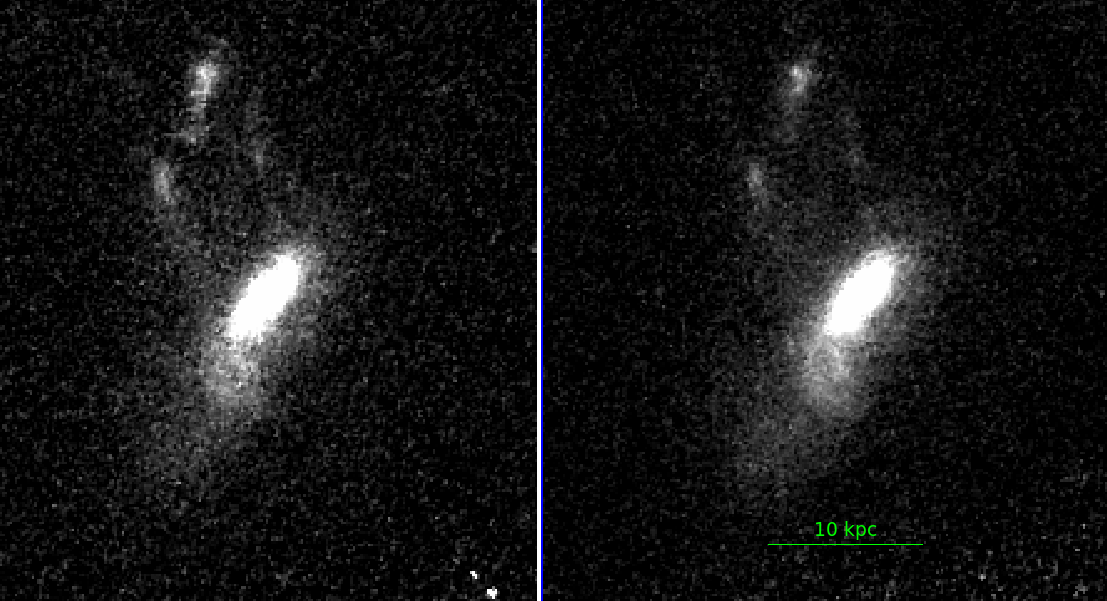}
  \includegraphics[width=8cm]{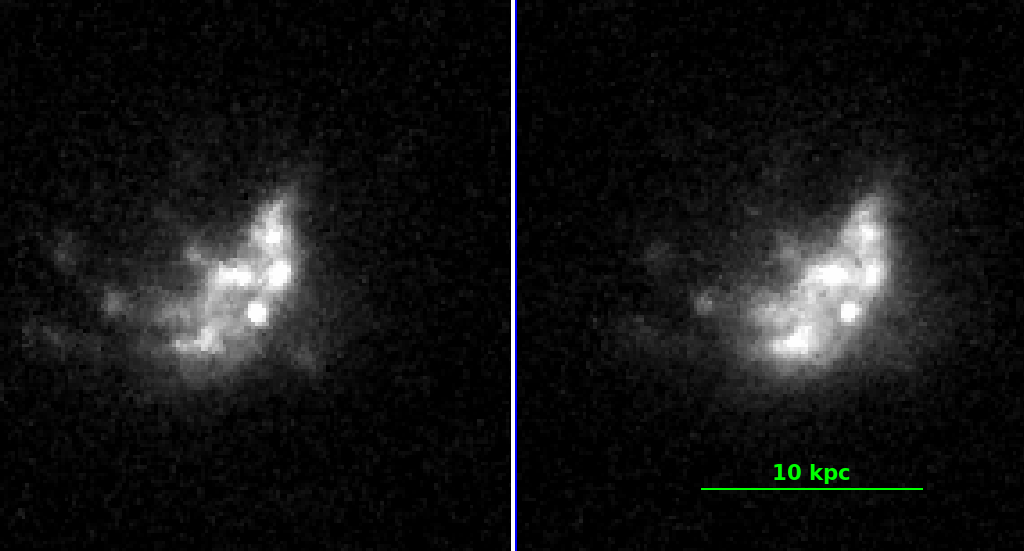}
\end{center}
\caption{MACS0429 (z=0.399) in F606W and F814W.
  Left: galaxy a, right: galaxy b. }
\label{fig:macs0429}
\end{figure*}

\begin{figure*}[h]
\begin{center}
  \includegraphics[width=5cm]{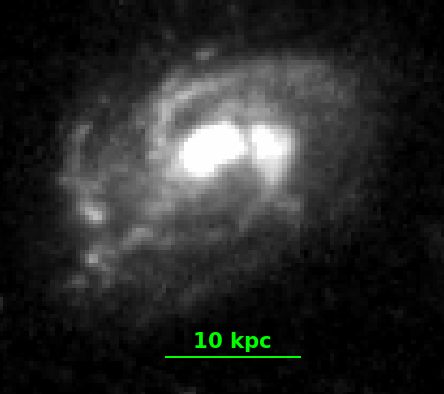}
  \includegraphics[width=5cm]{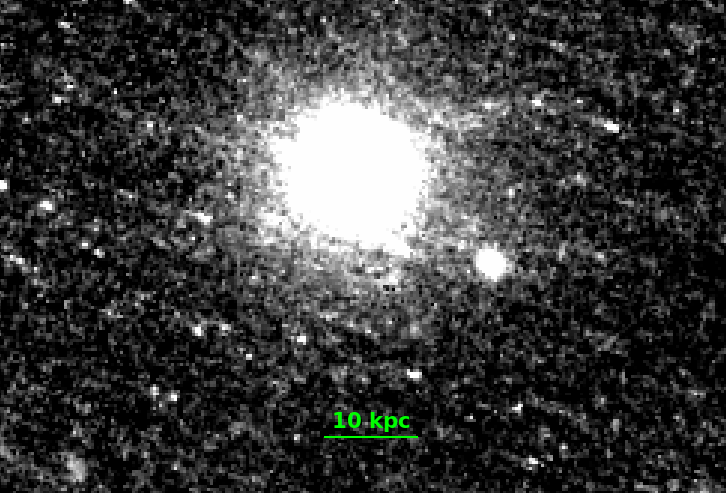}
  \includegraphics[width=5cm]{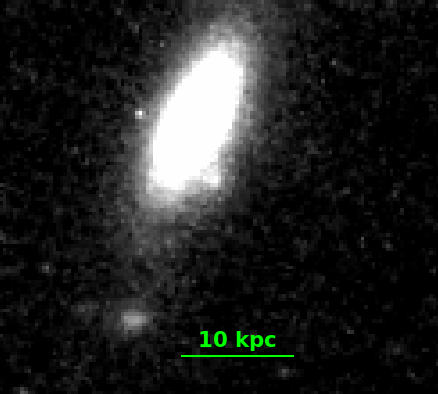}
  \includegraphics[width=5cm]{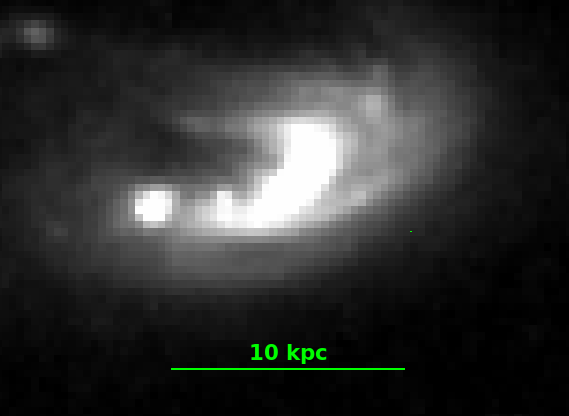}
  \includegraphics[width=5cm]{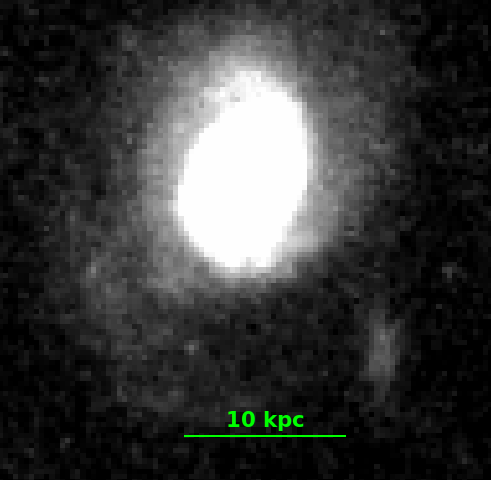}
\end{center}
\caption{MACS0454 (z=0.5377). All galaxies are in F814W. Row 1:
  galaxies a, b, and c. Row 2: galaxies d and e.  }
\label{fig:macs0454}
\end{figure*}

\begin{figure*}[h]
\begin{center}
  \includegraphics[width=5cm]{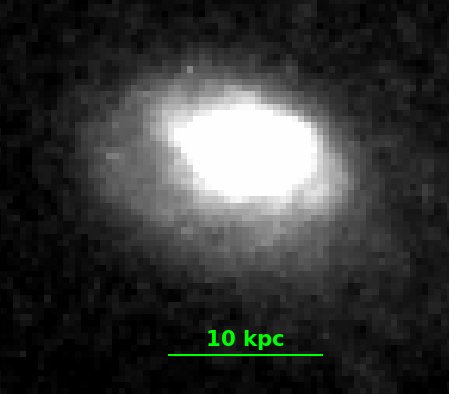}
  \includegraphics[width=5cm]{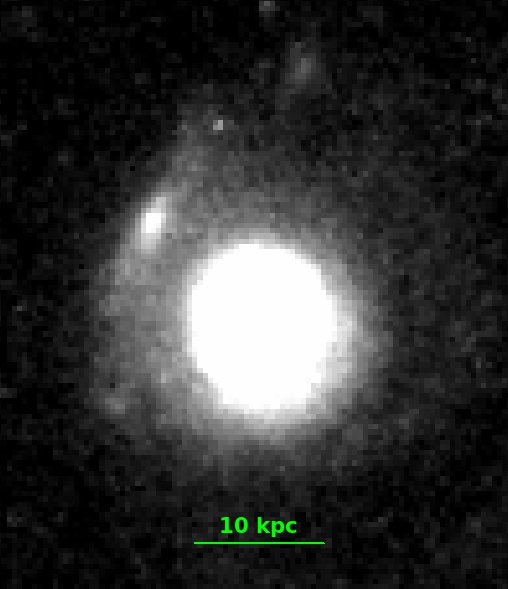}
  \includegraphics[width=5cm]{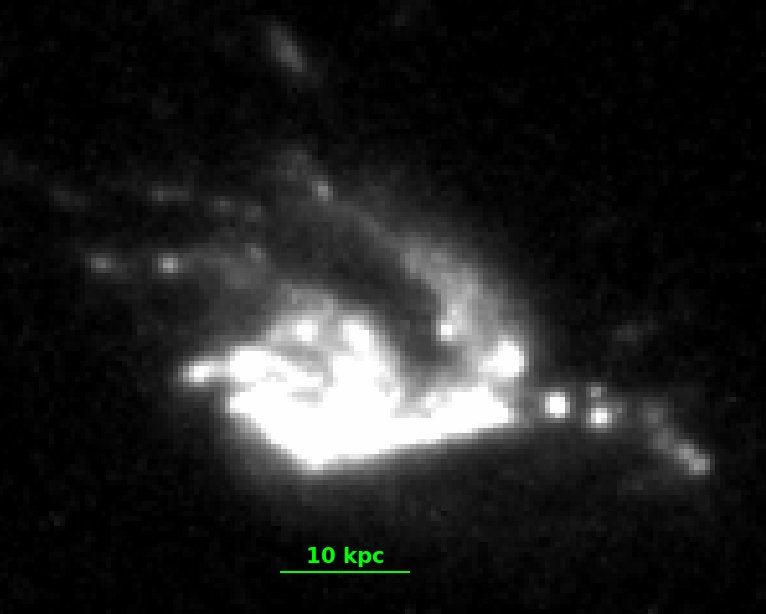}
  \includegraphics[width=5cm]{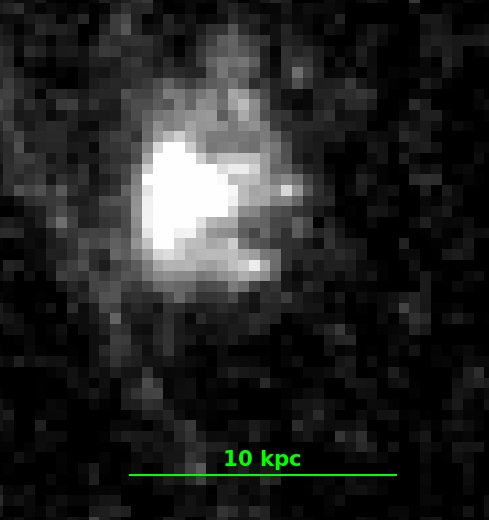}
  \includegraphics[width=5cm]{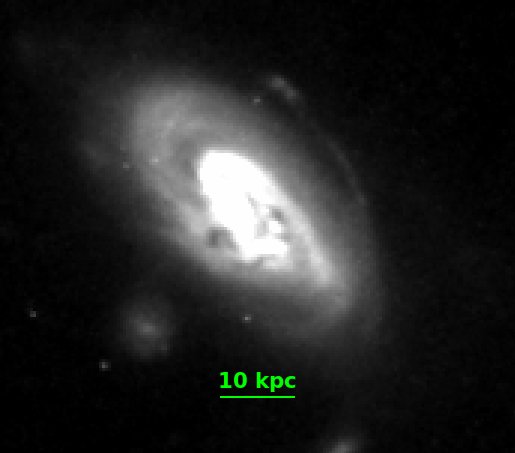}
  \includegraphics[width=5cm]{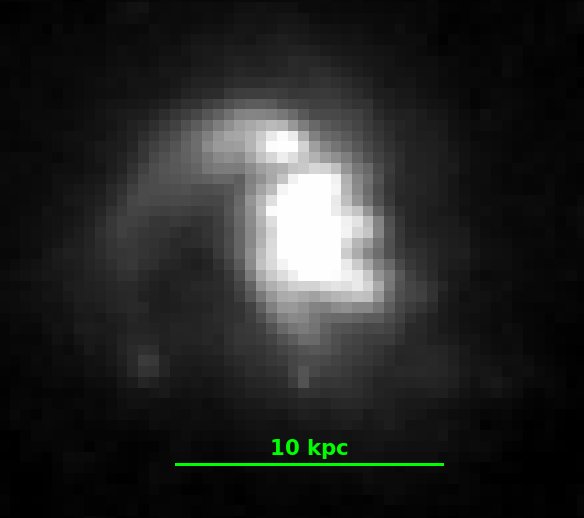}
  \includegraphics[width=5cm]{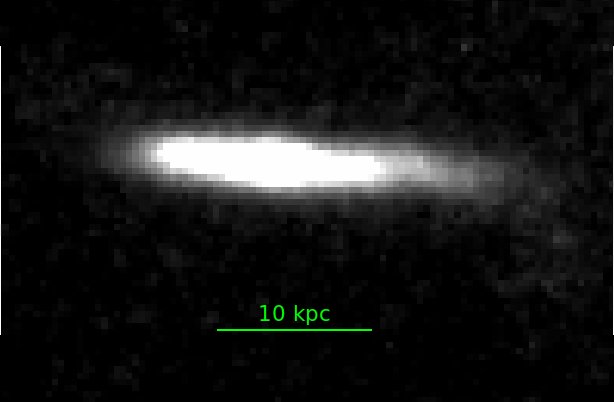}
  \includegraphics[width=5cm]{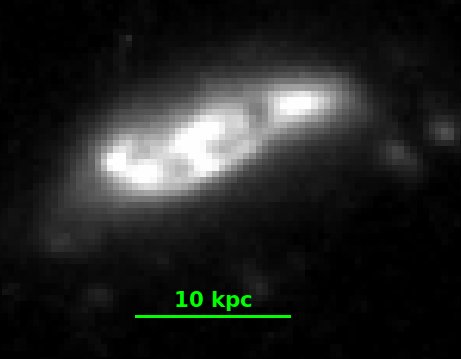}
  \includegraphics[width=5cm]{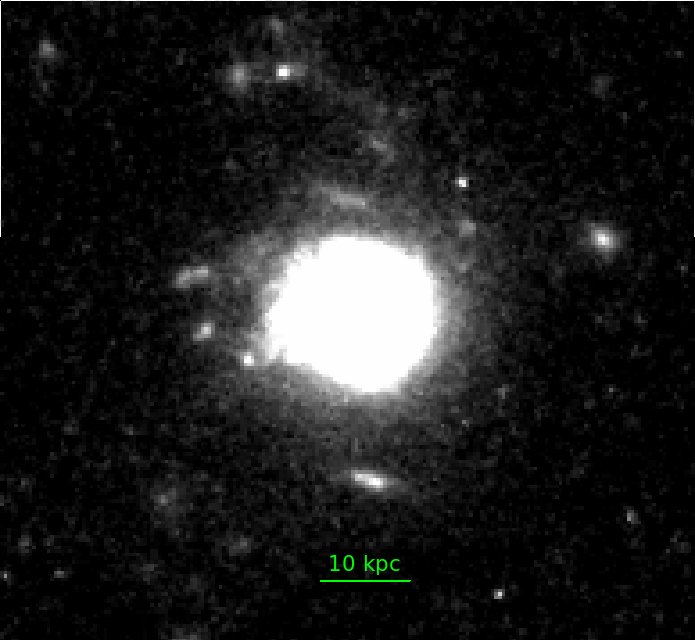}
  \includegraphics[width=5cm]{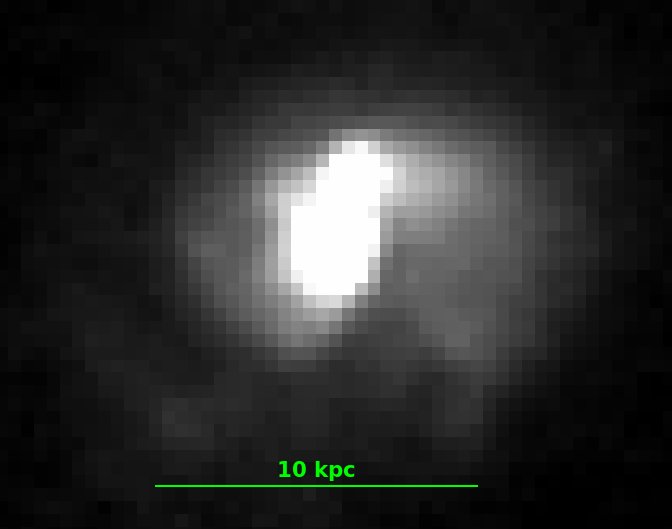}
  \includegraphics[width=5cm]{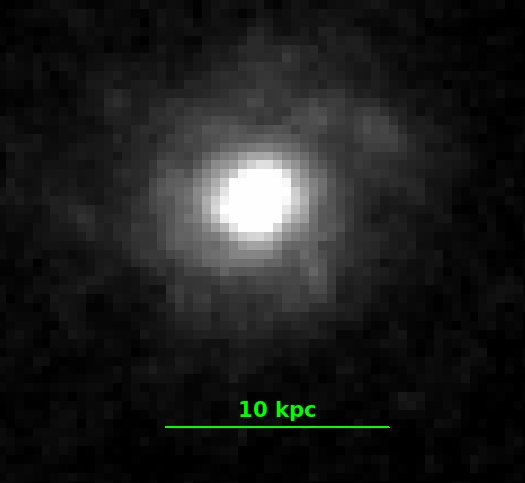}
\end{center}
\caption{A851 (z=0.4069). All the images are in F814W. Row 1: galaxies
  a, b and c. Row 2: galaxies d, e and f. Row 3: galaxies g, h, and
  i. Row 4: galaxies j and k.}
\label{fig:a851}
\end{figure*}

\begin{figure*}[h]
\begin{center}
  \includegraphics[width=5cm]{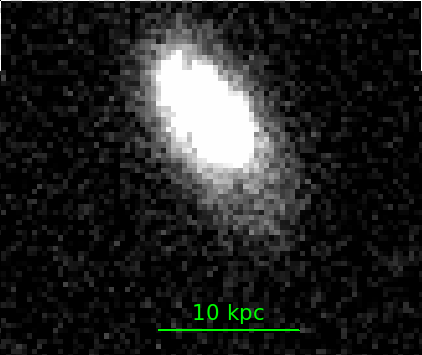}
  \includegraphics[width=5cm]{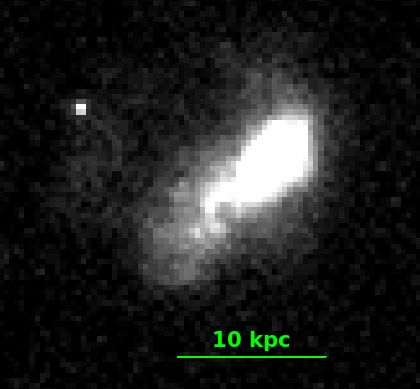}
  \includegraphics[width=5cm]{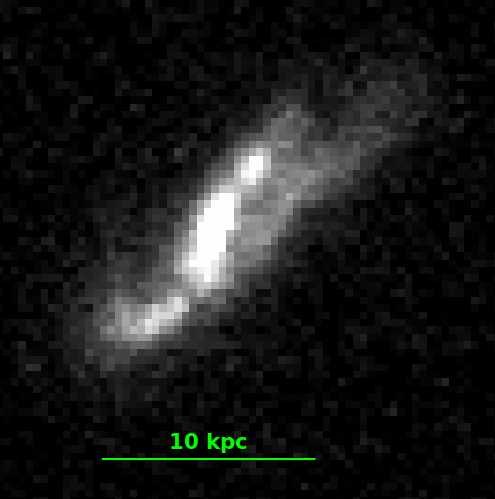}
  \includegraphics[width=5cm]{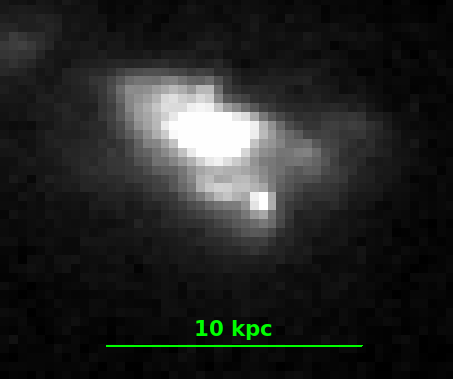}
  \includegraphics[width=5cm]{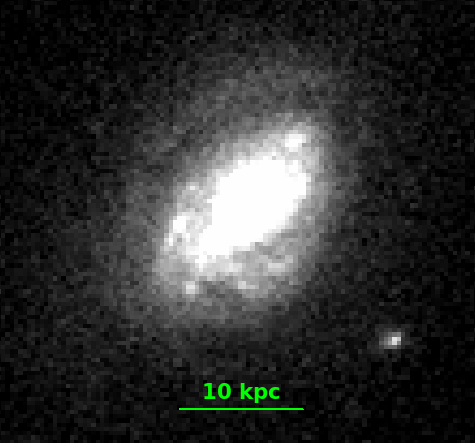}
  \includegraphics[width=5cm]{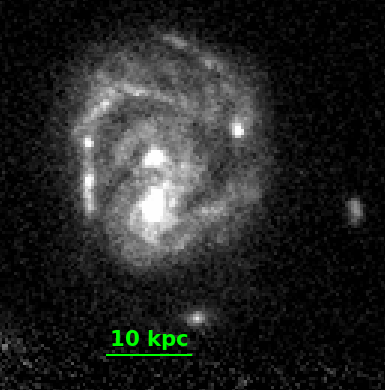}
  \includegraphics[width=5cm]{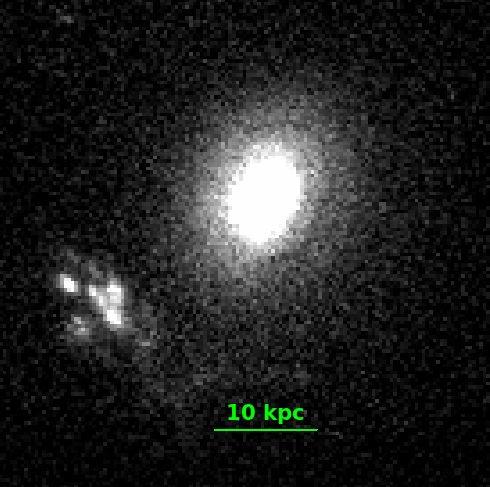}
  \includegraphics[width=5cm]{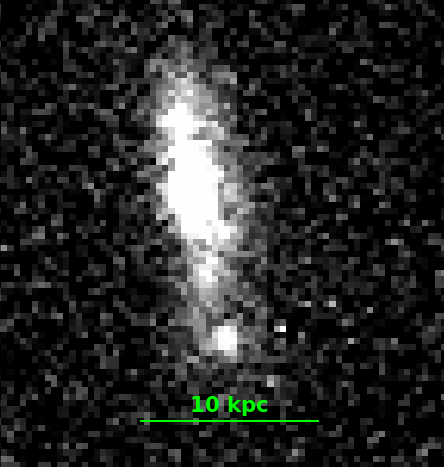}
  \includegraphics[width=5cm]{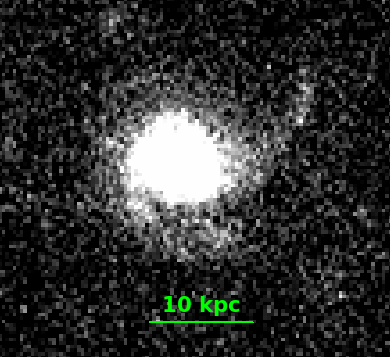}
\end{center}
\caption{LCDCS0172 (z=0.6972). All the images are in F814W. Row 1: galaxies
  a, b, and c. Row 2: galaxies d, e, and f. Row 3: galaxies g, h and i.}
\label{fig:lcdcs0172}
\end{figure*}

\begin{figure*}[h]
\begin{center}
  \includegraphics[width=8cm]{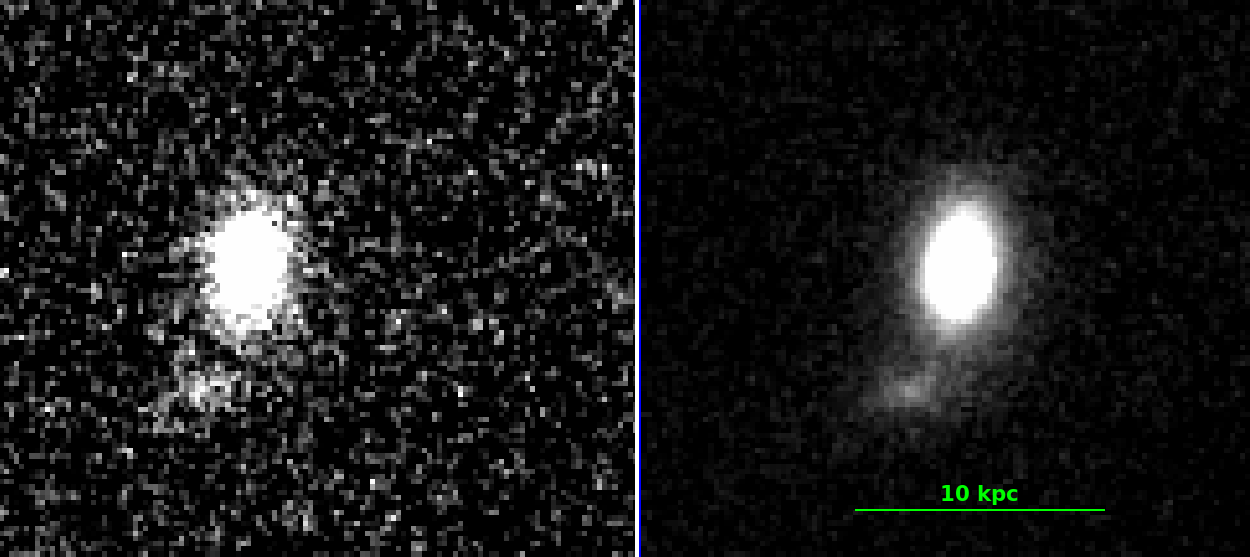}
  \includegraphics[width=8cm]{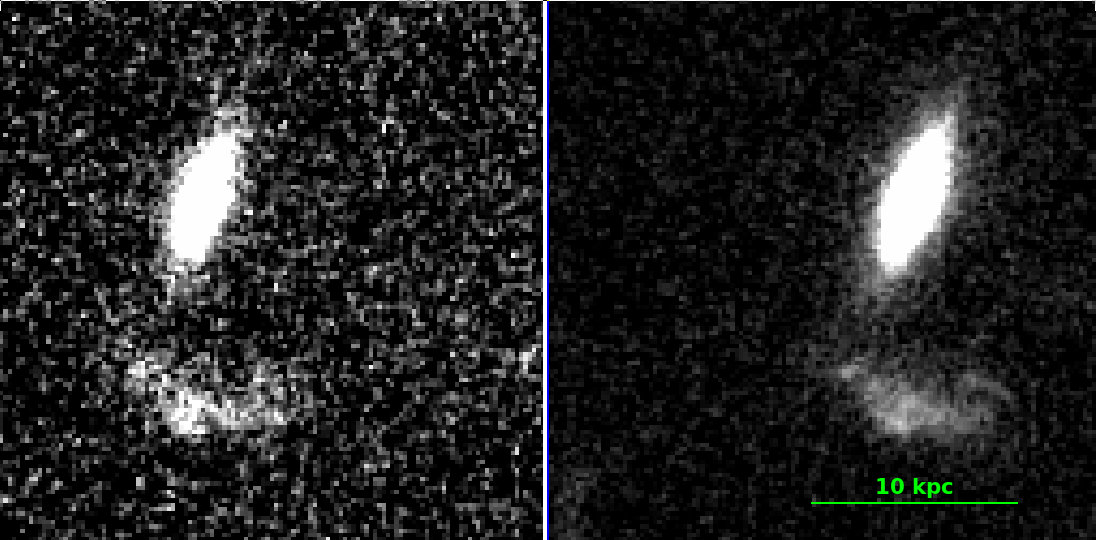}
  \includegraphics[width=8cm]{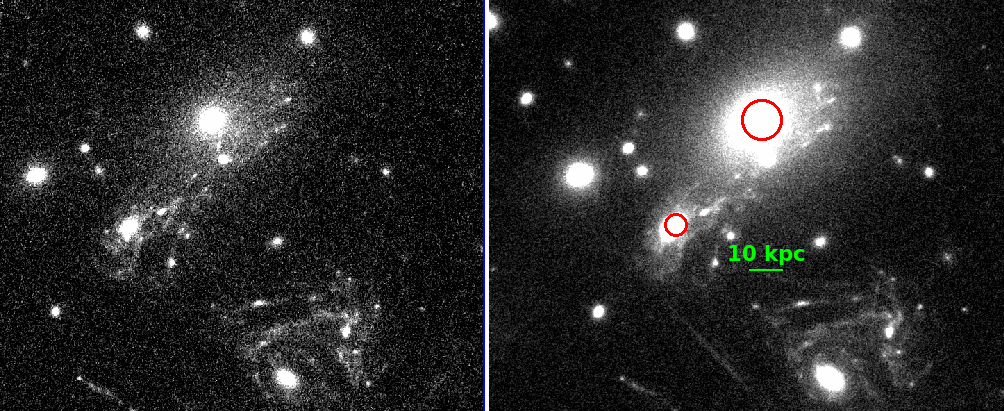}
\end{center}
\caption{MACS1149 (z=0.544). All the images are in F606W and F814W.
  Row 1: galaxies a and b. Row 2: galaxies c and d (both are at the
  cluster redshift, c is to the right).}
\label{fig:macs1149}
\end{figure*}

\begin{figure*}[h]
\begin{center}
  \includegraphics[width=8cm]{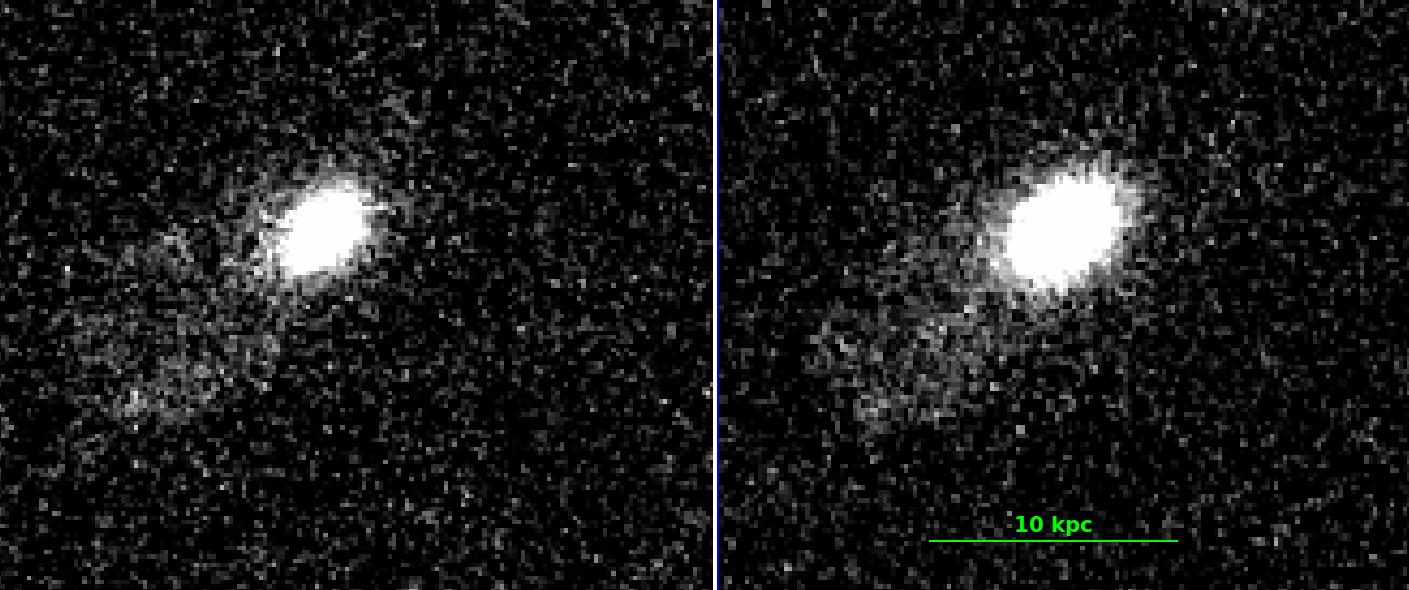}
  \includegraphics[width=8cm]{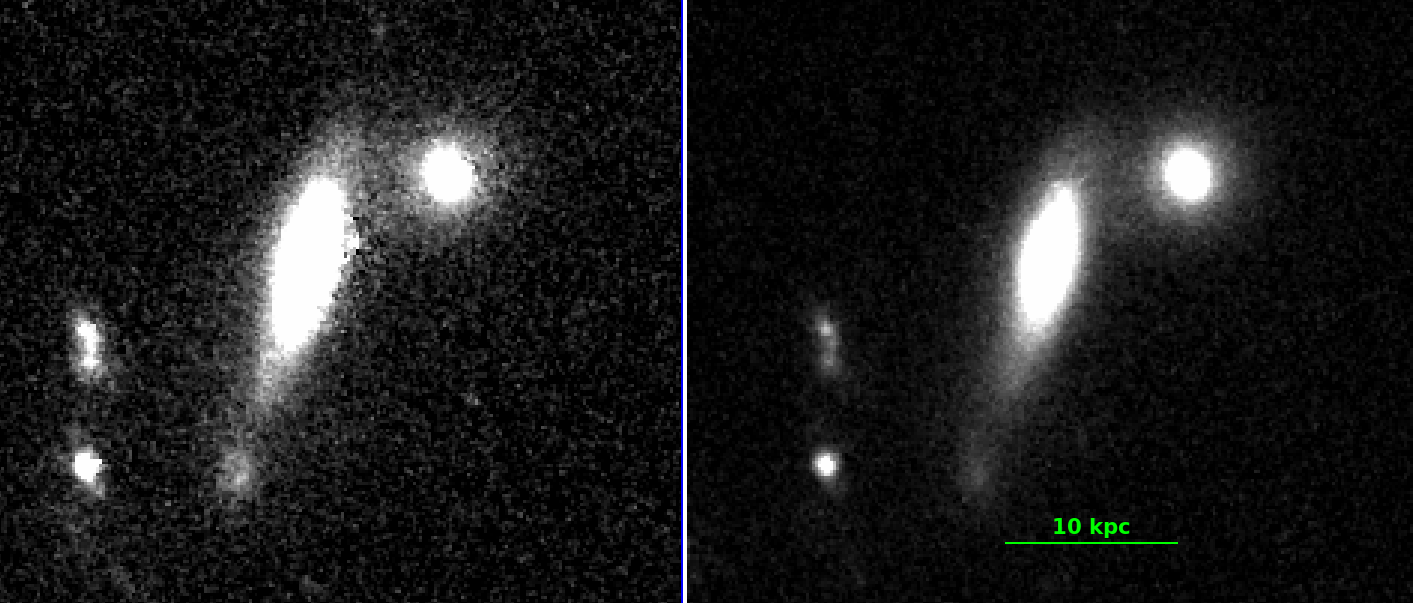}
  \includegraphics[width=8cm]{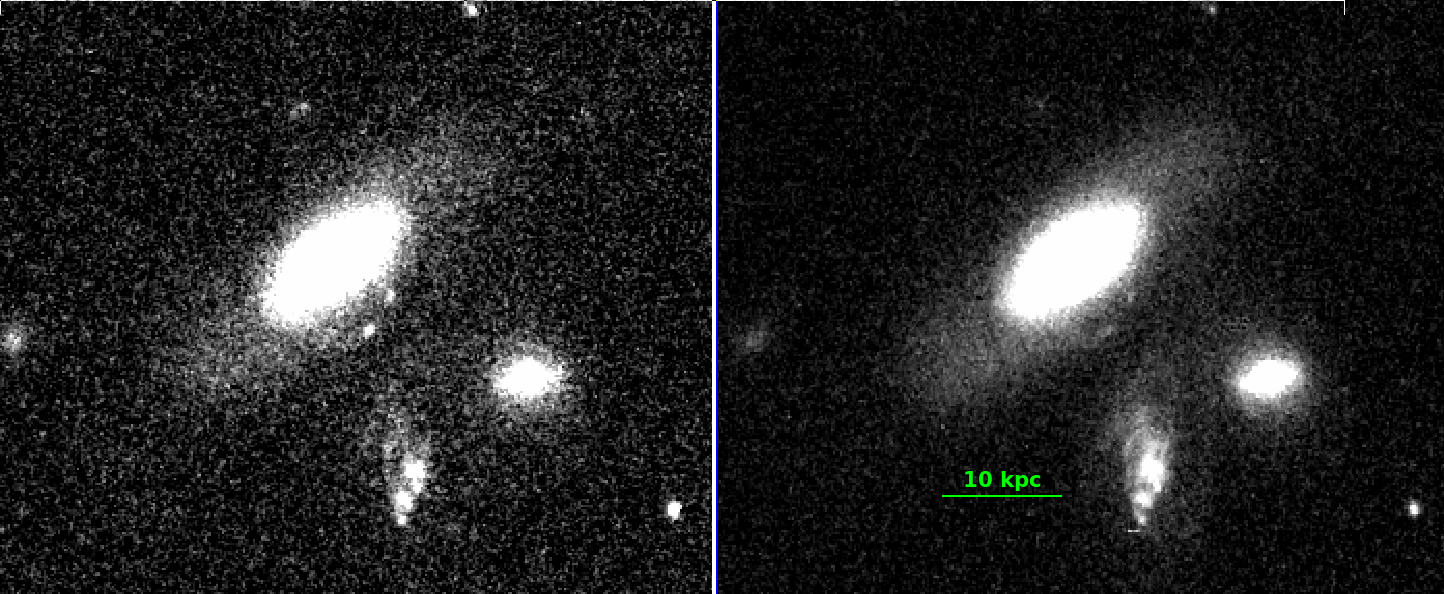}
\end{center}
\caption{MACS1206 (z=0.44). All the images are in F606W and F814W.
  Row 1: galaxies a and b. Row 2: galaxy c. }
\label{fig:macs1206}
\end{figure*}

\begin{figure*}[h]
\begin{center}
  \includegraphics[width=5cm]{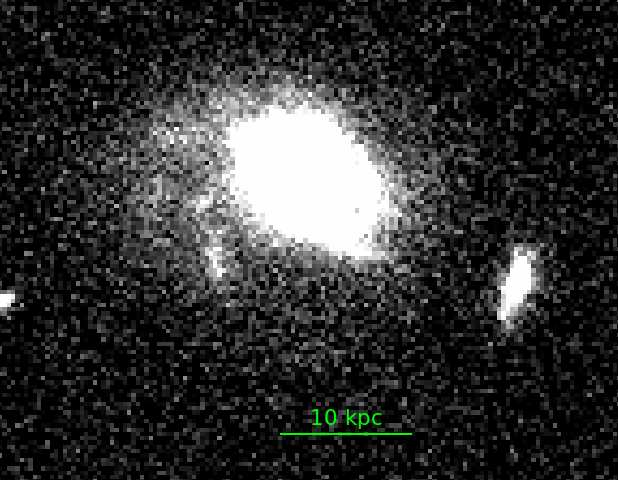}
  \includegraphics[width=5cm]{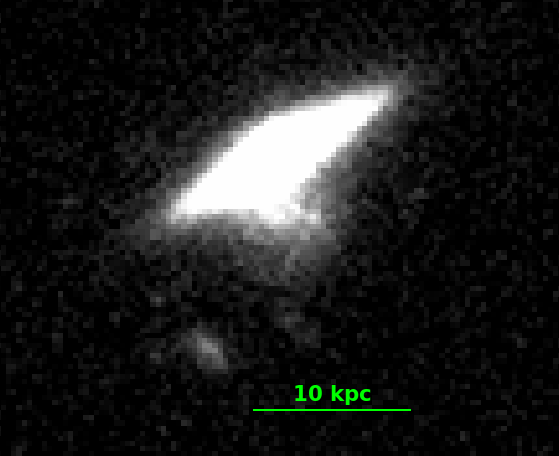}
  \includegraphics[width=5cm]{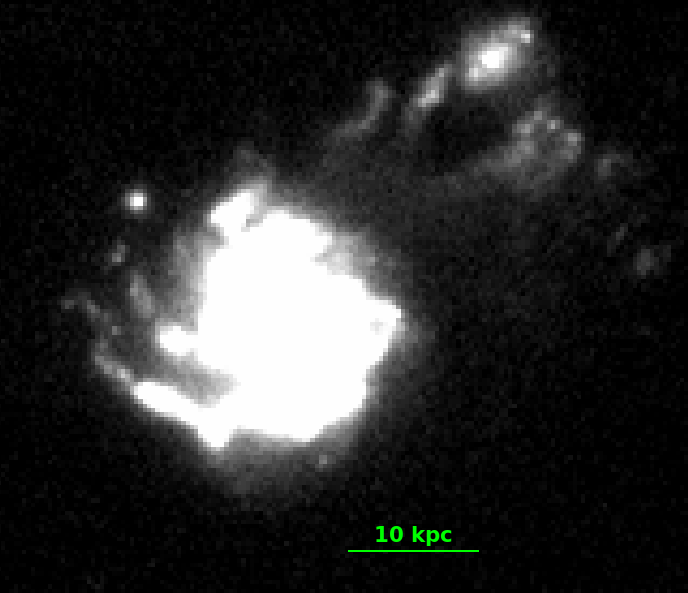}
  \includegraphics[width=5cm]{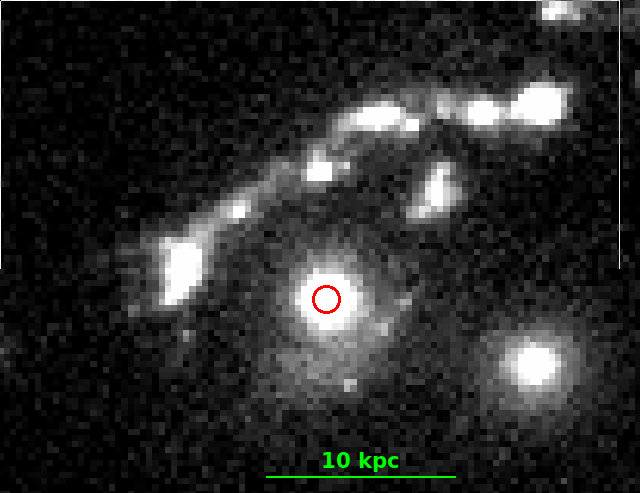}
  \includegraphics[width=5cm]{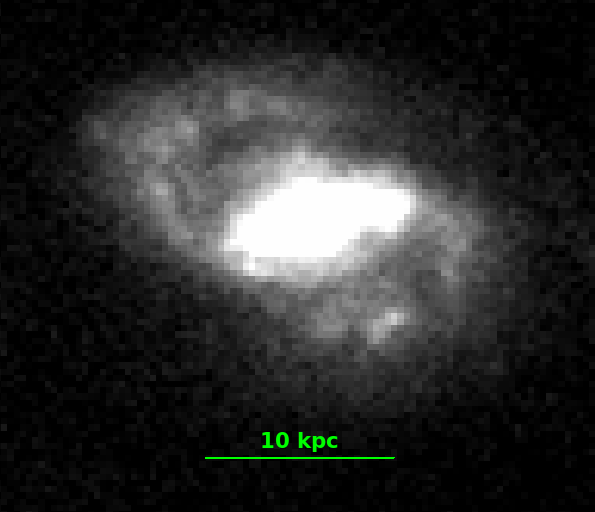}
  \includegraphics[width=5cm]{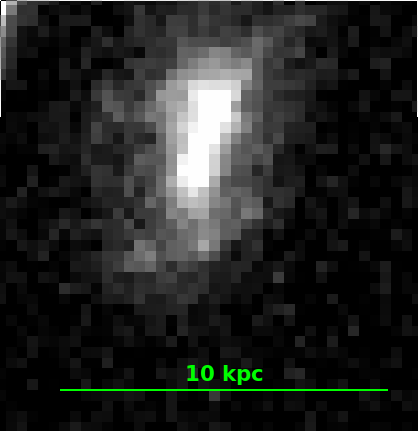}
  \includegraphics[width=5cm]{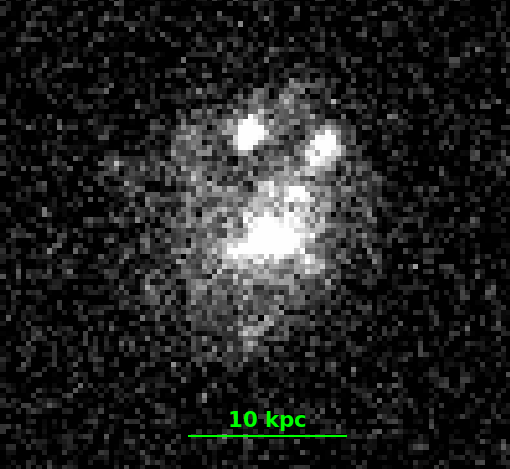}
\end{center}
\caption{LCDCS0541 (z=0.5414). All galaxies are in F814W.  Row 1:
  galaxies a, b, and c. Row 2: galaxies d (red circle), e and f. Row 3: galaxy g.}
\label{fig:lcdcs0541}
\end{figure*}

\begin{figure*}[h]
\begin{center}
  \includegraphics[width=5cm]{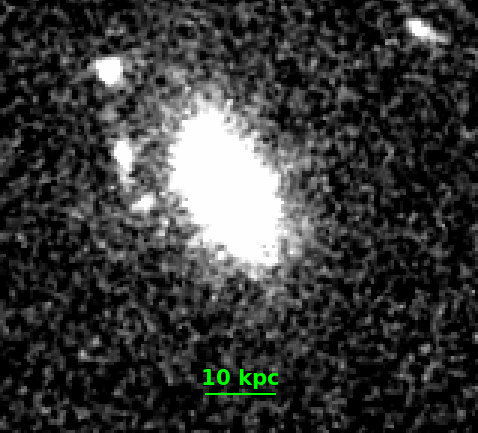}
\end{center}
\caption{MJM98\_034 (z=0.595). Galaxy a in F712W.}
\label{fig:mjm98}
\end{figure*}

\begin{figure*}[h]
\begin{center}
  \includegraphics[width=10cm]{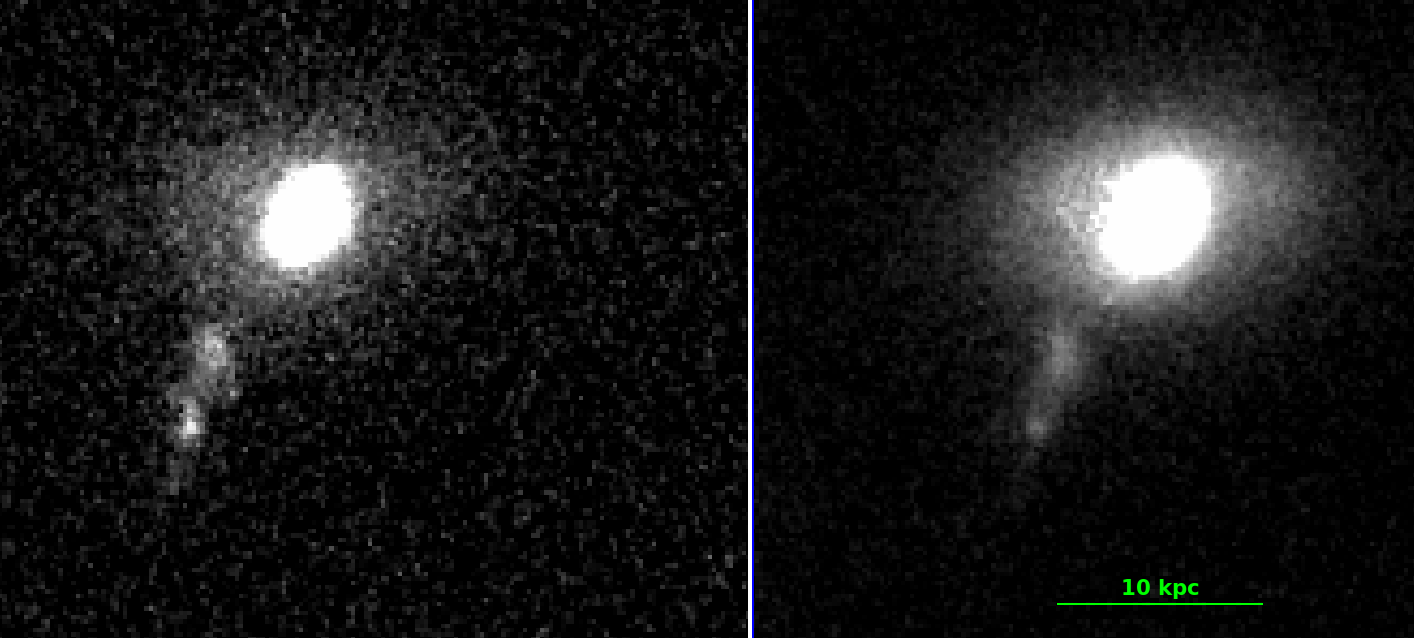}
\end{center}
\caption{LCDCS0829 (z=0.451). Galaxy a in F606W and F814W.}
\label{fig:lcdcs0829}
\end{figure*}

\begin{figure*}[h]
\begin{center}
  \includegraphics[width=5cm]{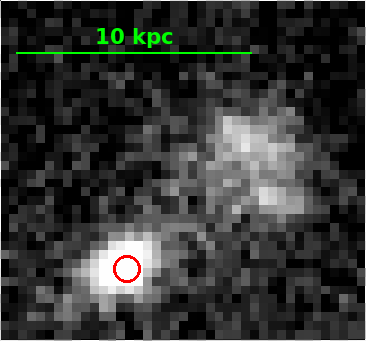}
  \includegraphics[width=5cm]{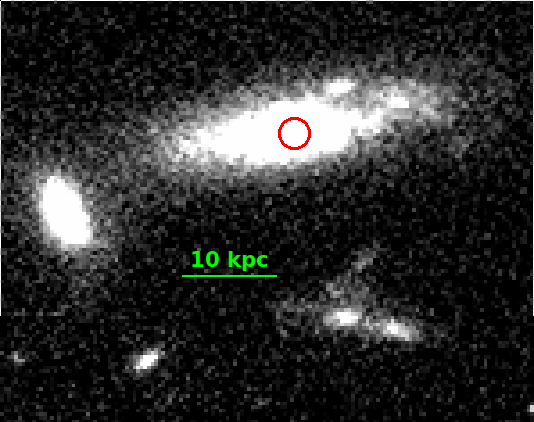}
  \includegraphics[width=5cm]{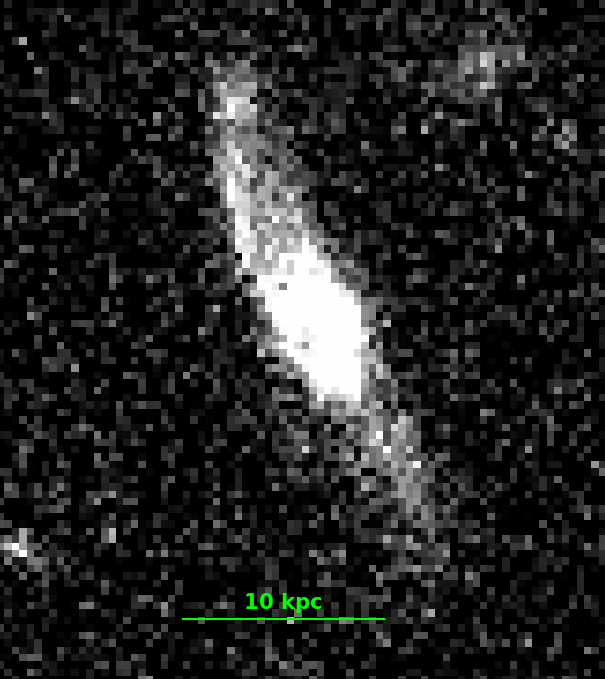}
  \includegraphics[width=5cm]{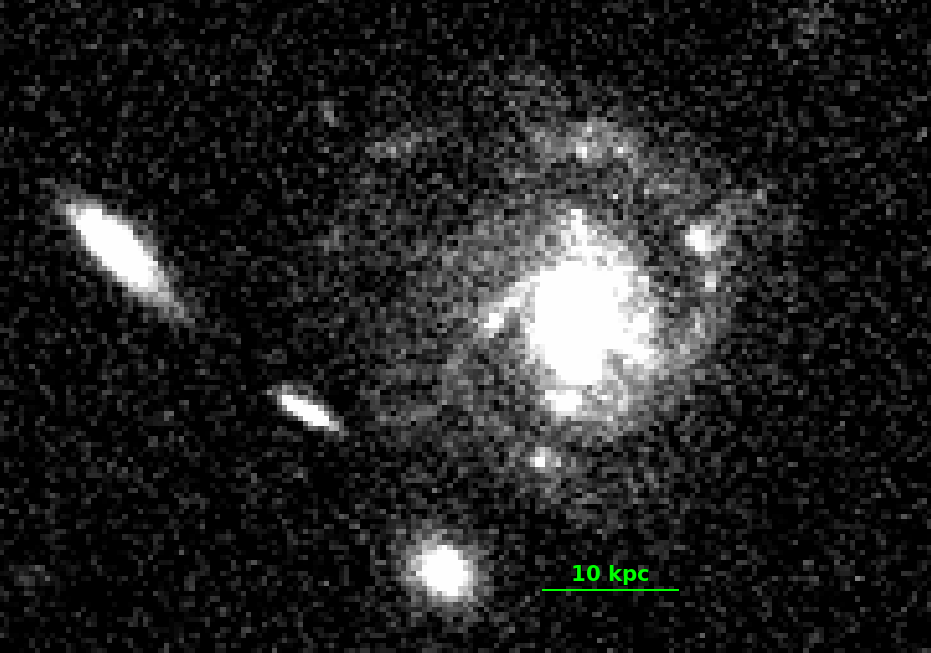}
  \includegraphics[width=5cm]{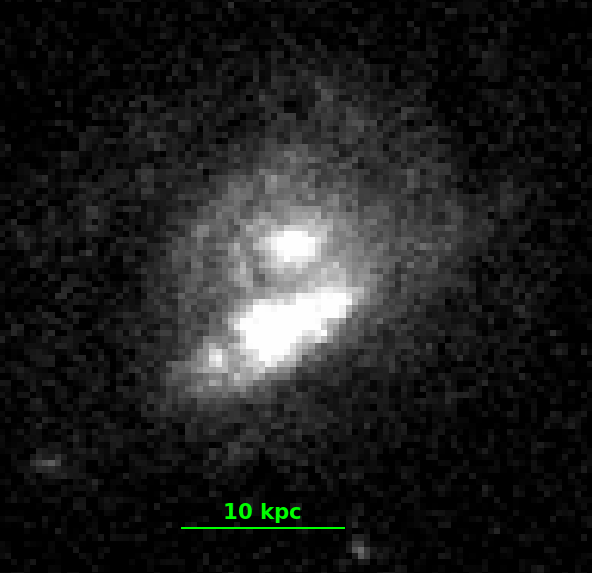}
  \includegraphics[width=5cm]{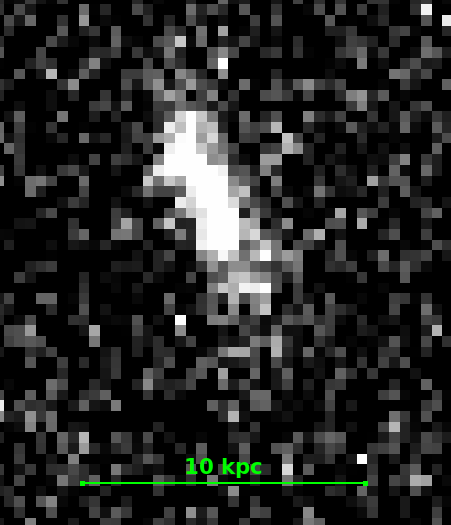}
  \includegraphics[width=5cm]{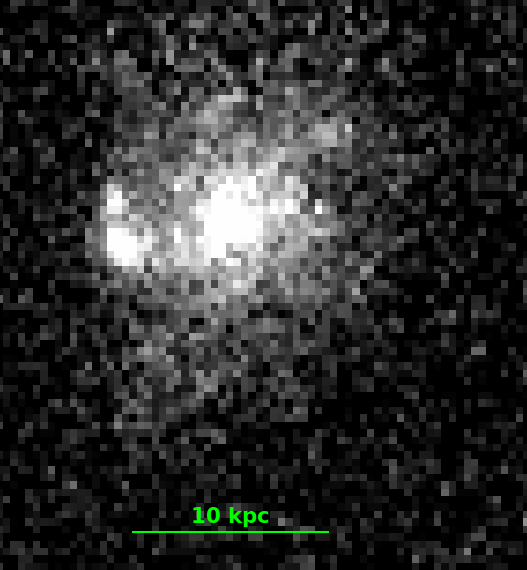}
\end{center}
\caption{LCDCS0853 (z=0.7627). All galaxies are in F814W.  Row 1:
  galaxies a (red circle), b (red circle), and c. Row 2: galaxies d,
  e, and f. Row 3: galaxy g.}
\label{fig:lcdcs0853}
\end{figure*}

\begin{figure*}[h]
\begin{center}
  \includegraphics[width=8cm]{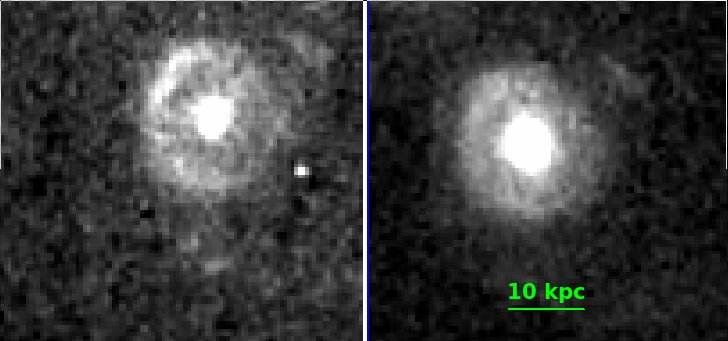}
  \includegraphics[width=5cm]{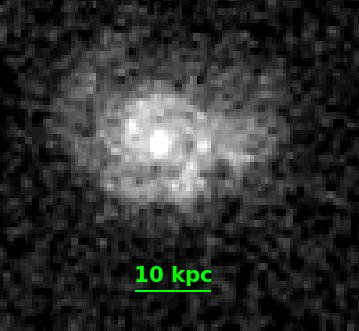}
  \includegraphics[width=8cm]{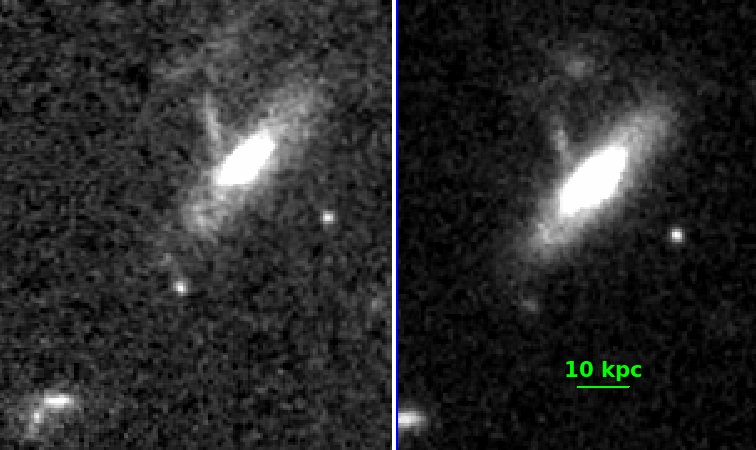}
  \includegraphics[width=8cm]{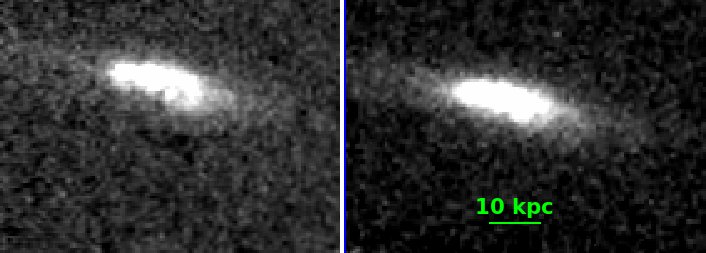}
  \includegraphics[width=5cm]{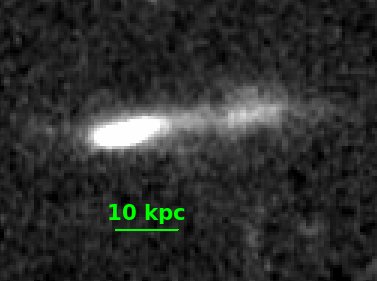} \kern0.1cm%
  \includegraphics[width=5cm]{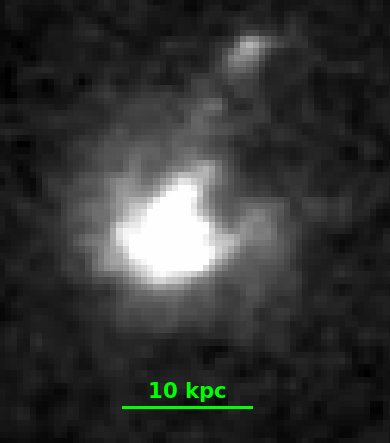}  
  \includegraphics[width=5cm]{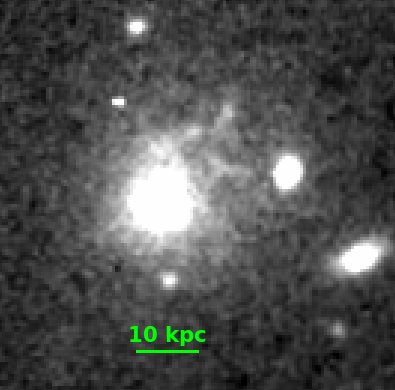} \kern0.1cm%
  \includegraphics[width=5cm]{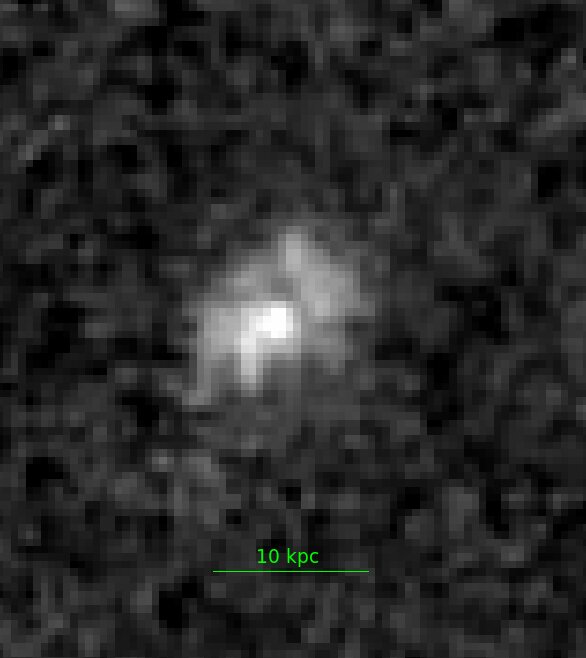}  
\end{center}
\caption{3C295 (z=0.46). Row 1: galaxy a in F555W and F814W, and
  galaxy b in F814W. Row 2: galaxies c and d in F555W and F814W. Row 3:
  galaxies e, f, and g in F555W. Row 4: galaxy h in F555W.}
\label{fig:3c295}
\end{figure*}

\begin{figure*}[h]
\begin{center}
  \includegraphics[width=8cm]{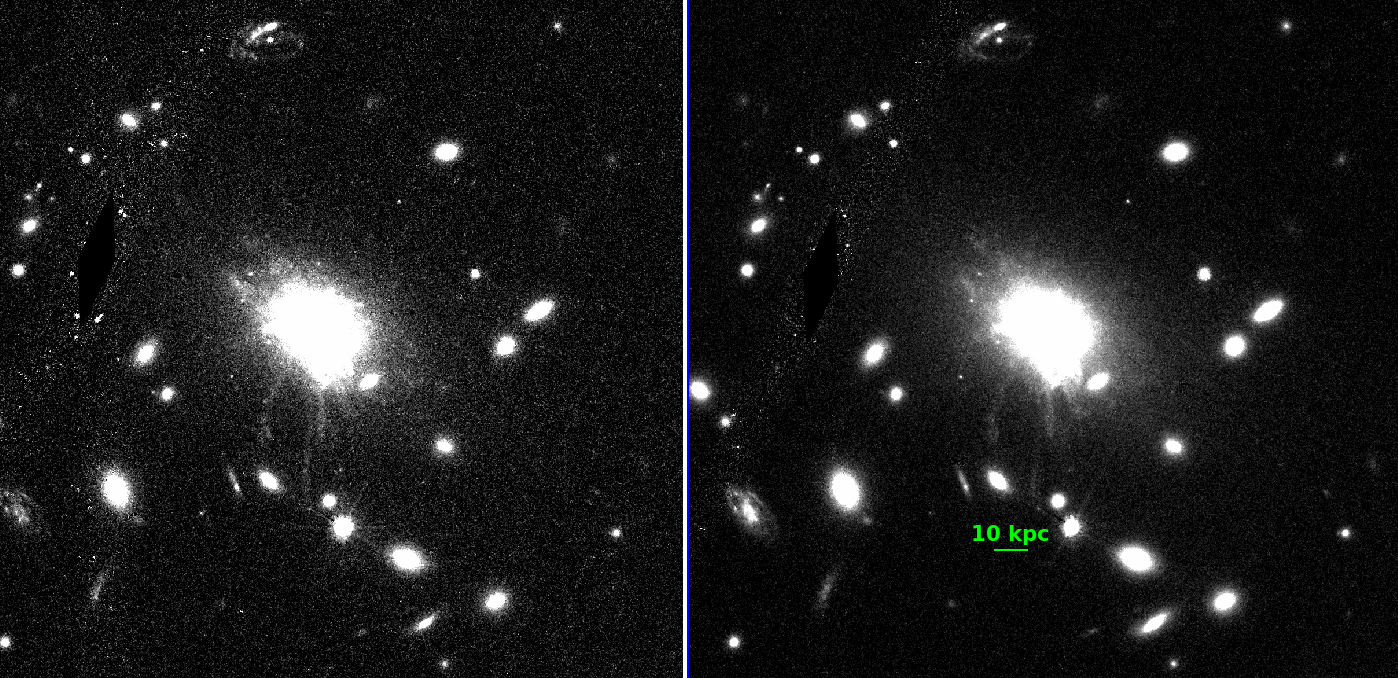}
  \includegraphics[width=8cm]{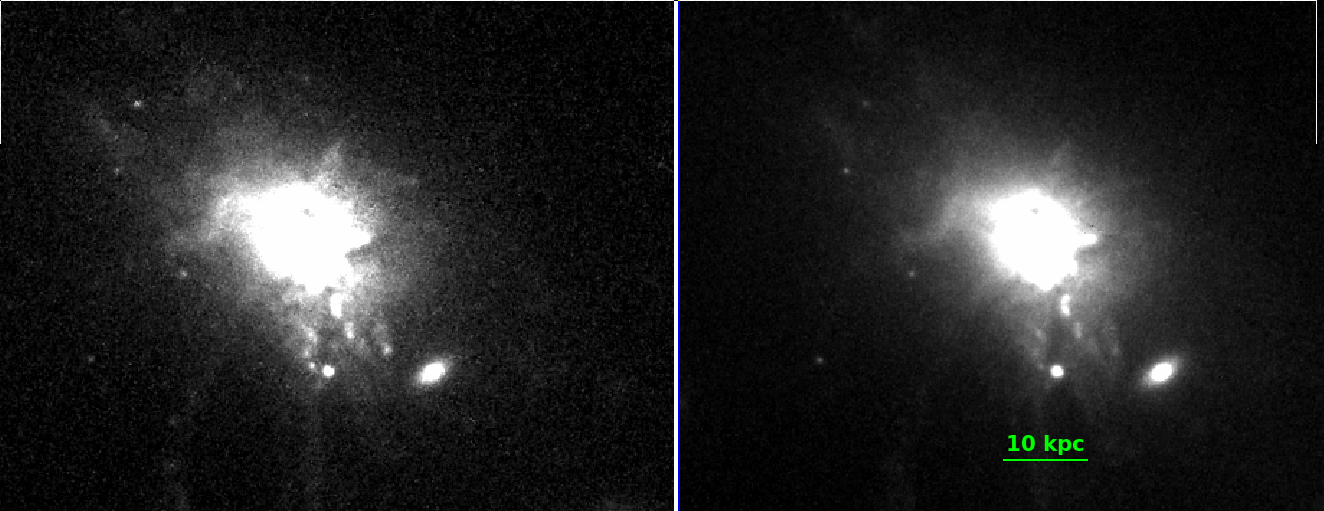}
\end{center}
\caption{RX1532 (z=0.345). Images are in F606W and F814W. Left:
  galaxy a, which is the BCG, showing filaments reminiscent of the
  Perseus cluster BCG.  Right: zoom on galaxy a.}
\label{fig:rx1532}
\end{figure*}

\begin{figure*}[h]
\begin{center}
  \includegraphics[width=5cm]{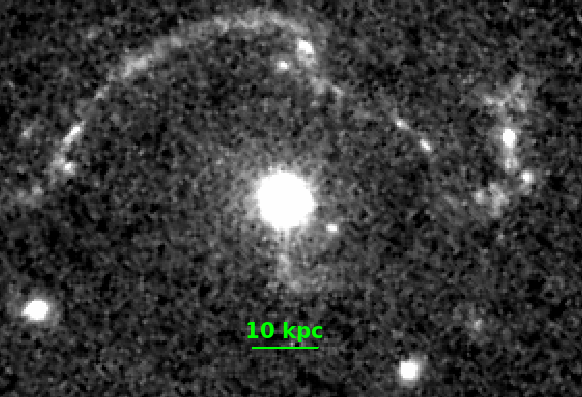}
  \includegraphics[width=5cm]{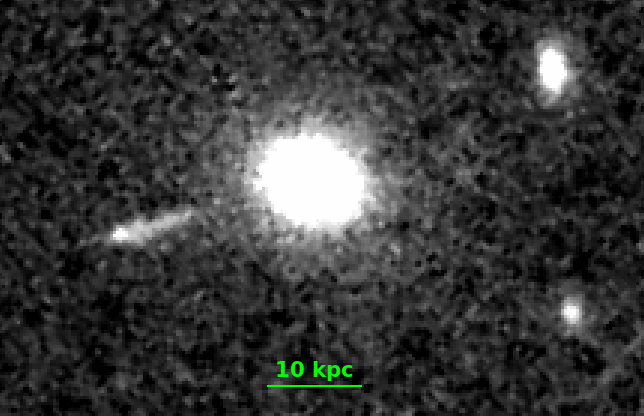} \kern0.1cm%
  \includegraphics[width=5cm]{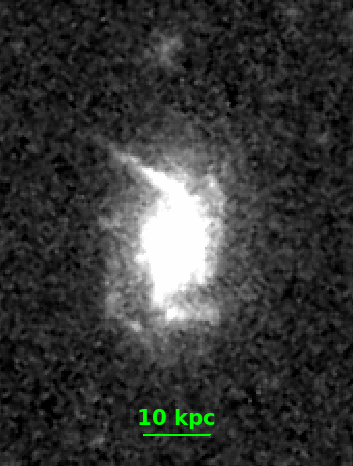}
\end{center}
\caption{MS1621 (z=0.426). Galaxies a, b, and c in F814W.}
\label{fig:ms1621}
\end{figure*}

\begin{figure*}[h]
\begin{center}
  \includegraphics[width=5cm]{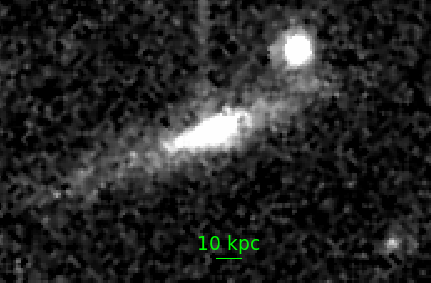}
\end{center}
\caption{RX1716 (z=0.813), galaxy a in F814W.}
\label{fig:rx1716}
\end{figure*}

\begin{figure*}[h]
\begin{center}
  \includegraphics[width=8cm]{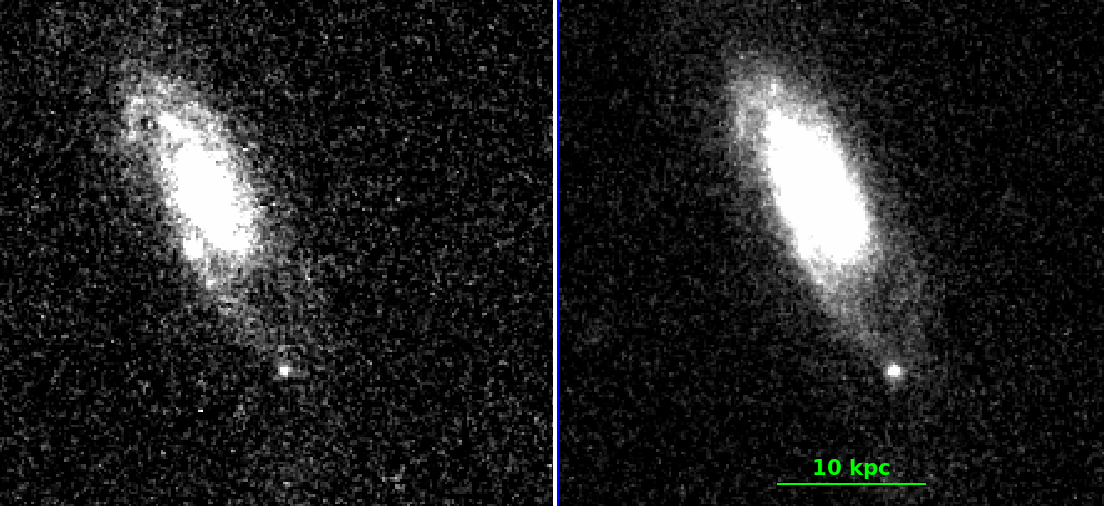}
  \includegraphics[width=8cm]{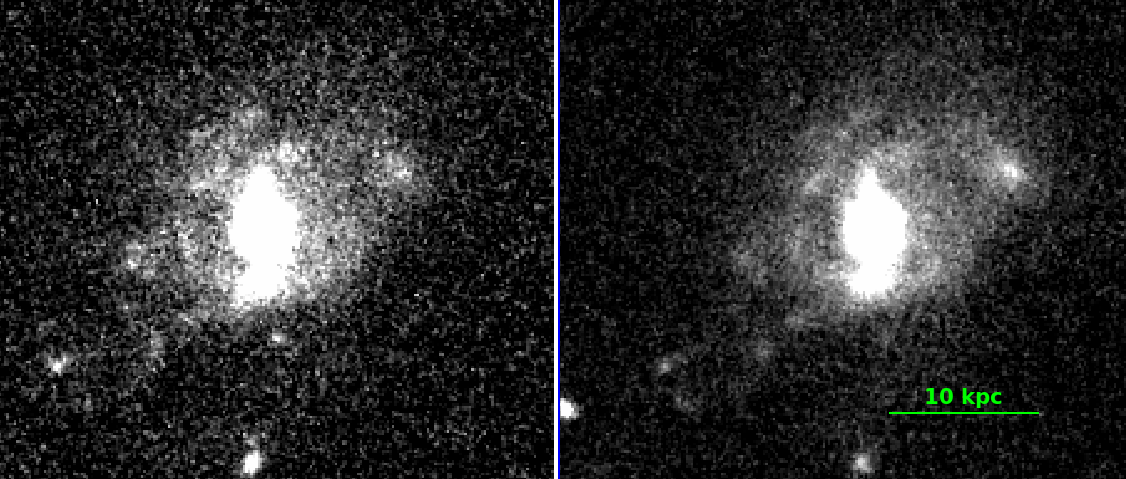}
\end{center}
\caption{MACS1931 (z=0.352). Images are in F606W and F814W.
  Left: galaxy a, right: galaxy b.}
\label{fig:macs1931}
\end{figure*}

\begin{figure*}[h]
\begin{center}
  \includegraphics[width=8cm]{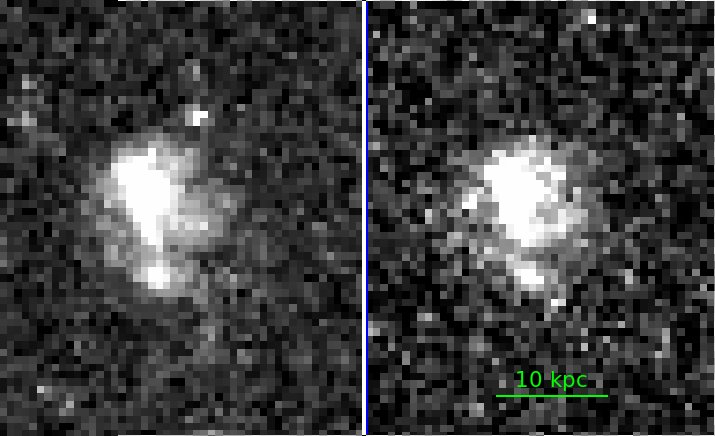}
\end{center}
\caption{MS~2053.7-0449 (z=0.583). Images of galaxy a in F606W (left) and F814W (right).}
\label{fig:ms2053}
\end{figure*}

\begin{figure*}[h]
\begin{center}
  \includegraphics[width=8cm]{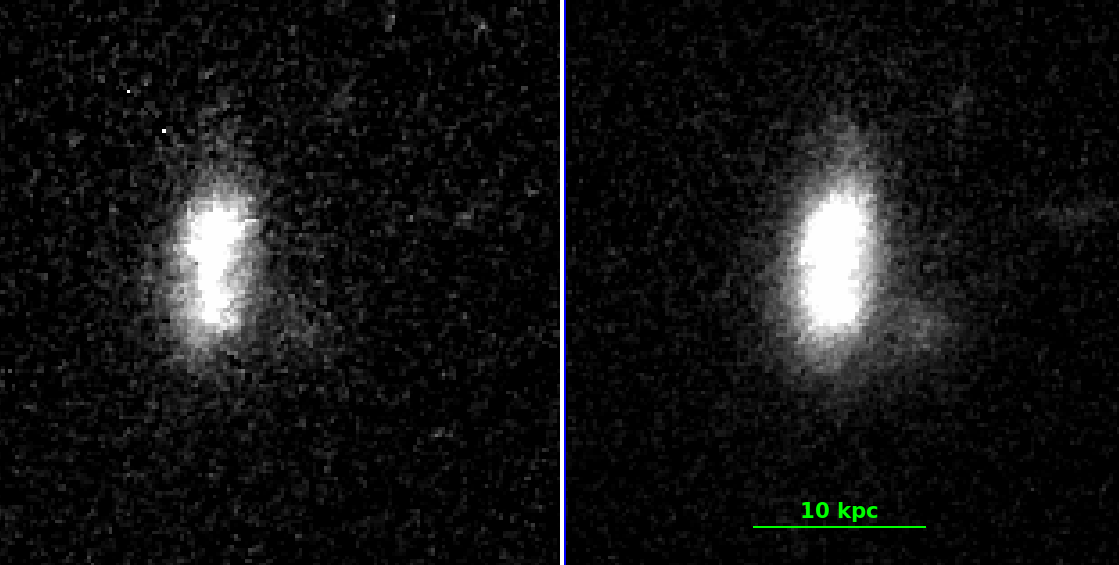}
  \includegraphics[width=8cm]{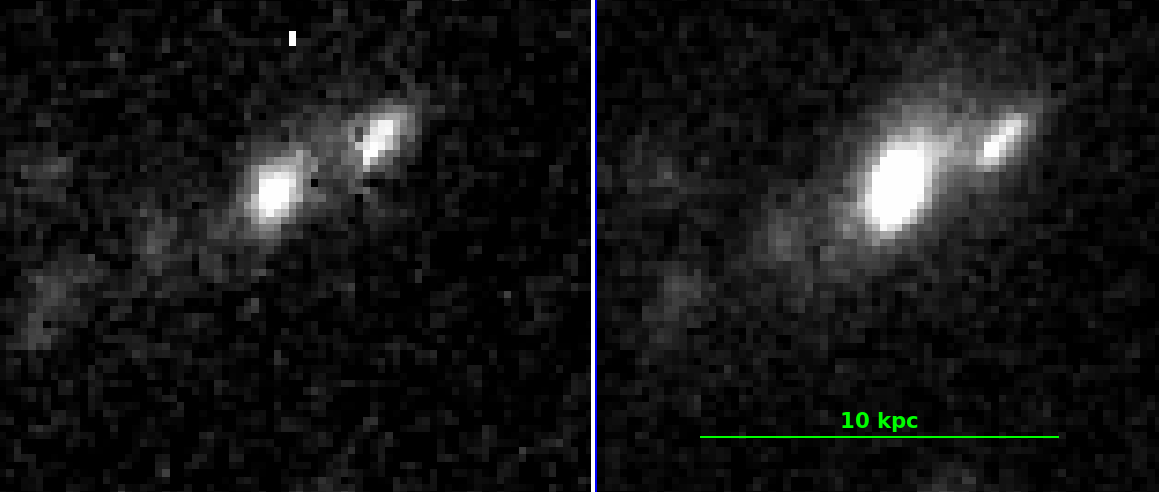}
  \includegraphics[width=8cm]{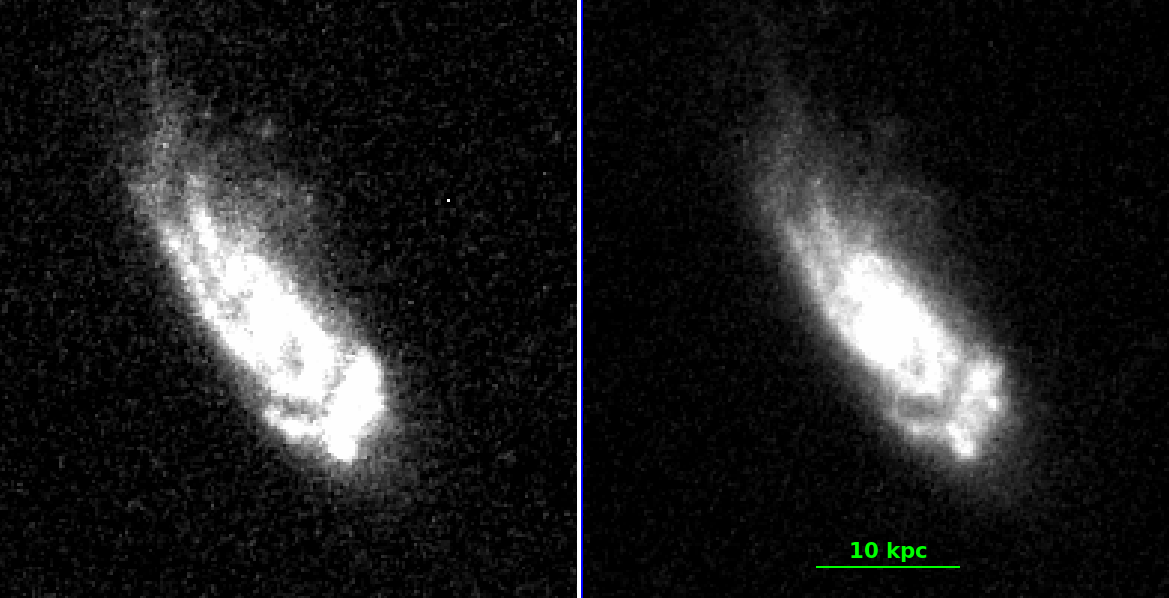}
  \includegraphics[width=8cm]{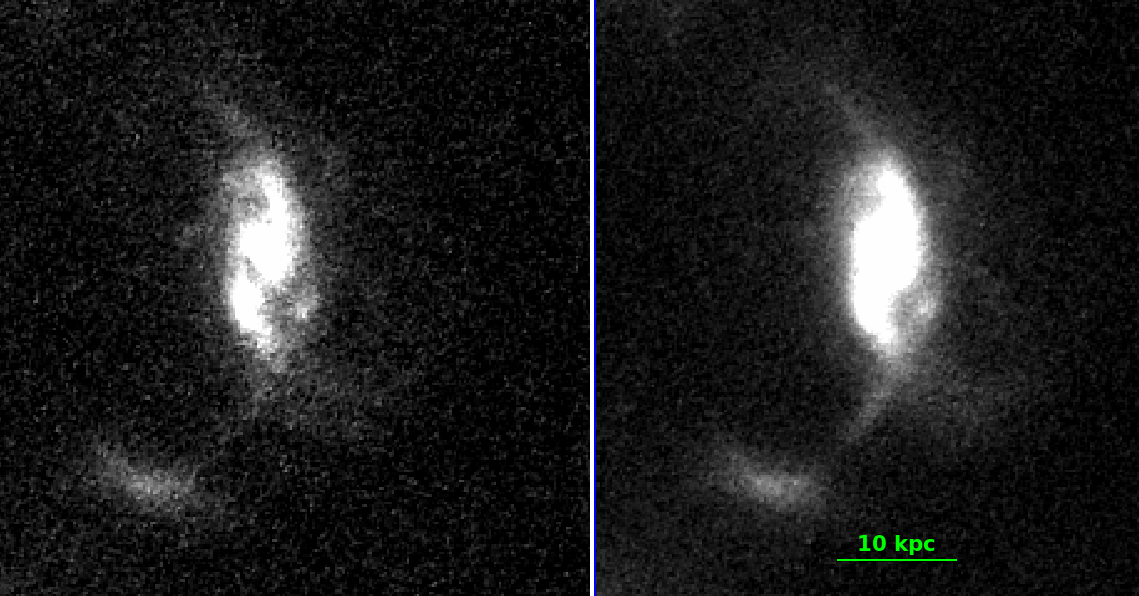}
  \includegraphics[width=8cm]{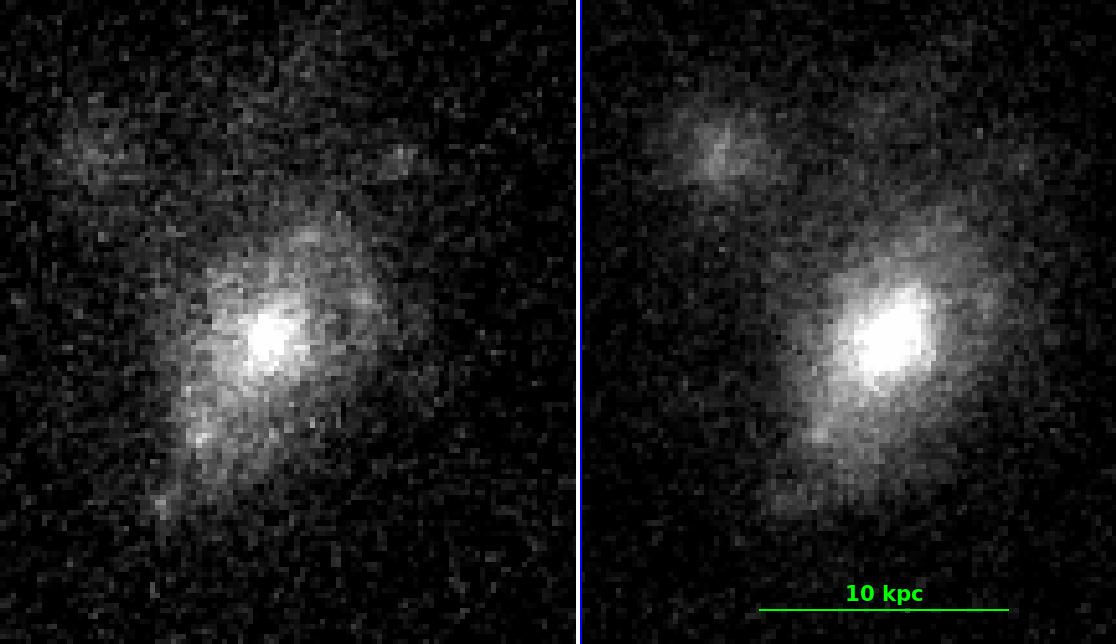}
\end{center}
\caption{RX2248 (z=0.348). Images are in F606W and F814W.
  Row 1: galaxies a and b. Row 2: galaxies c and d. Row 3: galaxy e.}
\label{fig:rx2248}
\end{figure*}

\end{appendix}

\end{document}